\documentclass[useAMS,referee,usenatbib]{mn2e}
\usepackage{epsfig}
\usepackage{graphicx}

\def\jnl@style{\rm}
\def\aaref@jnl#1{{\jnl@style#1}}
\def\aj{\aaref@jnl{AJ}}                   % Astronomical Journal
\def\araa{\aaref@jnl{ARA\&A}}             % Annual Review of Astron and Astrophys
\def\apj{\aaref@jnl{ApJ}}                 % Astrophysical Journal
\def\apjl{\aaref@jnl{ApJ}}                % Astrophysical Journal, Letters
\def\apjs{\aaref@jnl{ApJS}}               % Astrophysical Journal, Supplement
\def\ao{\aaref@jnl{Appl.~Opt.}}           % Applied Optics
\def\apss{\aaref@jnl{Ap\&SS}}             % Astrophysics and Space Science
\def\aap{\aaref@jnl{A\&A}}                % Astronomy and Astrophysics
\def\aapr{\aaref@jnl{A\&A~Rev.}}          % Astronomy and Astrophysics Reviews
\def\aaps{\aaref@jnl{A\&AS}}              % Astronomy and Astrophysics, Supplement
\def\azh{\aaref@jnl{AZh}}                 % Astronomicheskii Zhurnal
\def\baas{\aaref@jnl{BAAS}}               % Bulletin of the AAS
\def\jrasc{\aaref@jnl{JRASC}}             % Journal of the RAS of Canada
\def\memras{\aaref@jnl{MmRAS}}            % Memoirs of the RAS
\def\mnras{\aaref@jnl{MNRAS}}             % Monthly Notices of the RAS
\def\pra{\aaref@jnl{Phys.~Rev.~A}}        % Physical Review A: General Physics
\def\prb{\aaref@jnl{Phys.~Rev.~B}}        % Physical Review B: Solid State
\def\prc{\aaref@jnl{Phys.~Rev.~C}}        % Physical Review C
\def\prd{\aaref@jnl{Phys.~Rev.~D}}        % Physical Review D
\def\pre{\aaref@jnl{Phys.~Rev.~E}}        % Physical Review E
\def\prl{\aaref@jnl{Phys.~Rev.~Lett.}}    % Physical Review Letters
\def\pasp{\aaref@jnl{PASP}}               % Publications of the ASP
\def\pasj{\aaref@jnl{PASJ}}               % Publications of the ASJ
\def\qjras{\aaref@jnl{QJRAS}}             % Quarterly Journal of the RAS
\def\skytel{\aaref@jnl{S\&T}}             % Sky and Telescope
\def\solphys{\aaref@jnl{Sol.~Phys.}}      % Solar Physics
\def\sovast{\aaref@jnl{Soviet~Ast.}}      % Soviet Astronomy
\def\ssr{\aaref@jnl{Space~Sci.~Rev.}}     % Space Science Reviews
\def\zap{\aaref@jnl{ZAp}}                 % Zeitschrift fuer Astrophysik
\def\nat{\aaref@jnl{Nature}}              % Nature
\def\iaucirc{\aaref@jnl{IAU~Circ.}}       % IAU Cirulars
\def\aplett{\aaref@jnl{Astrophys.~Lett.}} % Astrophysics Letters
\def\apspr{\aaref@jnl{Astrophys.~Space~Phys.~Res.}}
\def\bain{\aaref@jnl{Bull.~Astron.~Inst.~Netherlands}}
\def\fcp{\aaref@jnl{Fund.~Cosmic~Phys.}}  % Fundamental Cosmic Physics
\def\gca{\aaref@jnl{Geochim.~Cosmochim.~Acta}}   % Geochimica Cosmochimica Acta
\def\grl{\aaref@jnl{Geophys.~Res.~Lett.}} % Geophysics Research Letters
\def\jcp{\aaref@jnl{J.~Chem.~Phys.}}      % Journal of Chemical Physics
\def\jgr{\aaref@jnl{J.~Geophys.~Res.}}    % Journal of Geophysics Research
\def\jqsrt{\aaref@jnl{J.~Quant.~Spec.~Radiat.~Transf.}}
\def\memsai{\aaref@jnl{Mem.~Soc.~Astron.~Italiana}}
\def\nphysa{\aaref@jnl{Nucl.~Phys.~A}}   % Nuclear Physics A
\def\physrep{\aaref@jnl{Phys.~Rep.}}   % Physics Reports
\def\physscr{\aaref@jnl{Phys.~Scr}}   % Physica Scripta
\def\planss{\aaref@jnl{Planet.~Space~Sci.}}   % Planetary Space Science
\def\procspie{\aaref@jnl{Proc.~SPIE}}   % Proceedings of the SPIE

\title[Spectroscopic Binaries from the AAPS]{The Observed Distribution of Spectroscopic Binaries from the Anglo-Australian Planet Search}
\author[J.S. Jenkins et al.]
       {J.S. Jenkins,$^1$\thanks{E-mail: jjenkins@das.uchile.cl}
        M. D\'iaz,$^1$ H.\,R.\,A. Jones,$^2$ R.\,P. Butler,$^3$ C.\,G. Tinney,$^{4,5}$
  \newauthor% starts a new line in the
S.\,J. O'Toole,$^6$ B.\,D. Carter,$^7$ R.\,A. Wittenmyer,$^{4,5}$
D.\,J. Pinfield$^2$
   \\\\
  $^1$Departamento de Astronom\'ia, Universidad de Chile, Casilla 36-D, Las Condes, Santiago, Chile\\
  $^2$Centre for Astrophysics Research, University of Hertfordshire, College Lane, Hatfield, Herts, AL10 9AB, UK\\
  $^3$Carnegie Institution of Washington, DTM, 5241 Broad Branch Road NW, Washington DC, 20015-1305, USA\\
  $^4$Exoplanetary Science at UNSW, School of Physics, UNSW Australia, Sydney, NSW 2052, Australia\\
  $^5$Australian Centre for Astrobiology, UNSW Australia, Sydney, NSW 2052, Australia\\
  $^6$Australian Astronomical Observatory, PO Box 915, North Ryde 1670, Australia\\
  $^7$Computational Engineering and Science Research Centre, University of Southern Queensland, Springfield QLD 4300, Australia}
\pagerange{\pageref{firstpage}--\pageref{lastpage}}
\begin{document}
\maketitle
\label{firstpage}

\begin{abstract}
We report the detection of sixteen binary systems from the
Anglo-Australian Planet Search. Solutions to the radial velocity data
indicate that the stars have companions orbiting with a wide range of
masses, eccentricities and periods. Three of the systems potentially
contain brown-dwarf companions while another two have eccentricities
that place them in the extreme upper tail of the eccentricity
distribution for binaries with periods less than 1000\,d. For 
periods up to 12\,years, the distribution of our stellar companion masses is
fairly flat, mirroring that seen in other radial velocity surveys, and
contrasts sharply with the current distribution of candidate planetary masses,
which rises strongly below 10\,M$_{\rm J}$.   When looking at a larger sample of  
binaries that have FGK star primaries as a function of the primary star metallicity, we 
find that the distribution maintains a binary fraction of $\sim$43$\pm$4\% between -1.0 to +0.6~dex in metallicity.  
This is in stark contrast to the giant exoplanet distribution.  
This result is in good agreement with binary formation models that 
invoke fragmentation of a collapsing giant molecular cloud, suggesting 
this is the dominant formation mechanism for close binaries and not fragmentation 
of the primary star's remnant proto-planetary disk.

\end{abstract}


\section{Introduction}

Binary systems, in their various guises, yield vital measures for a range of fundamental parameters - mass, radius, luminosity - for the component stars. 
Studies of the binary population distribution, correlation of orbital elements, and the frequencies
of the various forms of multiplicity, can be used to shed light on star-formation processes and evolutionary mechanisms. Systems comprising two or more stars are common. Surveys suggest the incidence of multiplicity is around 45\% \citep{raghavan10}, perhaps higher than 70\% among the more massive stars \citep*{abt90,mason98,preib99}, and somewhat lower ($\sim$30-40\%) for M dwarfs \citep{fischer92}.

Doppler searches for extrasolar planets have refined the art of single-lined spectroscopic binary analysis to the point where {\it relative} radial velocities (RVs) of the primaries can be measured with precisions at the $\sim$1 m s$^{-1}$ level using both the absorption cell and spectrograph stablisation methods (e.g. \citealp{vogt10,jenkins13a,jenkins14,wittenmyer14,anglada-escude14}). RV\,measurements that are not tied to an absolute zero point can achieve high internal precision by explicitly removing the need to quantify such effects as convective blue-shift and stellar
gravitational red-shift; a clear account of these and other effects is given in \citet{nidever2002}. The target stars for such planetary searches are generally solar-type and are selected to be
chromospherically `quiet' in order to minimise the potential for `noise' in any velocity measurement due to starspot activity (see \citealp{jenkins06}). They are also selected to have no resolvable companions to avoid flux contamination.

The Anglo-Australian Planet Search (AAPS) selection criteria for its initial group of stars are discussed in \citet{jones_targets}. The sample considered here comprises 178 F, G and K dwarf stars with declinations south of $\sim$\,--20$^\circ$ and is complete to {\it V}\,$<$\,7.5. There is a requirement that the activity seen in Ca\,{\small II}\,H\&K absorption lines has an index (measured by log\,$R'_{\rm{HK}}$ - hereafter referred to as $R'_{\rm{HK}}$; for details see \citealp{jenkins06,jenkins08,jenkins11}) of less than -4.5, and for there to be no known companions within 2~arcsec. The spectroscopic binaries presented in this paper are drawn from this sample.

\section{Observations \& Data Reduction}
\subsection{The Primaries: Stellar Characteristics}

A summary of the characteristics and masses for the primaries are given in Table \ref{primaries}. Spectral types, {\it B-V} colours, magnitudes, and parallaxes for all the stars are taken from the Simbad and HIPPARCOS databases.  Metallicities for the stars in this sample are drawn from two sources; spectroscopic metallicities were extracted from \citet{bond06} and photometric values were taken from the catalogue of \citet{casagrande11}.  For five of the binary stars we report there are no Bond et al. spectroscopic values, however, these have Casagrande et al. metallicities.  In fact, all but one of the primaries, HD39213, have Casagrande et al. measurements, allowing for a uniform and self-consistent set of [Fe/H] estimates to be generated for the sample. 

$R'_{\rm{HK}}$ data are used with the activity-age relation given in \citet*{soderblom91} to provide secondary age estimates for the stars.  Note that since our stars were preselected to have $R'_{\rm{HK}}$ values below -4.5, the Soderblom et al. relations are quantitatively the same in this regime to the updated work of \citet{mamajek08} due to the sparse activity-age data for older dwarf stars. The activity values have been drawn from the studies of \citet{henrytHK}, \citet{tinney02}, \citet{jenkins06}, and \citet{jenkins11}, yet even considering these four works, there are still three stars with unknown activities, highlighting the lack of chromospheric activity studies in the southern hemisphere compared to the north. The primary stellar ages, along with the stellar masses, are determined through interpolation of the Yonsei-Yale isochrones \citep{yi2001} and uncertainties can be found to reach 100\% for the ages of these types of old and Sun-like dwarf stars. Given the $\sim$0.06-0.10\,dex uncertainty in metallicities, natural variations in stellar activity (for example solar $R'_{\rm{HK}}$ activity variation between --4.75 and --5.10 translates to an age variation from 2.2 and 8.0\,Gyr -- \citeauthor{henrytHK}), uncertainty in the precise form of the age-activity relationship, along with the possibility of flux contamination from the secondary, a number of isochronal mass/age/metallicity points can equally account for a star's colour and magnitude.  In fact, an offset is found between ages derived from the activity indices and those measured from isochrone fitting, whereby the activity derived ages are generally significantly younger than those measured from fitting the isochrones.  This result highlights that more work is needed to make ages derived from stellar activity relations, or gyrochronology, and those derived from evolutionary models, more consistent for old field stars.   A consideration of these uncertainties enables us to determine a consistent mass range for each star.

\begin{table*}
\centering
 \begin{minipage}{140mm}
  \caption{The Primaries: Stellar Characteristics}
  \begin{tabular}{@{}cllrcccrrrc@{}}
  \\
  \hline
   &Star & {\it B-V} & {\it V}& Parallax & {\it M$_{V}$}&Spectral&[Fe/H]&[Fe/H]& log$R'_{\rm{HK}}$& Mass\\
  && & &  (mas) & & Type&Casa & Bond && (M$_{\sun}$)\\
  \hline
1&HD18907&0.79&5.9&31.1&3.4&K2V&-0.46&-0.50$\pm$0.07&-5.11&1.05 $\pm$ 0.15\\
2&HD25874&0.67&6.7&38.6&4.6&G2V&-0.02&-&-4.95&1.00 $\pm$ 0.05\\
3&HD26491&0.64&6.4&42.3&4.5&G1V&-0.11&-0.08$\pm$0.07&-4.95&0.97 $\pm$ 0.05\\
4&HD39213&0.81&9.0&16.3&5.1&K0V&-&0.20$\pm$0.07&-5.10&0.93 $\pm$ 0.05\\
5&HD42024&0.55&7.2&18.2&3.5&F7V&0.19&-&-&1.30 $\pm$ 0.05\\
6&HD64184&0.68&7.5&30.0&4.9&G3V&-0.18&-0.23$\pm$0.07&-4.88&0.93 $\pm$ 0.05\\
7&HD120690&0.70&6.4&51.4&5.0&G5+V&-0.08&-0.10$\pm$0.06&-4.78&0.98 $\pm$ 0.05\\
8&HD121384&0.78&6.0&25.8&3.1&G8V&-0.39&-0.40$\pm$0.07&-5.22&0.98 $\pm$ 0.10\\
9&HD131923&0.71&6.3&41.9&4.4&G4V&0.06&-0.05$\pm$0.08&-4.90&1.05 $\pm$ 0.05\\
10&HD145825&0.65&6.6&46.4&4.9&G3V&0.12&-0.04$\pm$0.07&-4.74&1.03 $\pm$ 0.05\\
11&HD150248&0.65&7.0&37.5&4.9&G3V&-0.13&-0.11$\pm$0.07&-4.88&0.93 $\pm$ 0.05\\
12&HD156274B&0.76&5.5&113.6&5.8&G8V&-0.40&-&-4.95&0.83 $\pm$ 0.06\\
13&HD158783&0.67&7.1&23.7&4.0&G4V&0.05&-0.05$\pm$0.07&-4.91&1.04 $\pm$ 0.05\\
14&HD162255&0.66&7.2&24.9&4.2&G3V&0.17&-0.01$\pm$0.08&-&1.12 $\pm$ 0.08\\
15&HD169586&0.55&6.8&21.4&3.5&G0V&0.32&-&-4.92&1.25 $\pm$ 0.05\\
16&HD175345&0.57&7.4&21.3&4.0&G0V&-0.16&-&-&1.05 $\pm$ 0.05\\
\hline \label{primaries}
\end{tabular}
\end{minipage}
\end{table*}

\subsection[]{Determination of Radial Velocities}

\subsubsection[]{UCLES Data}

Observations were made at the 3.9m Anglo-Australian Telescope using the University College London Echelle Spectrograph (UCLES), operated in its 31 lines/mm mode. High precision Doppler measurements are made possible by the use of an iodine absorption cell that permits detailed modeling of the spectrograph Point Spread Function (PSF). The reader is referred to Butler et al. (1996; 2001) for a detailed description, however, the procedure is outlined below.

Multi-epoch spectra at a resolution of {\it R} $\sim$\,45,000 are obtained for each star with the I$_{2}$ cell mounted behind the UCLES, imprinting the stellar spectra with thousands of iodine absorption lines in the 5000-6200~\AA\, region. Each spectrum can be synthesised from a product of a Doppler-shifted copy of a pure stellar spectrum for the star in question (a higher resolution stellar {\it template} spectrum from which the spectrograph PSF has been removed) and an iodine absorption spectrum, all of which is convolved with the spectrograph's PSF at the time of the observation. A least squares fitting process matches this synthetic spectrum with the observed spectrum, and determines up to 14 free parameters (one being the Doppler shift, one the wavelength dispersion, and the remainder associated with the detailed PSF profile). This fitting process is carried out on 2\,\AA\ chunks of the spectrum between 5000-6200\,\AA\, and the resulting velocities are weighted by the gradient ($\partial$F/$\partial\lambda$) of the spectral profile for each chunk. The mean of these weighted velocities, corrected for the Earth's motion relative to the Solar System barycentre \citep{chrisMS}, represents the RV for that observation.  A barycentric correction is also applied to the Julian dates.  The internal uncertainty is obtained from the standard deviation of the velocities. This technique has demonstrated consistently that 3 ms$^{-1}$ precision is achievable down to the {\it V}=7.5 magnitude limit of the survey for suitably inactive stars over the long term \citep{jones_precision}. The barycentric Julian dates and RV data of our sources are given in Table\,\ref{rvdata}.

\subsubsection[]{HARPS Data}

In order to supplement the velocities measured using the UCLES spectrograph, we performed a search of the ESO Archive Facility to determine if any of 
these stars had high precision ESO-HARPS measurements that could be used to increase the phase coverage of our orbits.  The search revealed that six of 
targets had been observed with HARPS multiple times, such that the inclusion of the velocities yielded much better constraints on the binary orbits.  The 
HARPS RV data is also shown in Table\,\ref{rvdata}.

At this point it is worth briefly discussing the HARPS strategy for measuring precision RVs from high resolution and high S/N echelle spectra.  The HARPS 
data is automatically processed by the HARPS-DRS version 3.5, with the reduced and analysis quality data on the Advanced Data Products page of the 
ESO Archive website.  The actual reduction and analysis method itself is based in general on the procedure explained in \citet{baranne96}.  Unlike the method 
employed by the AAPS using UCLES, no iodine cell is used by HARPS, but instead, precision RVs are measured by maintaining the highest stability possible 
over the long term, but placing the spectrograph in a vaccuum tank to maintain the pressure and temperature as stable as possible, and feeding the light 
to the spectrograph using optical fibres.  

The actual RV measurements are not performed in chunks like they are using UCLES, but each entire echelle order is used to measure the RV.  A weighted binary 
mask is constructed that synthetically mimics the position of an absorption line in the star (\citealp{pepe02}), and a weighted cross-correlation between 
the stellar spectrum and the binary mask gives the RV.  The mean of the RV for each order gives rise to the final absolute RV measurement from the star, with 
the uncertainty measured following the procedure in \citet{bouchy01}.  The stability of the spectrograph is maintained by using a calibration Thorium-Argon 
lamp, that is also simultaneously fed to the spectrograph using another fibre, allowing any drifts in the wavelength solution to be measured at the 0.1 ms$^{-1}$ 
level (\citealp{lovis07}).  The drift is then removed from the measured RV to get the most precise value, however a stability of less than 1 ms$^{-1}$ 
has been found for HARPS data over the long term (e.g. \citealp{locurto10}).

\subsection{Orbital Parameters}
For the analysis of spectroscopic binaries the task is to provide a set of orbital parameters (period, {\it P}, eccentricity, {\it e}, periastron angle, $\omega$, time of periastron passage, {\it T$_{p}$}, and the projected semi-major axis of the primary, {\it a$_{p}$\,$\sin{i}$}) - and a velocity offset, $\dot{z}_{o}$, that optimise the fit of the equation

\begin{equation}
    \dot{z}={2\pi
a_{p}\sin{i}\over{P\sqrt{1-e^{2}}}}(\cos(\upsilon+\omega)+e\cos\omega)+\dot{z}_{o}
\end{equation}

to the $n$ observations of line-of sight radial velocity, $\dot{z}$, at true anomalies $\upsilon$ (derived from the observed times, {\it t}, through iteration of  Kepler's equation). A least-squares minimisation procedure, invoking several IDL routines, is used to fit the equation.

In the search for an orbital solution, periods are initially identified via Lomb-Scargle periodogram analysis \citep{lomb,scargle}. Orbital solutions are plotted in Fig.\,\ref{rvcurve} and summarized in Table\,\ref{orbit} (i \& ii). Where the RV data have a monotonic variation, or only one extremum occurs without any clear inflection in the RV variation to constrain a second extremum (HD\,18907, HD\,25874, HD\,26491, HD\,131923, HD\,156274B), convergence is reached for a number of different periods. In these cases, where the period is clearly greater than the duration of the observations, and where periodogram analysis is least effective, we consider only the minimum orbital period in our solutions.
Periodograms are shown (Fig.\,\ref{pgram}) for all the targets where the RV phase coverage is nearly a cycle or more. Where the sampling is sparse, aliasing introduces spectral power over a range of frequencies and is particularly marked for HD\,64184 which has just eight RV observations and for HD\,121384 due to the highly eccentric nature of the orbit (see \citealp{otoole09}). This is perhaps a reminder that data sets should comprise more than a dozen observations for the periodogram technique to be properly effective \citep{horne86}. The period inferred from the orbital solution in each case is marked with a vertical dashed line. Identification of periods is least effective for HD\,131923 and HD\,156274B, where the phase coverage is less than one cycle.  Nevertheless, the inflections in the RV variation enable a robust Keplerian period to be found.

\begin{table*}
\centering
 \begin{minipage}{170mm}
 \caption{\textbf{(i)} Orbital Parameters 1: Where the RV data extend over more than one cycle or a clear inflection in the RV variation is seen to be able to constrain a second extremum, a single set of orbital parameters emerge and are listed  below. The quantity {\it a\,$\sin{i}$} represents the semimajor axis of the binary system.}
\scalebox{0.82}{
  \begin{tabular}{@{}lccccccccc@{}}
  \\
  \hline
   Star & {\it P}&$\omega$ & {\it e} &{\it T$_{p}$}&{\it K}&
   {\it a} & $\chi^{2}_{\upsilon}$ & rms\\
& days & deg & &JD-&m/s&au & & m/s\\ &&&&2450000\\
  \hline
\\
HD39213            & 1309 $\pm$ 159  & 265 $\pm$ 23    & 0.2 $\pm$ 0.1          & 1636.7 $\pm$ 2.8       & 1265 $\pm$ 273     & 2.3 $\pm$ 0.3       & 1.97        & 17.1 \\
HD42024            & 76.26 $\pm$ 0.03          & 347 $\pm$ 87    & 0.19$\pm$ 0.01           & 1154.8 $\pm$ 0.7       & 3228 $\pm$  133    & 0.38 $\pm$ 0.01   & 3.22            & 11.8 \\
HD64184             & 17.863 $\pm$ 0.002     & 184 $\pm$ 84     & 0.249 $\pm$ 0.004     & 949.50 $\pm$ 0.05         & 12910 $\pm$ 439   & 0.130 $\pm$ 0.002   &  0.93   &     2.39  \\
HD120690          & 3800 $\pm$ 18      & 339 $\pm$ 55     & 0.34 $\pm$ 0.01         & 1140 $\pm$ 11     & 6325 $\pm$ 100       &  4.73 $\pm$ 0.02        & 27.7      & 15.2  \\
HD121384           & 178.7 $\pm$ 0.1       & 182 $\pm$ 28     & 0.84 $\pm$ 0.01         & 852 $\pm$ 1         &10887 $\pm$ 2201  & 0.61 $\pm$ 0.02    & 7.19          & 15.1  \\
HD150248           & 3272  $\pm$ 29  & 356 $\pm$ 68       & 0.67 $\pm$ 0.04           & 2365 $\pm$ 13     & 1995 $\pm$ 12       & 4.36 $\pm$ 0.01   &   3.19         &  3.55 \\
HD158783          & 4535 $\pm$ 225  & 170 $\pm$ 46    &  0.05 $\pm$ 0.05         & 986 $\pm$ 27       & 2133 $\pm$ 527               & 5 $\pm$ 1  &   3.03          & 6.62  \\
HD162255            & 48 $\pm$ 1        & 51 $\pm$ 5        & 0.06 $\pm$ 0.01          & 1017.1 $\pm$ 0.1       & 16223 $\pm$ 3419 &  0.27 $\pm$ 0.01 & 2.49            & 7.81 \\
HD169586           &  2935$\pm$ 119 & 59 $\pm$ 10       & 0.4 $\pm$ 0.1           & 2653 $\pm$ 22     & 7857 $\pm$ 2708   & 4.3 $\pm$ 0.2      &   11.5      & 60.2  \\
HD175345          & 312.4 $\pm$ 0.1        & 277 $\pm$ 14   & 0.75 $\pm$ 0.05         & 1256 $\pm$ 11     & 16099 $\pm$ 3675 & 0.92 $\pm$ 0.01  &  3.44    &    13.9 \\
\hline
\label{orbit}
\end{tabular}
}
\end{minipage}
\end{table*}
\setcounter{table}{1}
\begin{table*}
 \begin{minipage}{170mm}
 \small
  \caption{\textbf{(ii)} Orbital Parameters 2: Where the RV data have a monotonic variation, or only one extremum occurs without any clear inflection in the RV variation to constrain a second extremum, a number of solutions emerge at different periods. For each system we list the orbital parameters for the ``best fitting'' solution having the shortest period. The associated uncertainties are for this fit.}
\scalebox{0.82}{
  \begin{tabular}{@{}lccccccccc@{}}
  \\
  \hline
   Star & {\it P$_{min}$}&$\omega$&{\it e}&{\it T$_{p}$}&{\it K}& {\it a}& $\chi^{2}_{\upsilon}$ & rms \\
  & days & deg & & JD-&m/s&au & & m/s\\ &&&&2450000\\
  \hline
\\
HD18907        &    13770 $\pm$ 3528           &      314 $\pm$ 34     &   0.28 $\pm$ 0.07     & 865 $\pm$ 9               & 3112 $\pm$ 1903          &  13 $\pm$ 4   & 4.31    & 9.94               \\
HD25874        &  71108 $\pm$ 2308                   &      274 $\pm$ 10   &   0.8 $\pm$ 0.1         & 4382 $\pm$ 2453       & 2578 $\pm$ 253  & 33 $\pm$ 2    & 4.16    &   4.67        \\
HD26491        &    9748  $\pm$ 1223        &      221 $\pm$ 38     &   0.57 $\pm$ 0.05     & 250.9 $\pm$ 0.5          & 4717 $\pm$ 2969          & 10 $\pm$ 2           & 3.16    & 9.71               \\
HD131923      &         7495 $\pm$ 334          &    15 $\pm$ 79            &   0.72 $\pm$ 0.02      & 528 $\pm$ 30          & 5784 $\pm$ 821  & 8.7 $\pm$ 0.4             & 21.8    & 8.60               \\
HD145825      &    6024 $\pm$ 163           &    149 $\pm$ 53      &  0.33 $\pm$ 0.02        & 1095 $\pm$ 13       & 669 $\pm$ 10        & 6.54 $\pm$ 0.04         &14.4   &   7.68         \\
HD156274B    &       191455 $\pm$ 13707          &      30 $\pm$ 85          &   0.0 $\pm$ 0.1      & 5222 $\pm$ 10588               & 692 $\pm$ 55                & 63.5 $\pm$ 0.2      & 11.4      & 5.95           \\
\hline

\label{orbit2}
\end{tabular}
}
\end{minipage}
\end{table*}

\begin{figure}
\centering
\epsfig{file=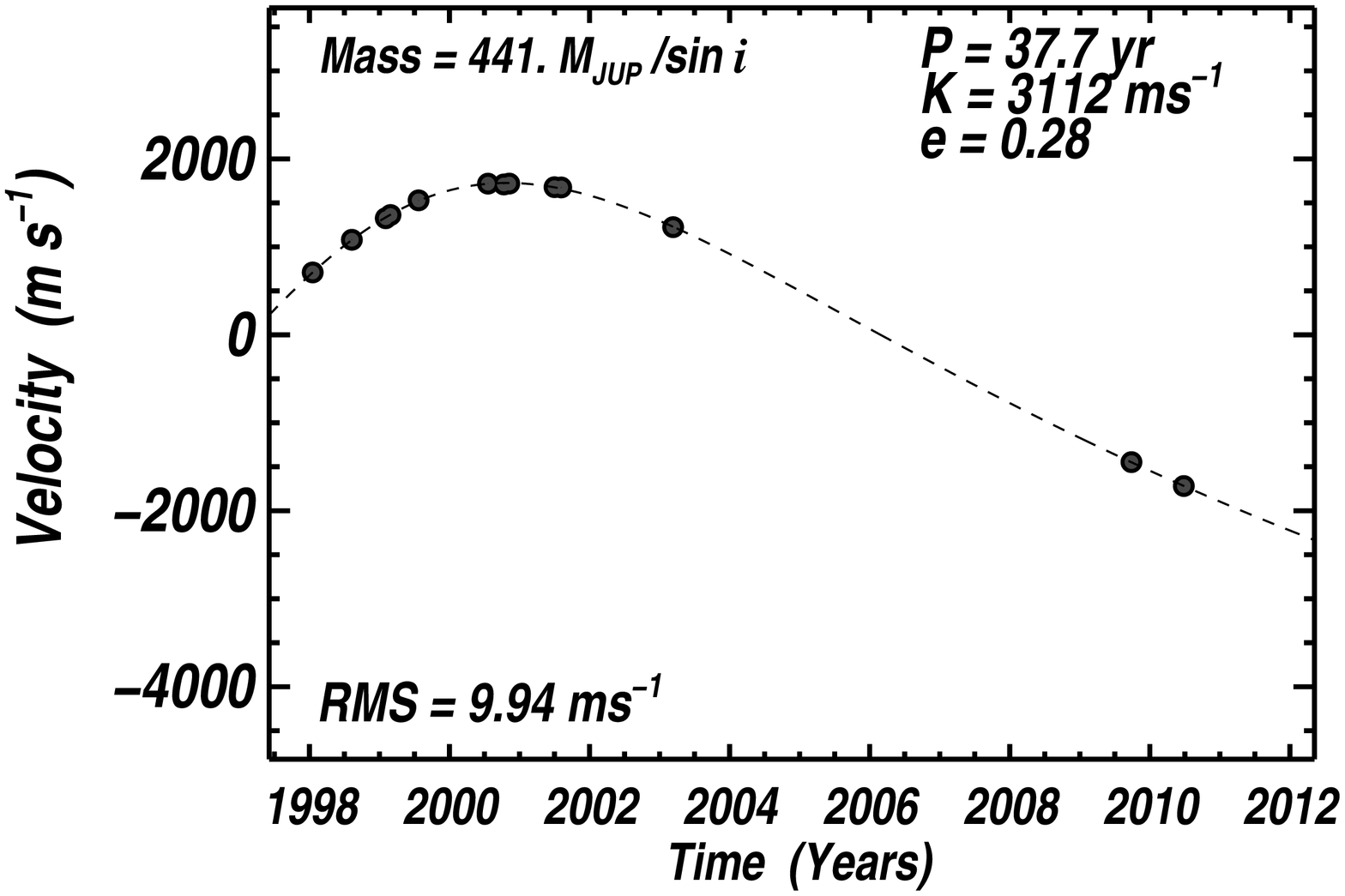,width=5.7cm,angle=90}
\hspace{0.1cm}
\epsfig{file=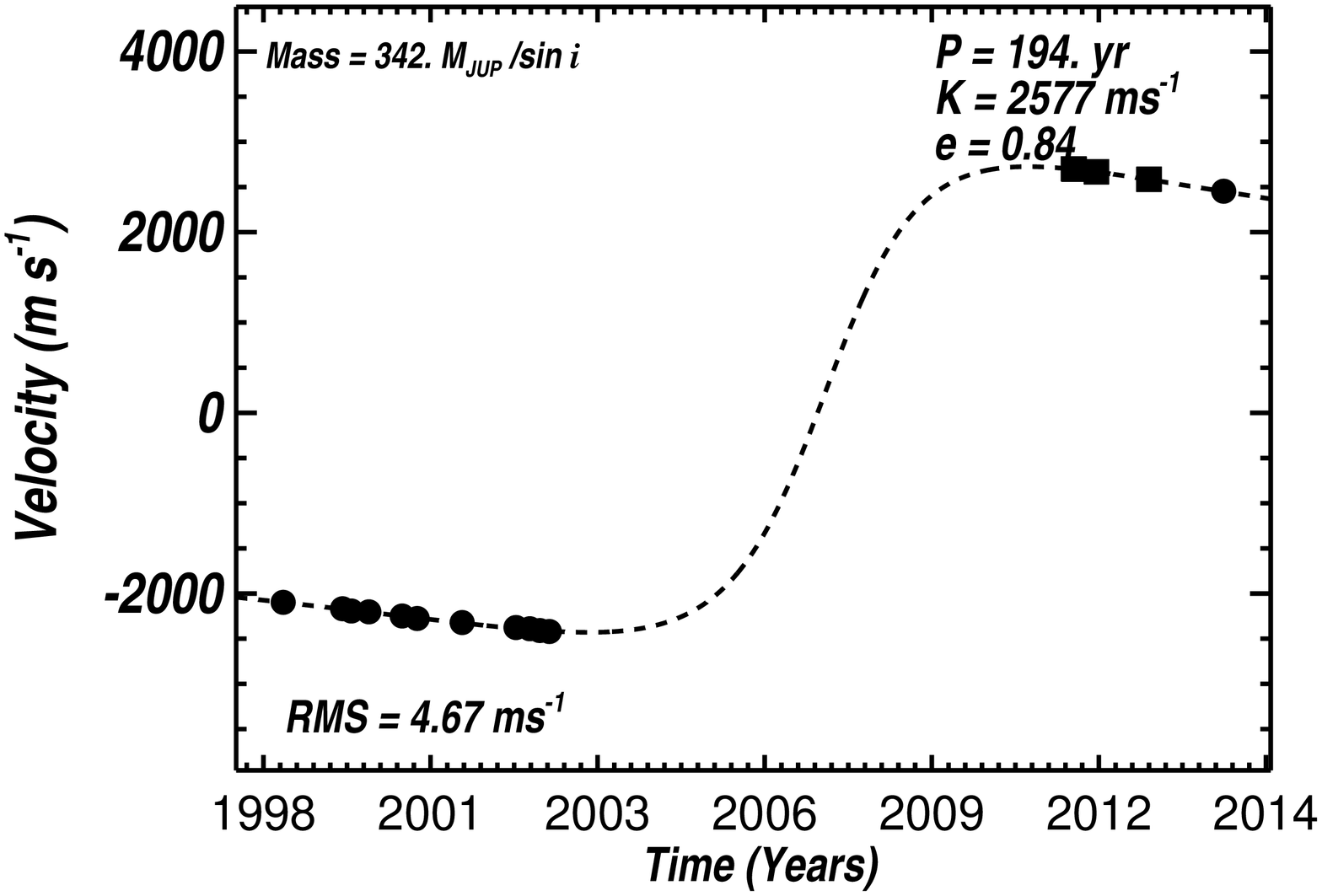,width=5.0cm,angle=90}
\epsfig{file=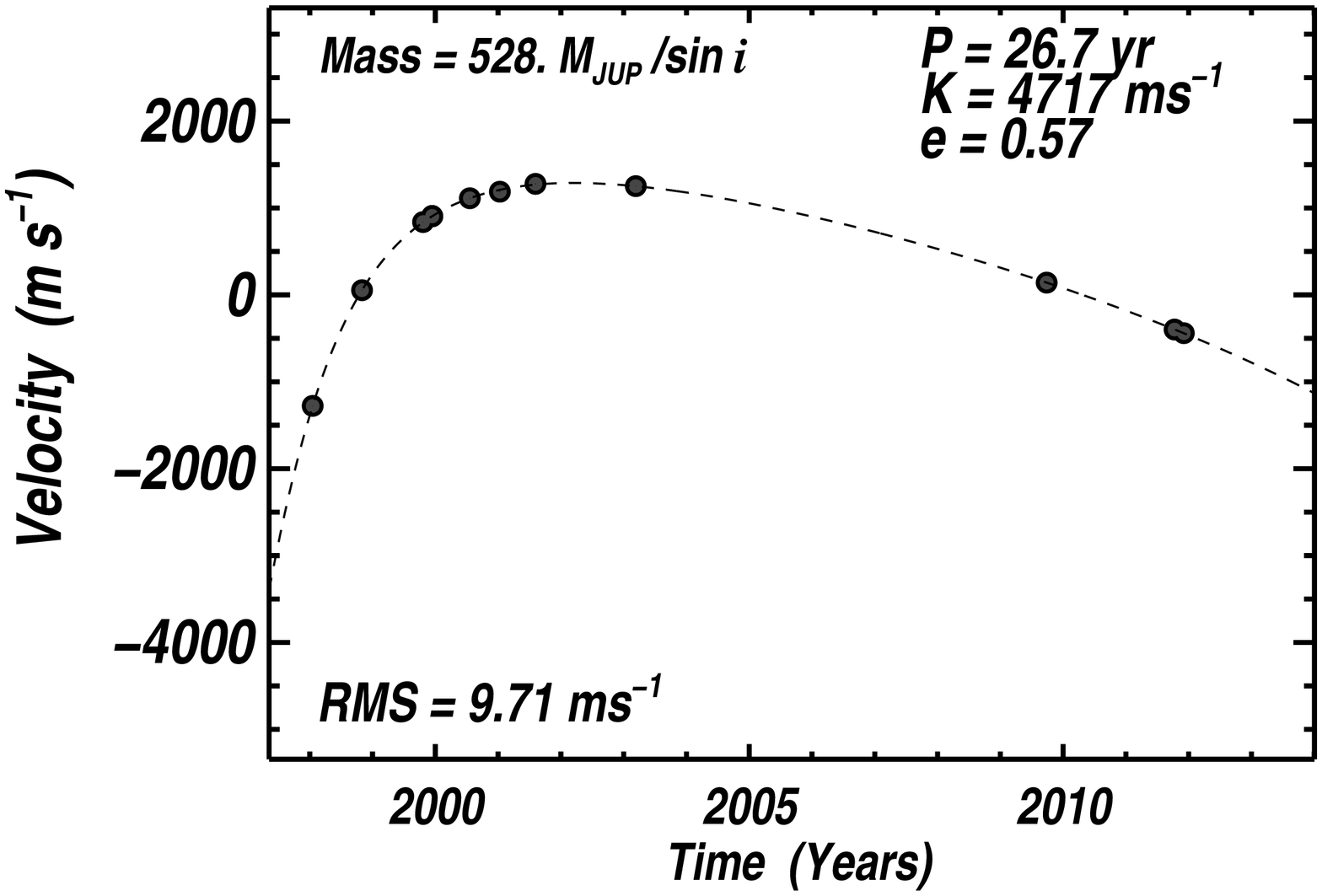,width=5.7cm,angle=90}
\epsfig{file=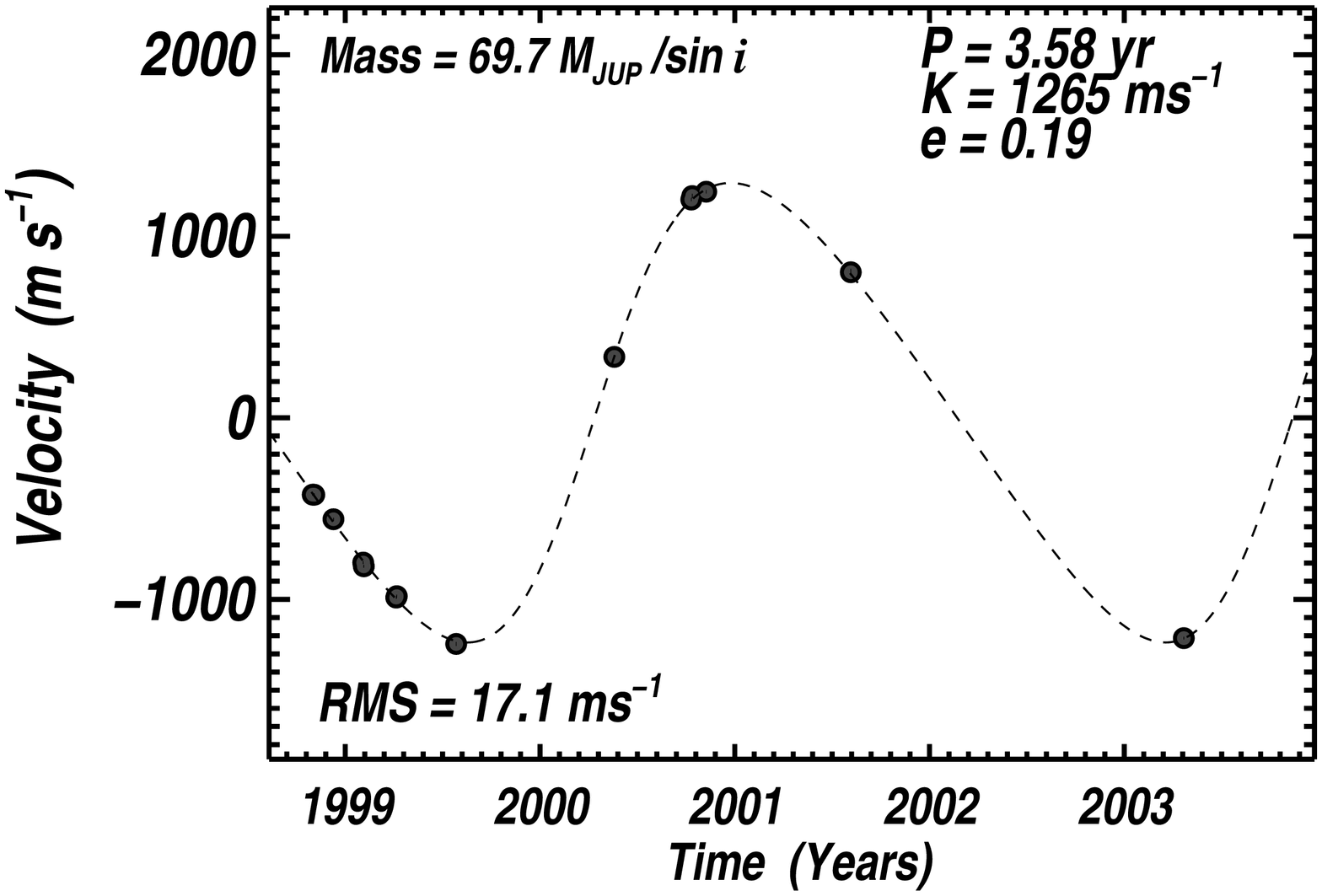,width=5.7cm,angle=90}
\epsfig{file=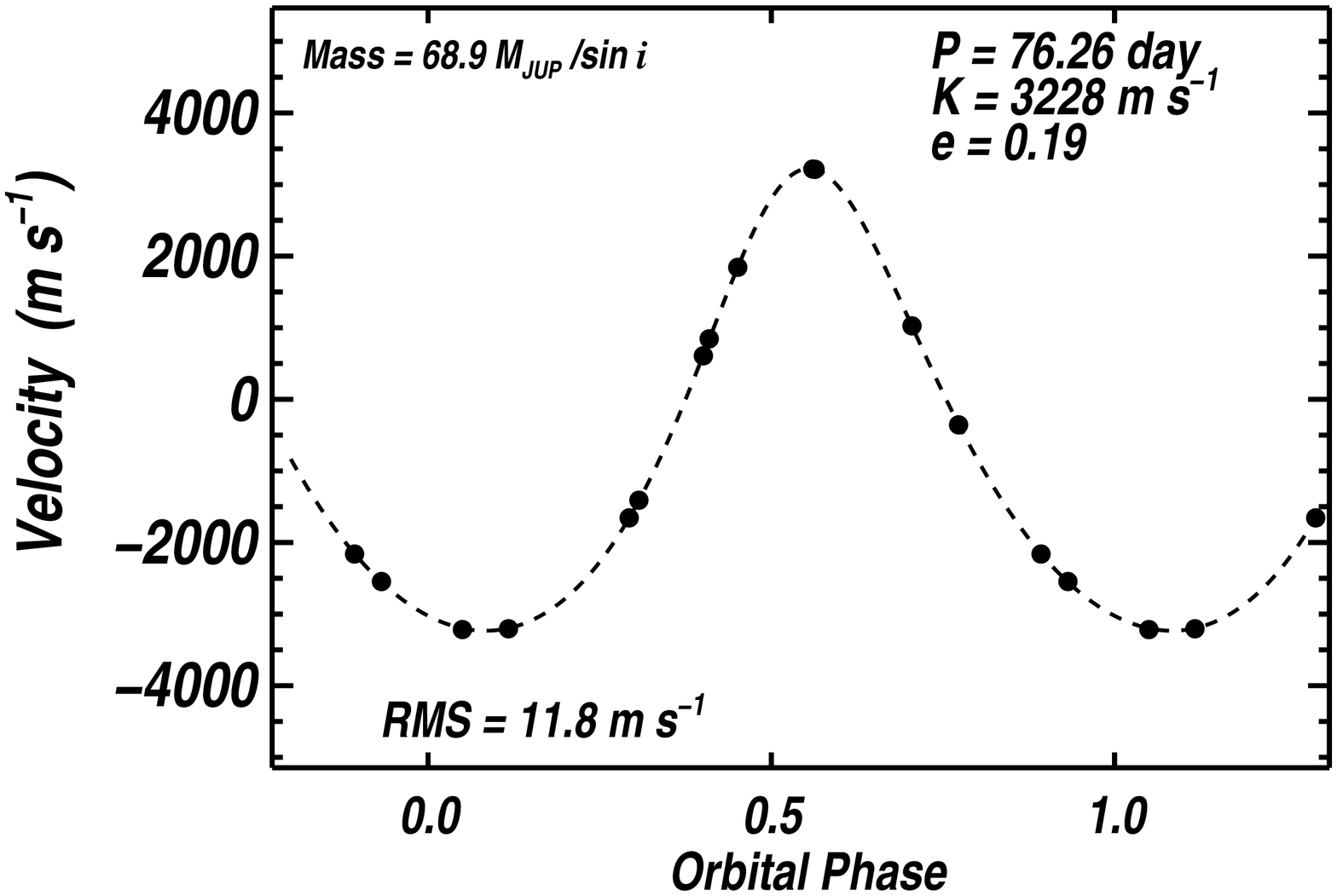,width=5.7cm,angle=90}
\epsfig{file=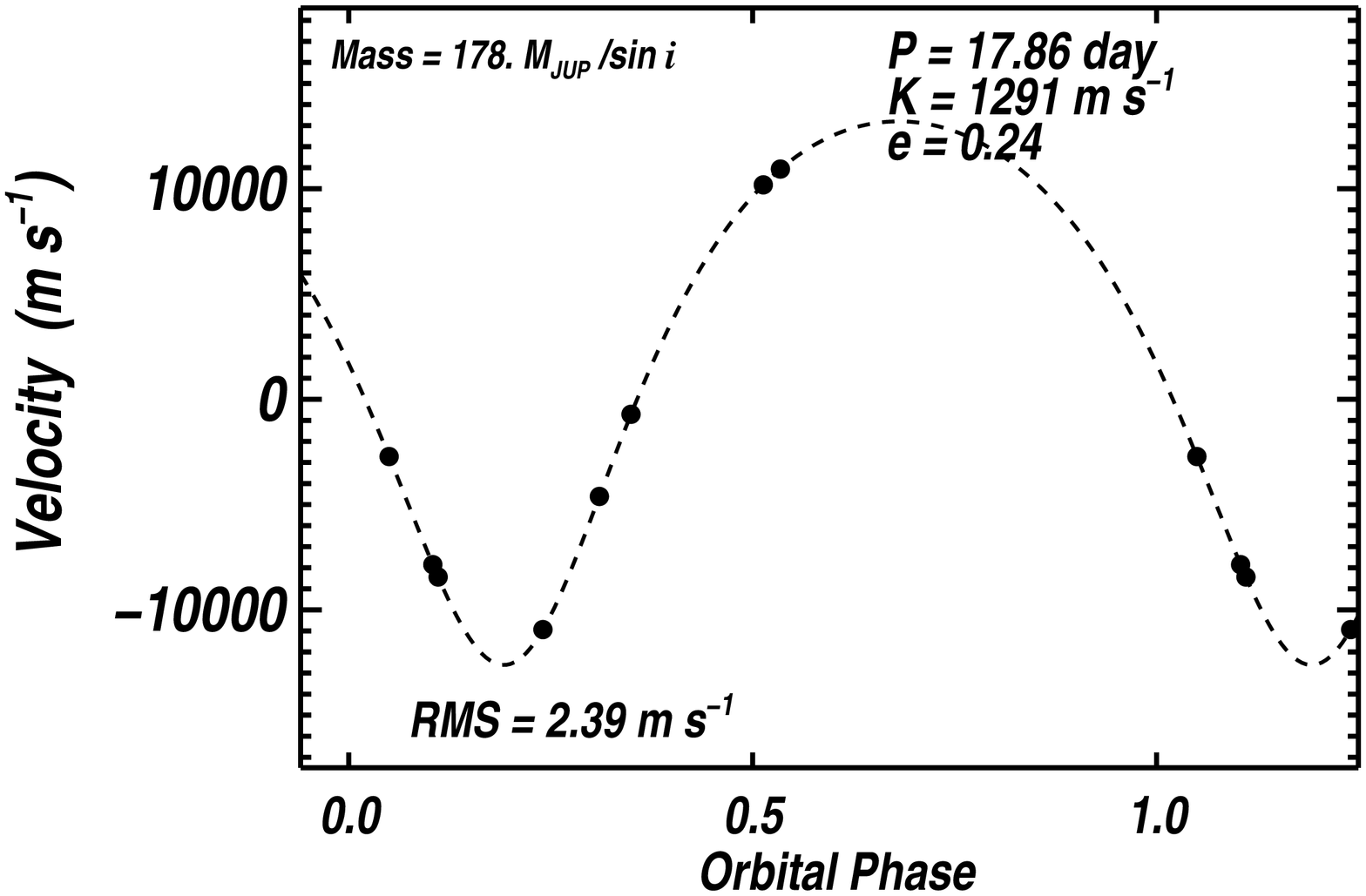,width=5.7cm,angle=90}
\epsfig{file=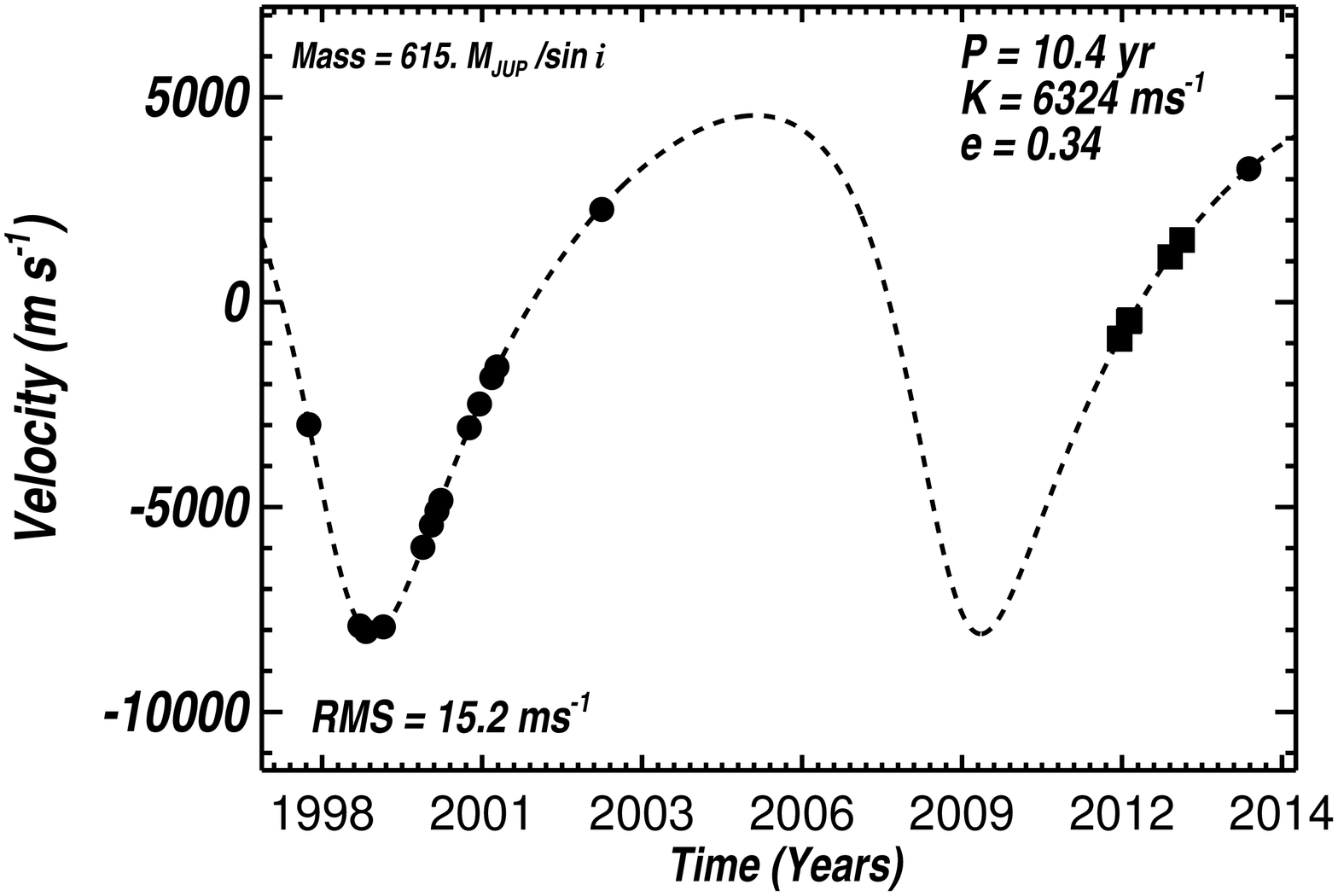,width=5.0cm,angle=90}
\hspace{0.5cm}
\epsfig{file=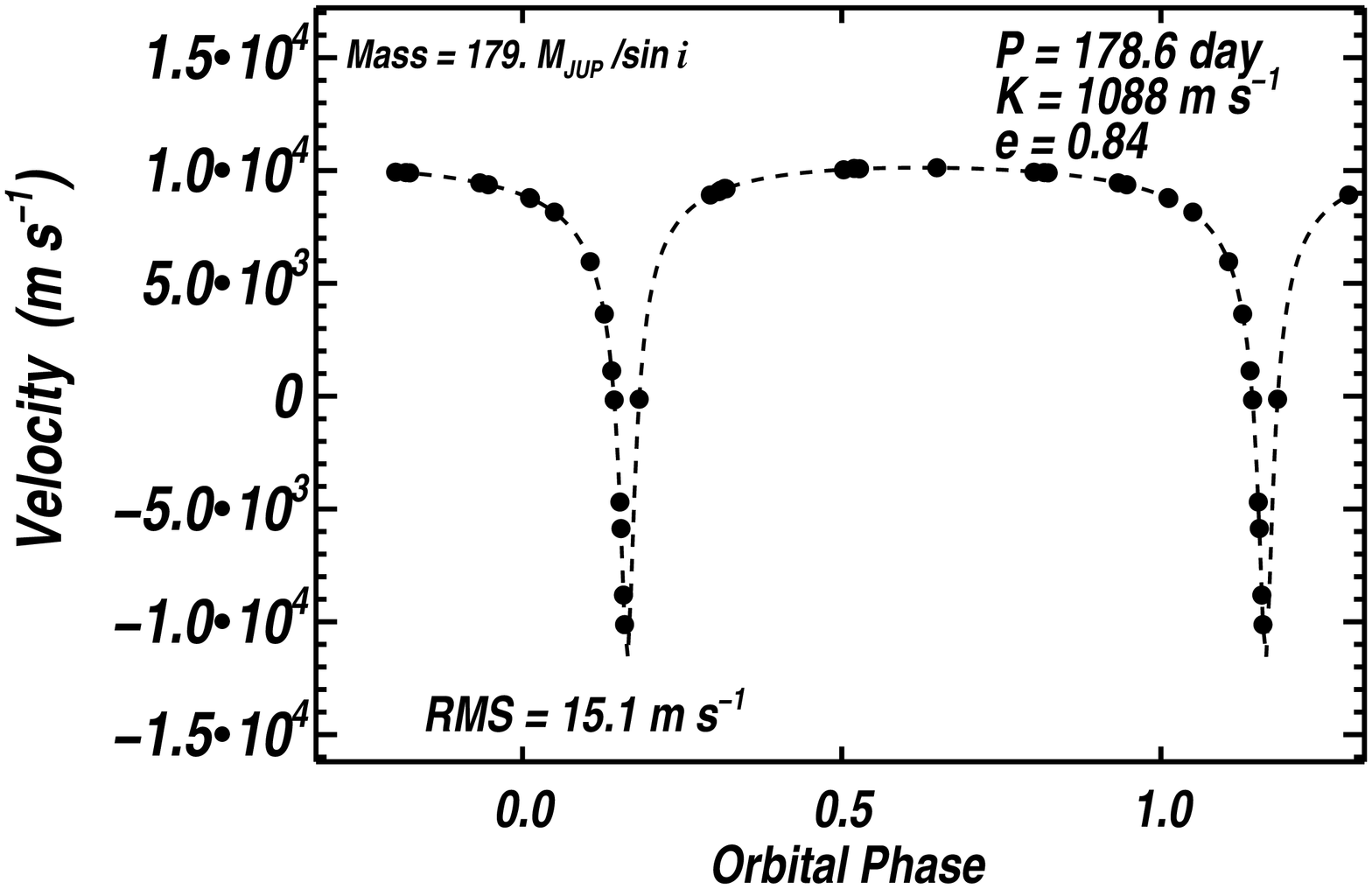,width=5.4cm,angle=90}
\caption{Radial velocity curves for our stars, with filled circles representing data from the AAPS and filled squares data from HARPS.  From top left to bottom right we show the stars in catalogue order, HD18907, HD25874, HD26491, HD39213, HD42024, HD64184, HD120690, HD121384, respectively.}\label{rvcurve}
\end{figure}

\begin{figure}
\setcounter{figure}{0}
\epsfig{file=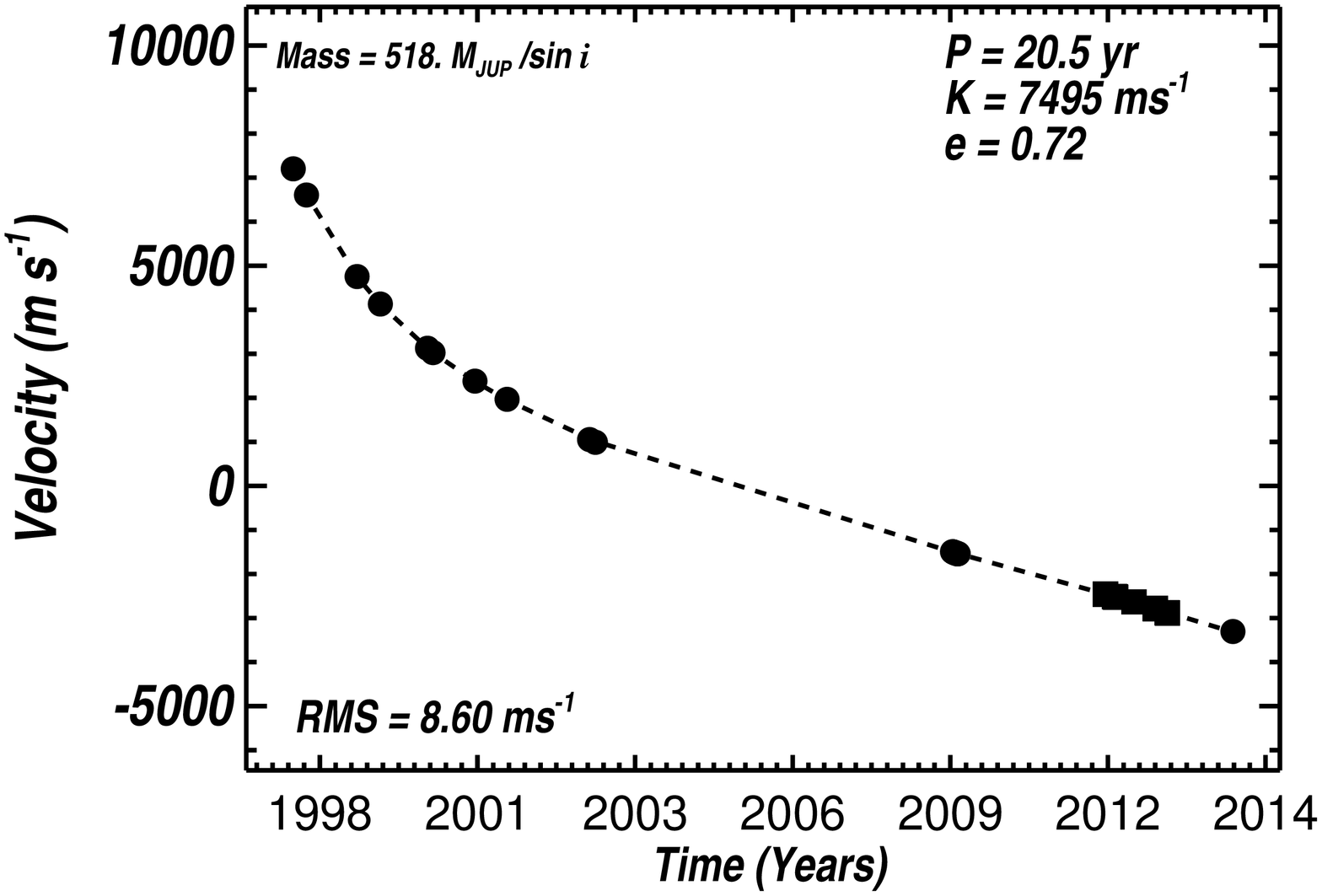,width=4.9cm,angle=90}
\vspace{0.3cm}
\epsfig{file=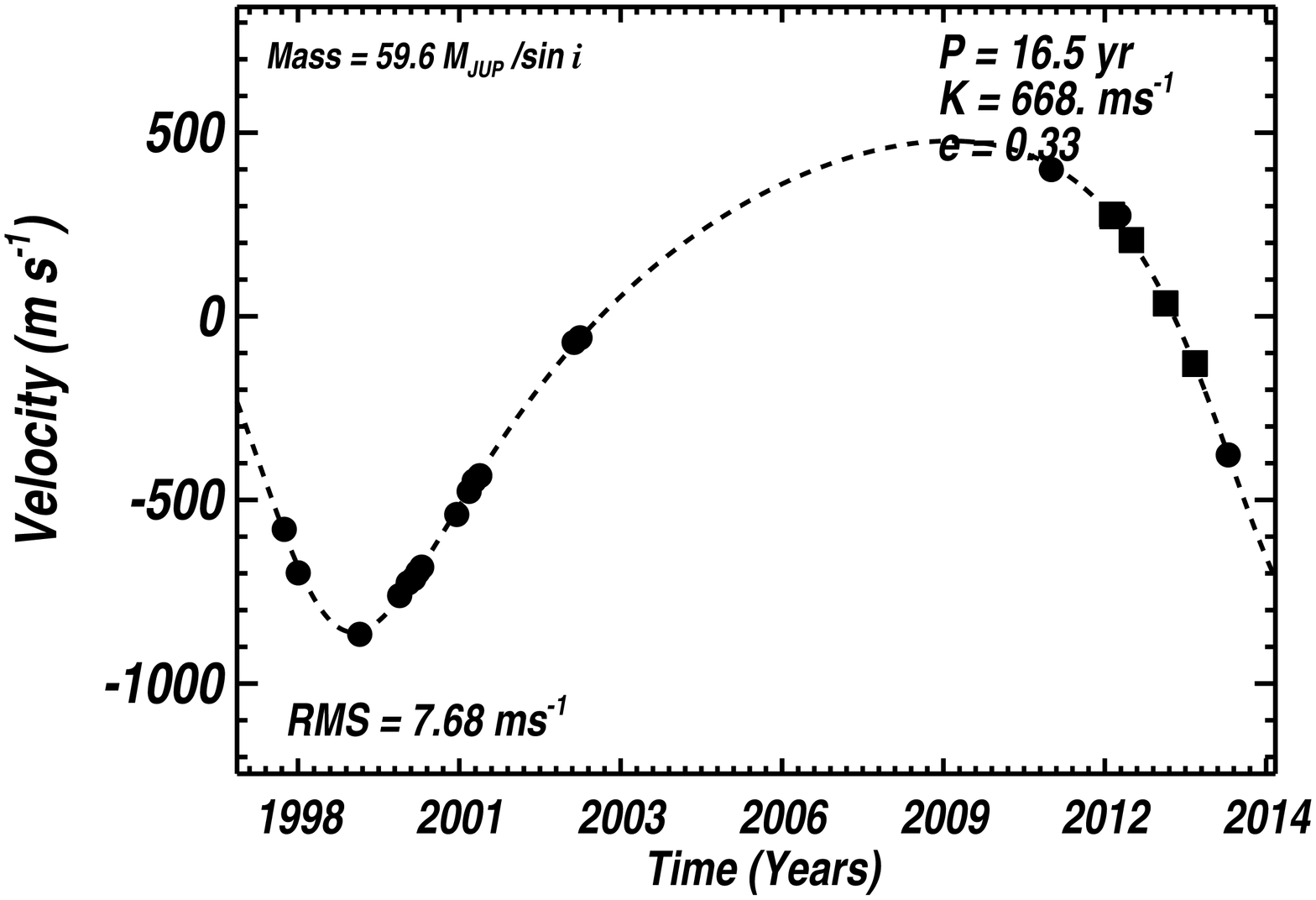,width=4.9cm,angle=90}
\epsfig{file=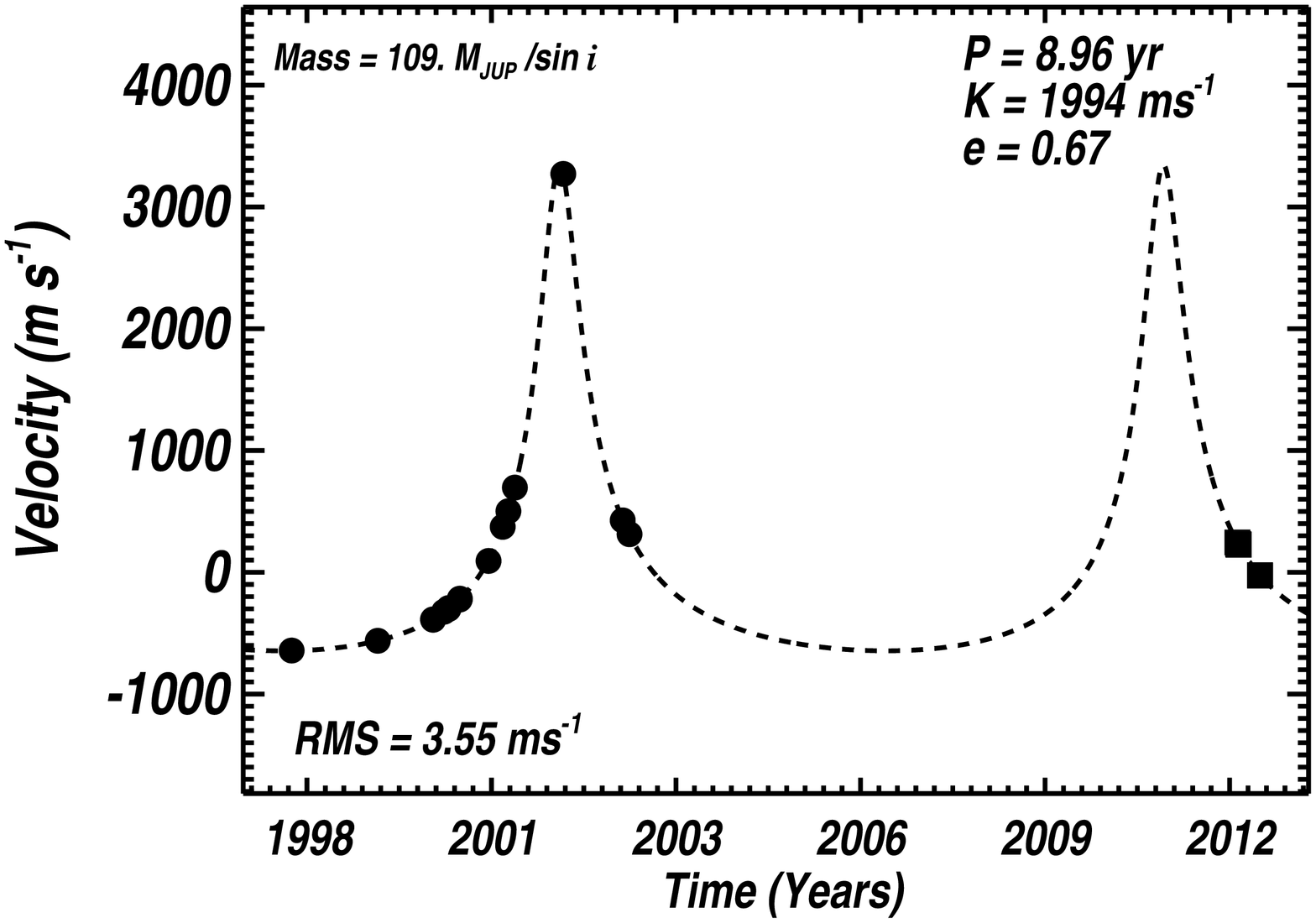,width=4.9cm,angle=90}
\epsfig{file=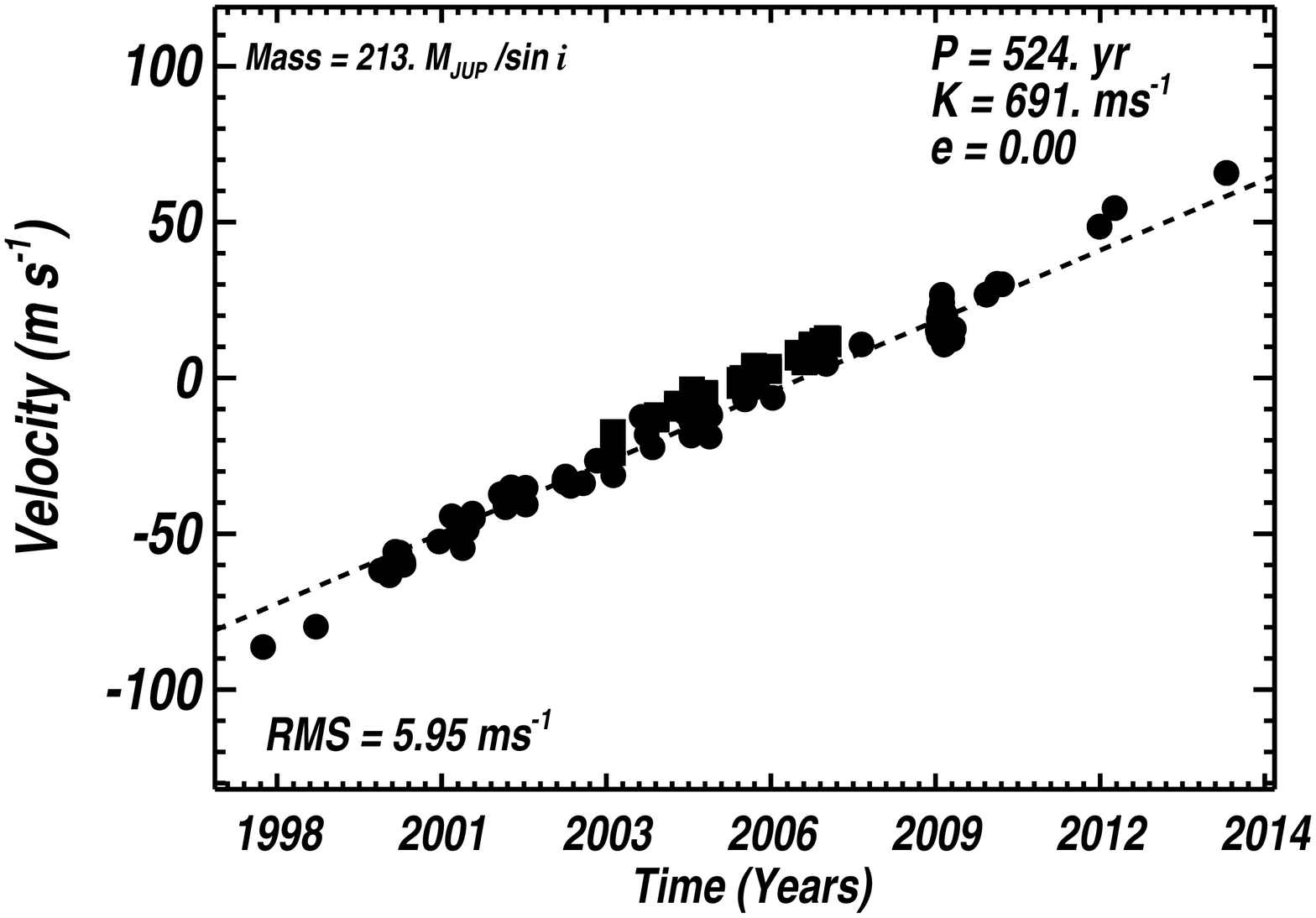,width=4.9cm,angle=90}
\epsfig{file=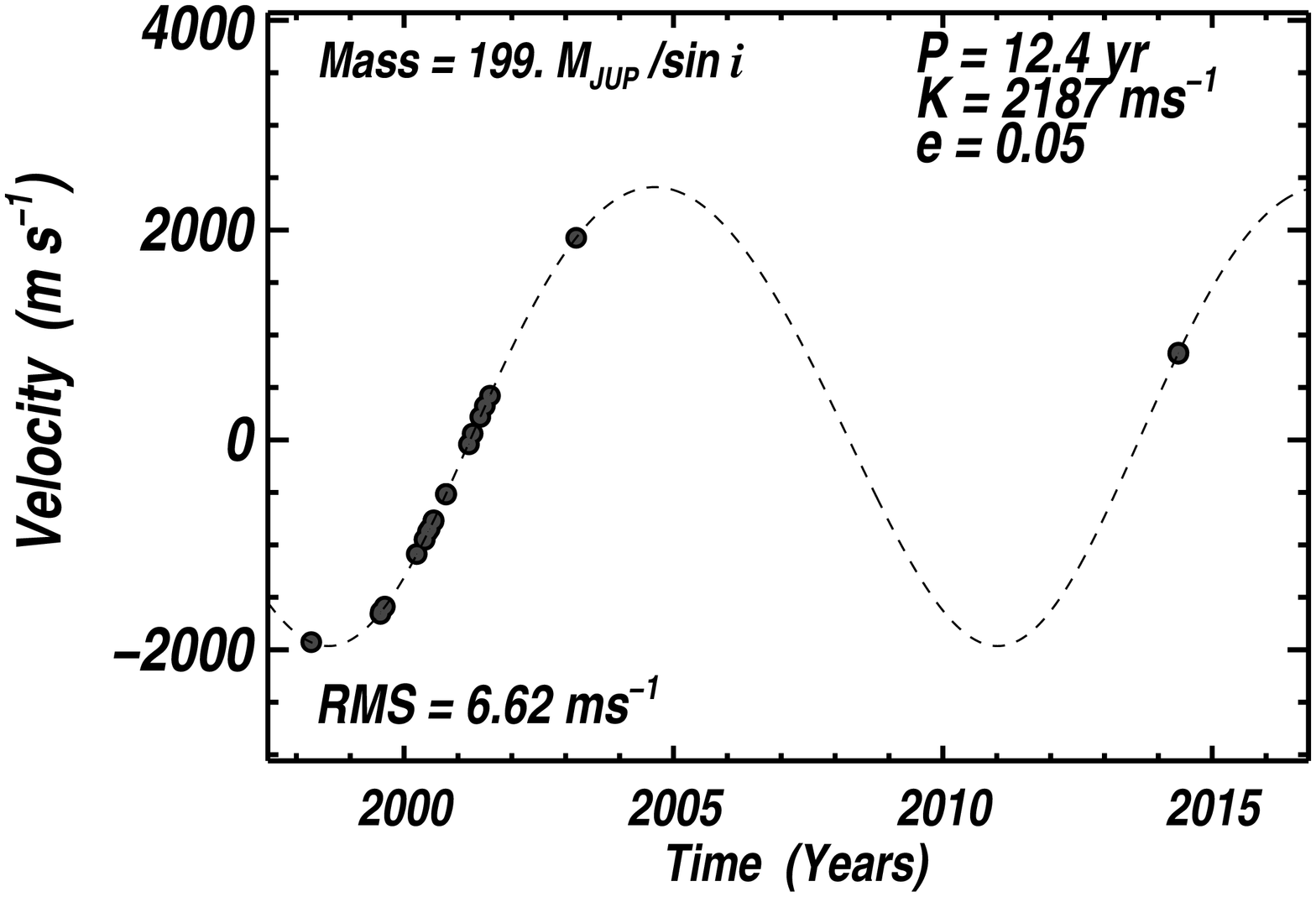,width=5.7cm,angle=90}
\epsfig{file=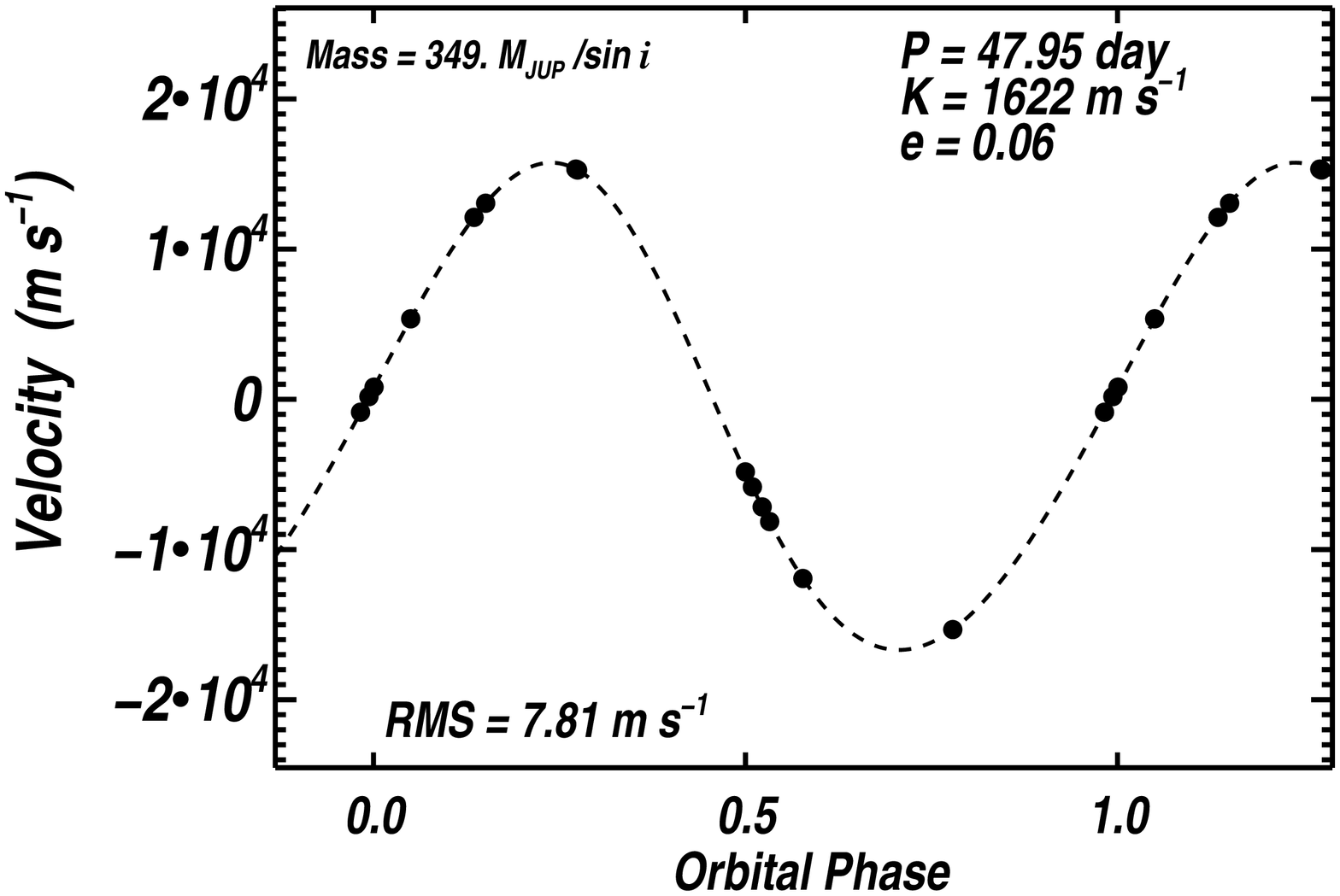,width=5.7cm,angle=90}
\epsfig{file=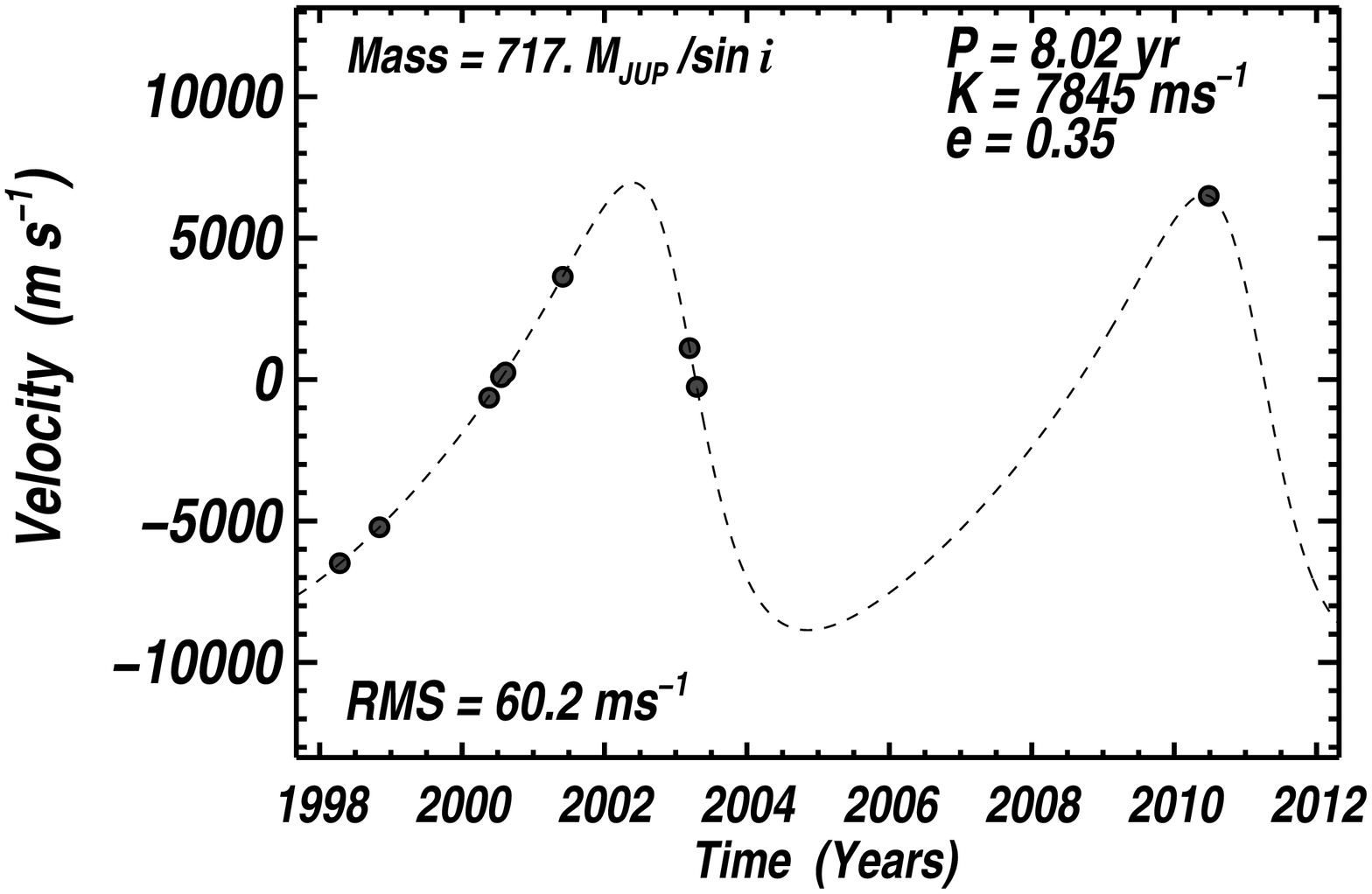,width=5.7cm,angle=90}
\hspace{0.5cm}
\epsfig{file=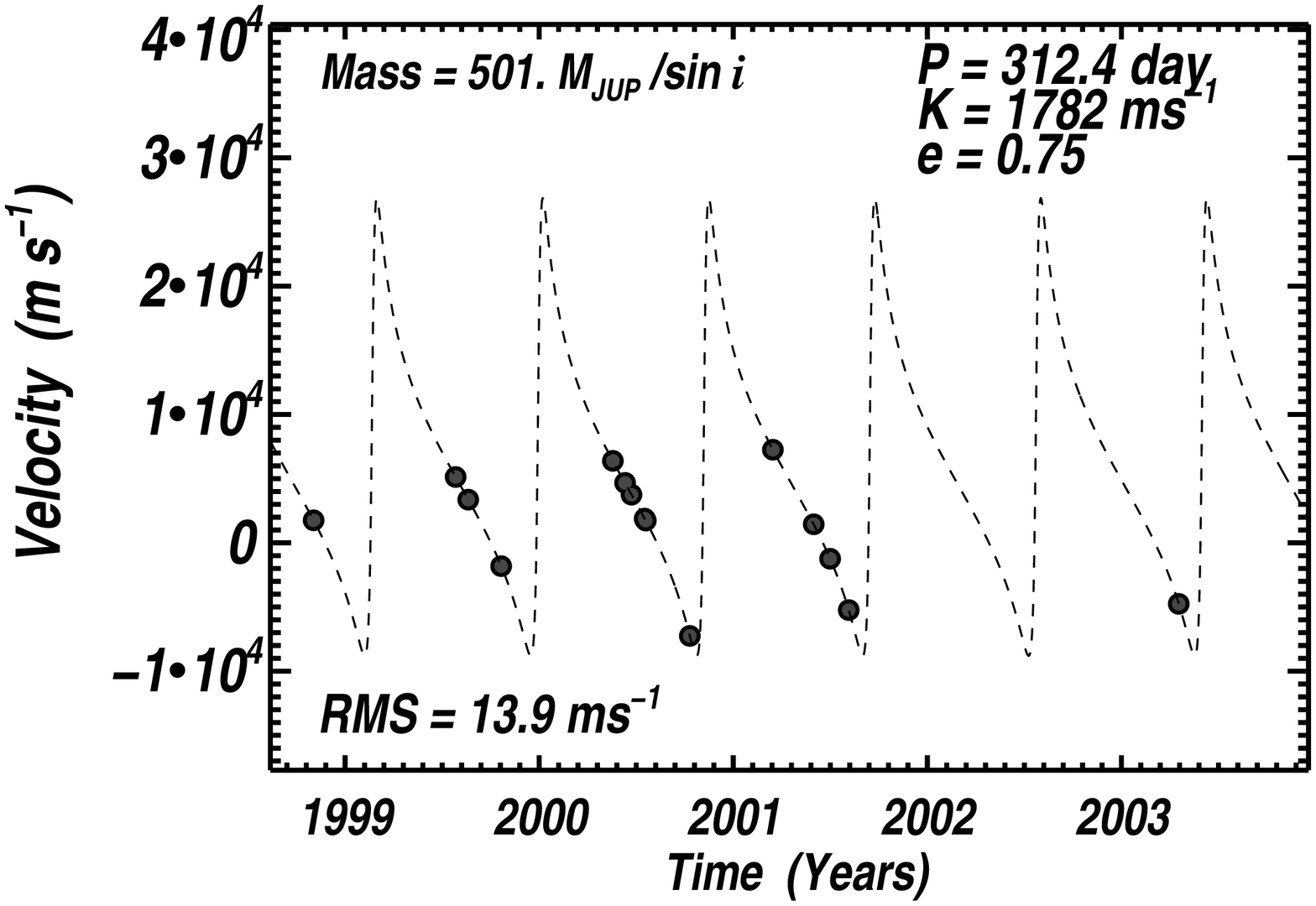,width=5.5cm,angle=90}
\caption{(cont.) From top left to bottom right we show the stars in catalogue order, HD131923, HD145825, HD150248, HD156274B, HD158783, HD162255, HD169586, and HD175345, respectively.}\label{rvcurve2}
\end{figure}

\begin{figure}
\vspace{0.4cm}
\epsfig{file=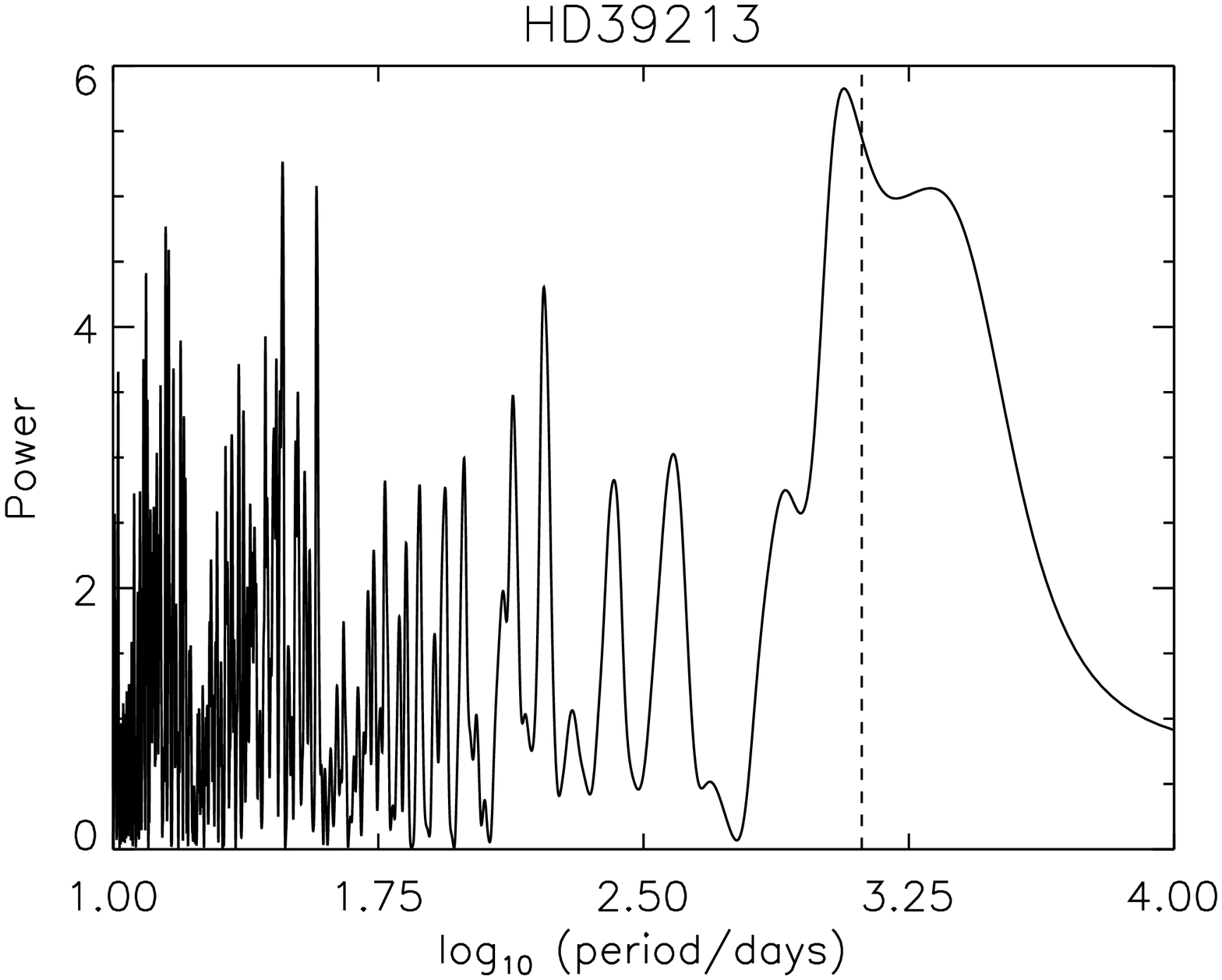,width=5.5cm,angle=90}
\epsfig{file=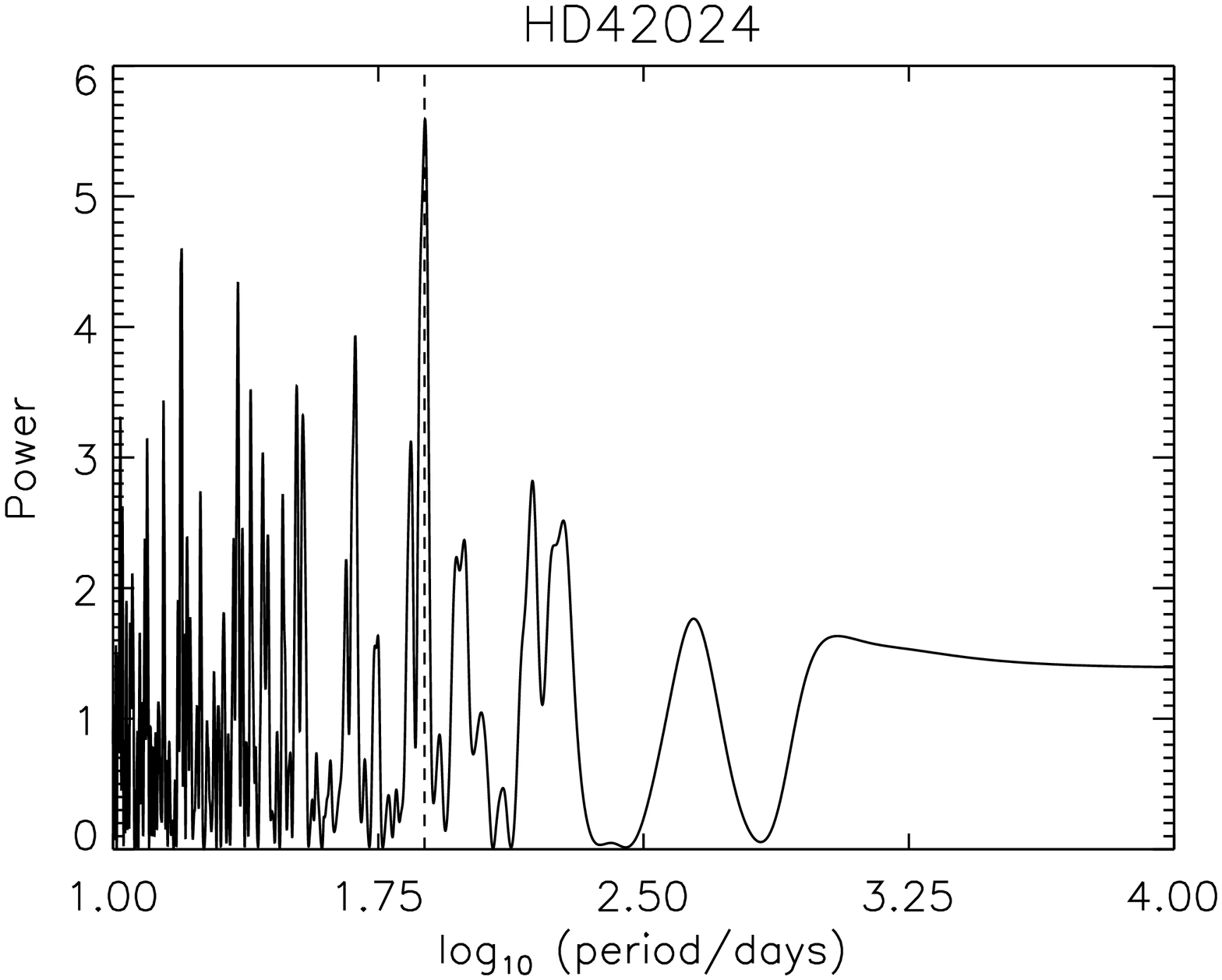,width=5.5cm,angle=90}
\vspace{0.8cm}

\epsfig{file=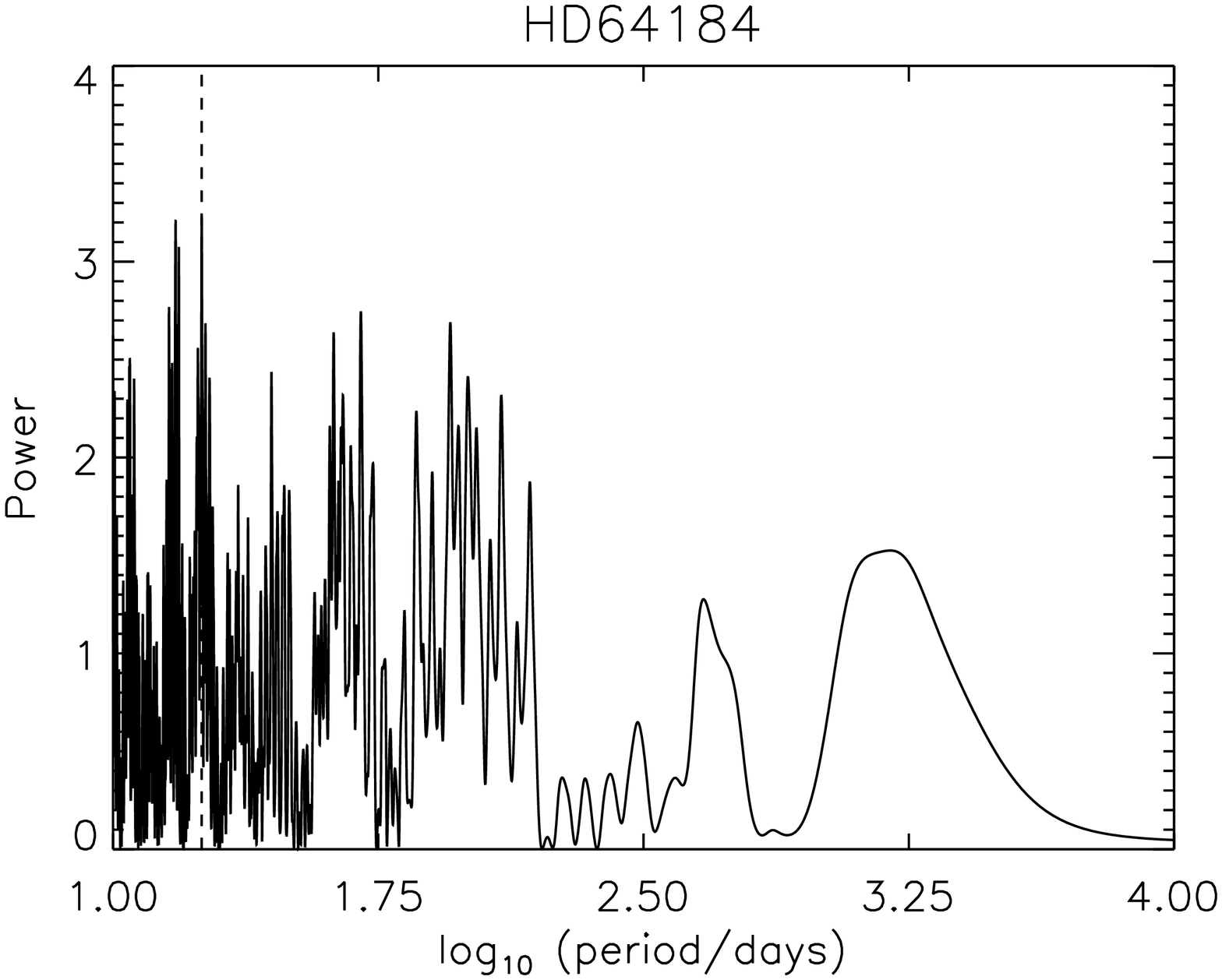,width=5.5cm,angle=90}
\epsfig{file=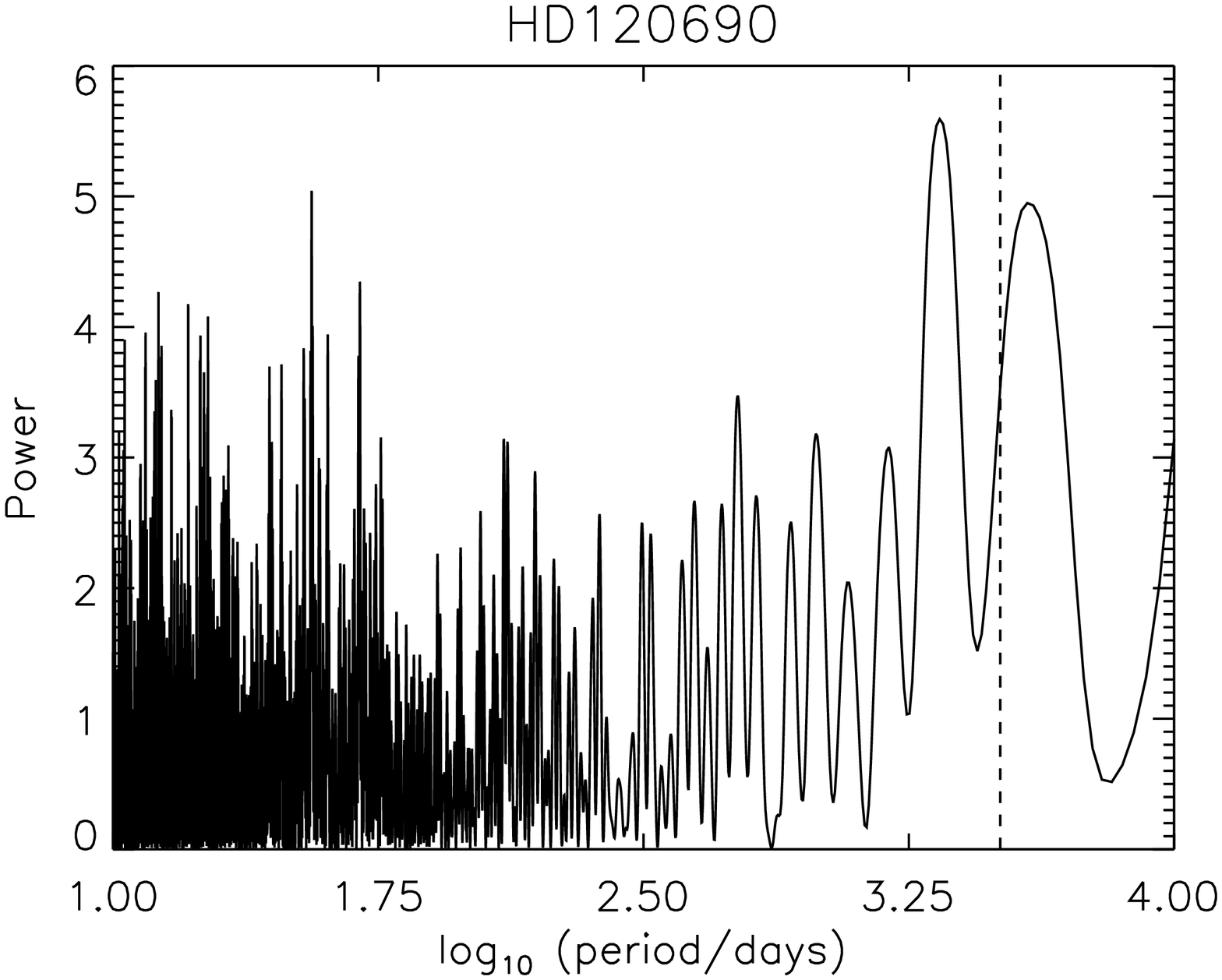,width=5.5cm,angle=90}
\vspace{0.8cm}

\epsfig{file=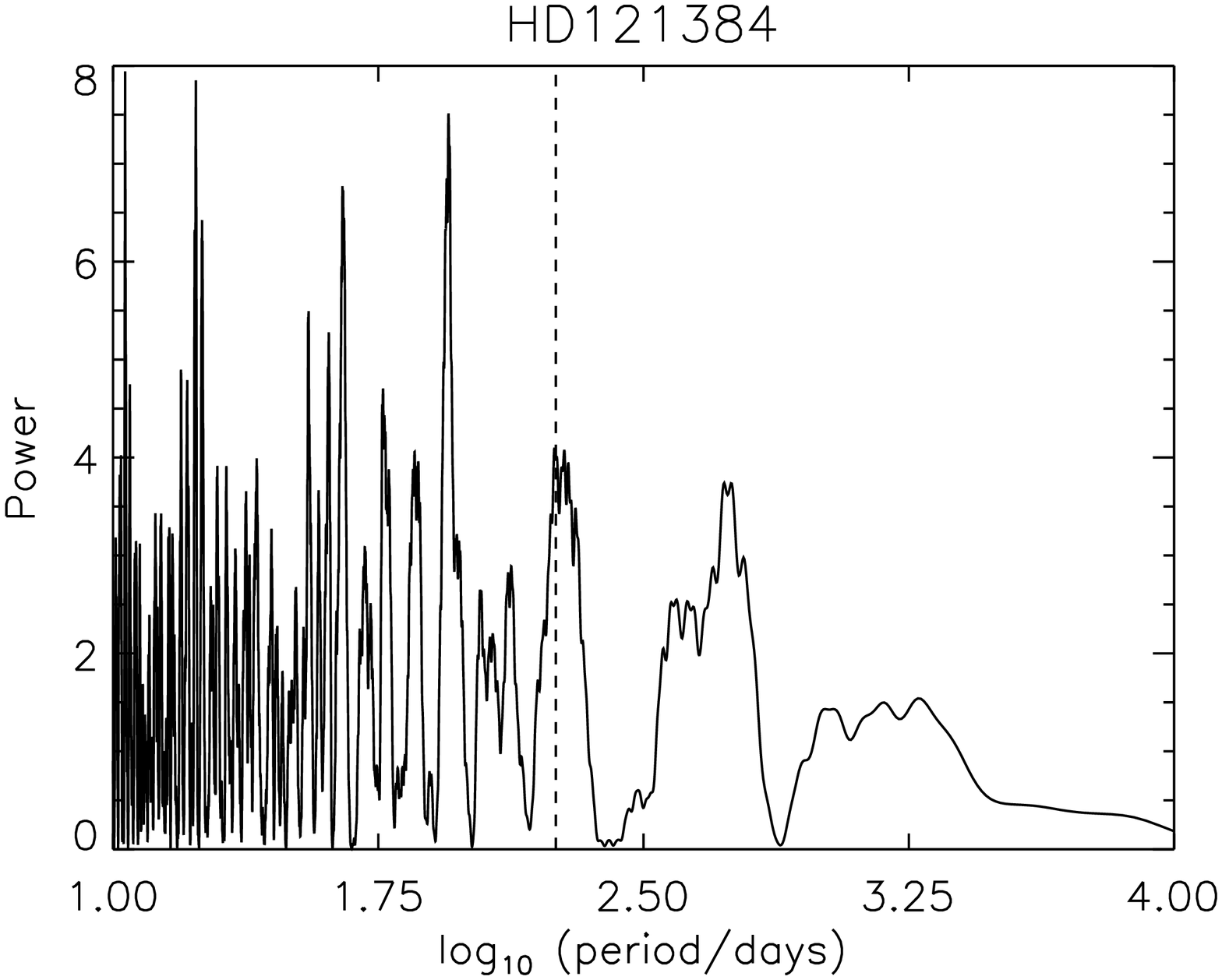,width=5.5cm,angle=90}
\epsfig{file=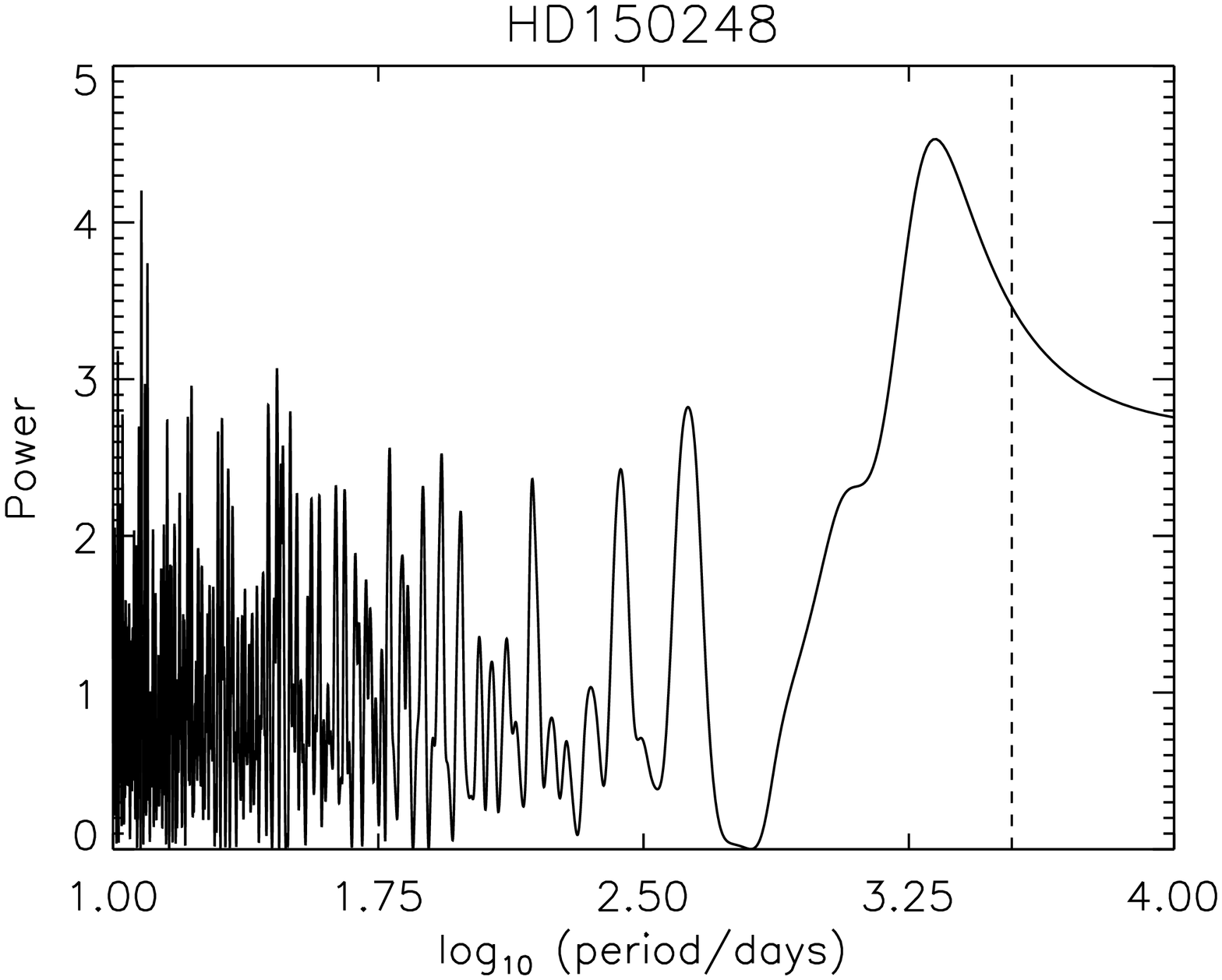,width=5.5cm,angle=90}
\caption{Lomb-Scargle periodograms. These are plots of spectral power
against {\it log} period and are shown for all the stars where the
phase coverage of the RV data is nearly a cycle or more. Where the
sampling is sparse, aliasing introduces spectral power over a range of
frequencies and is particularly marked for HD\,64184. The period inferred from the orbital solution in
each case is marked with a vertical dashed line.}\label{pgram}
\end{figure}

\begin{figure}
\setcounter{figure}{1} 
\vspace{0.4cm}
\epsfig{file=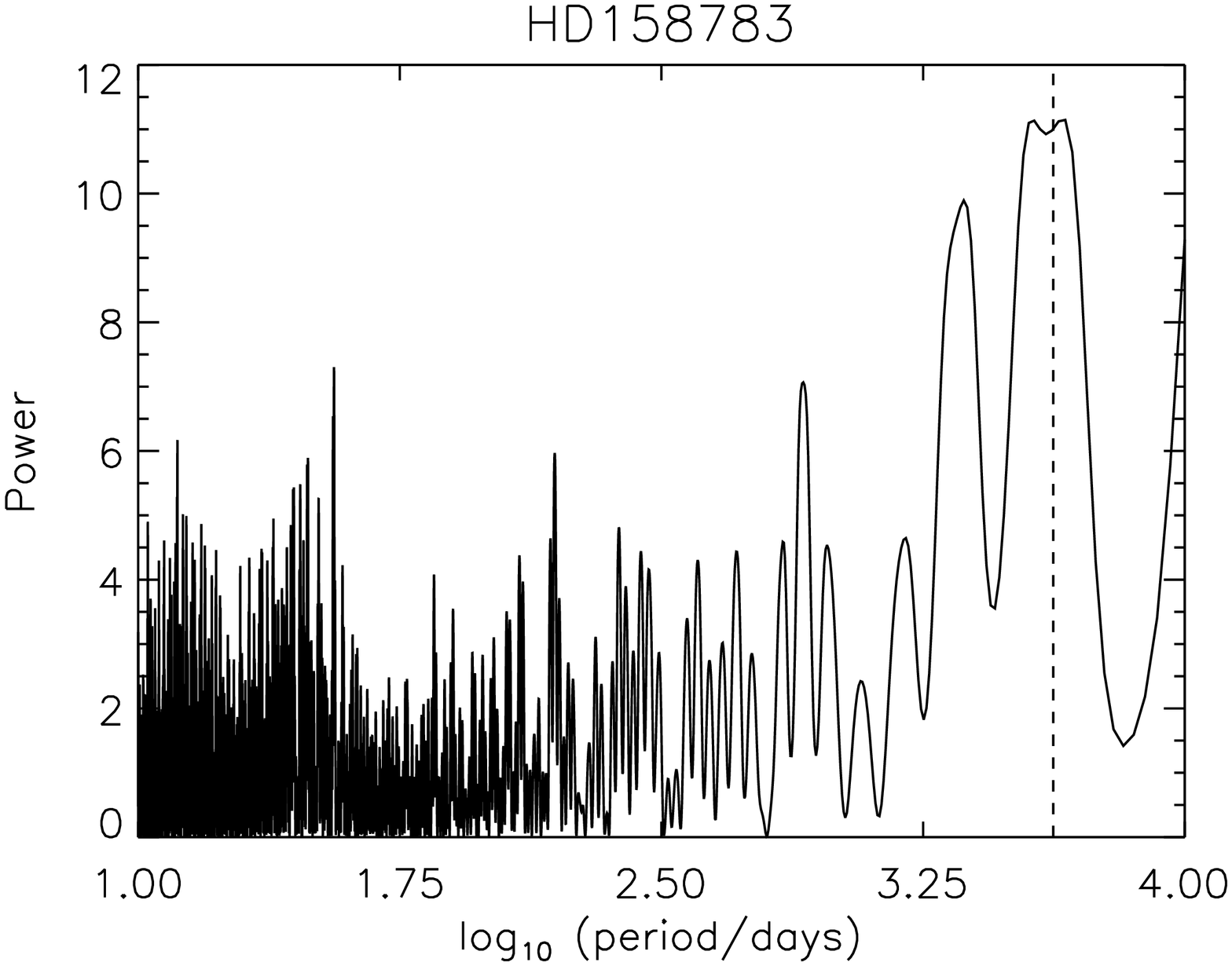,width=5.5cm,angle=90}
\epsfig{file=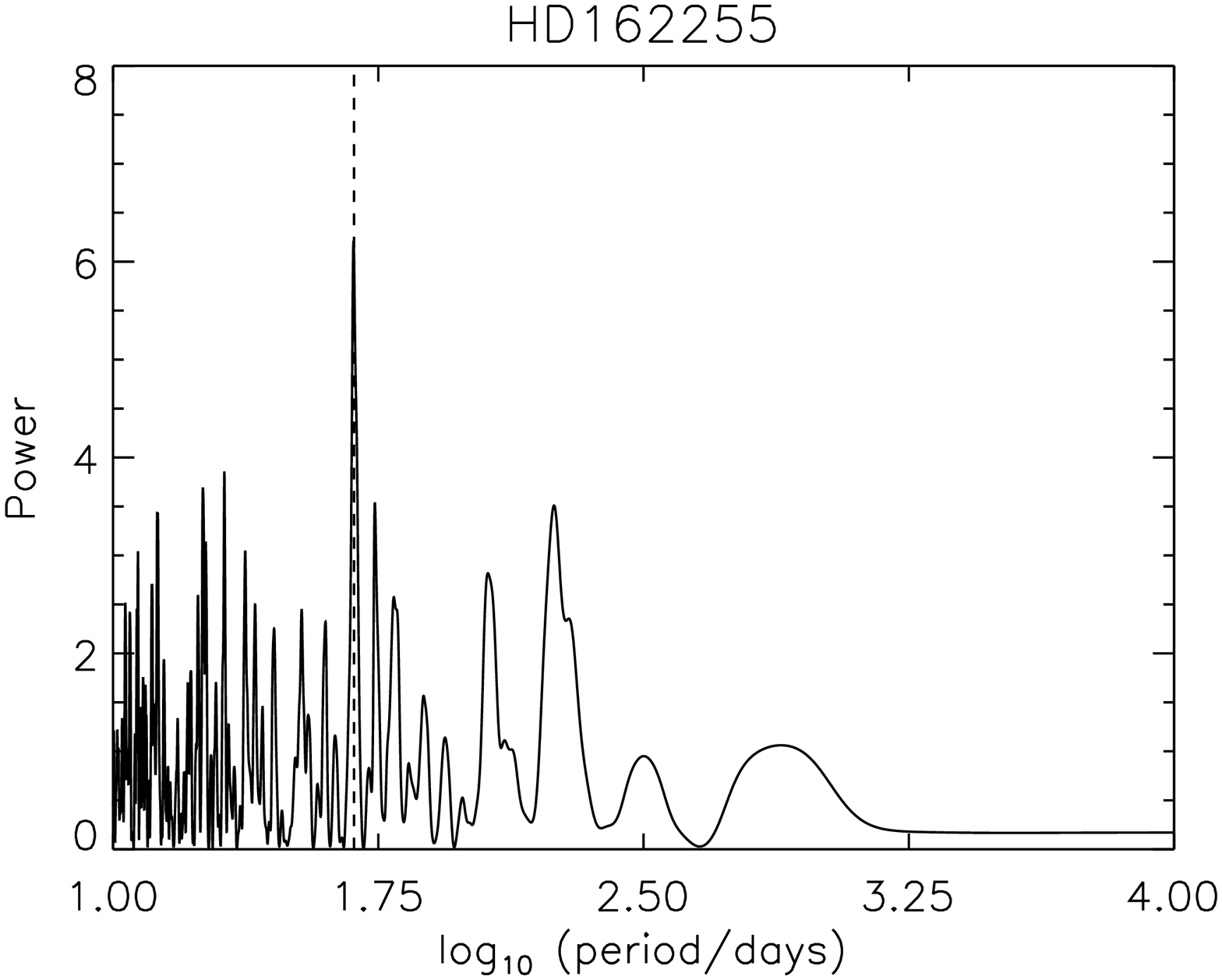,width=5.5cm,angle=90}

\vspace{0.8cm}
\epsfig{file=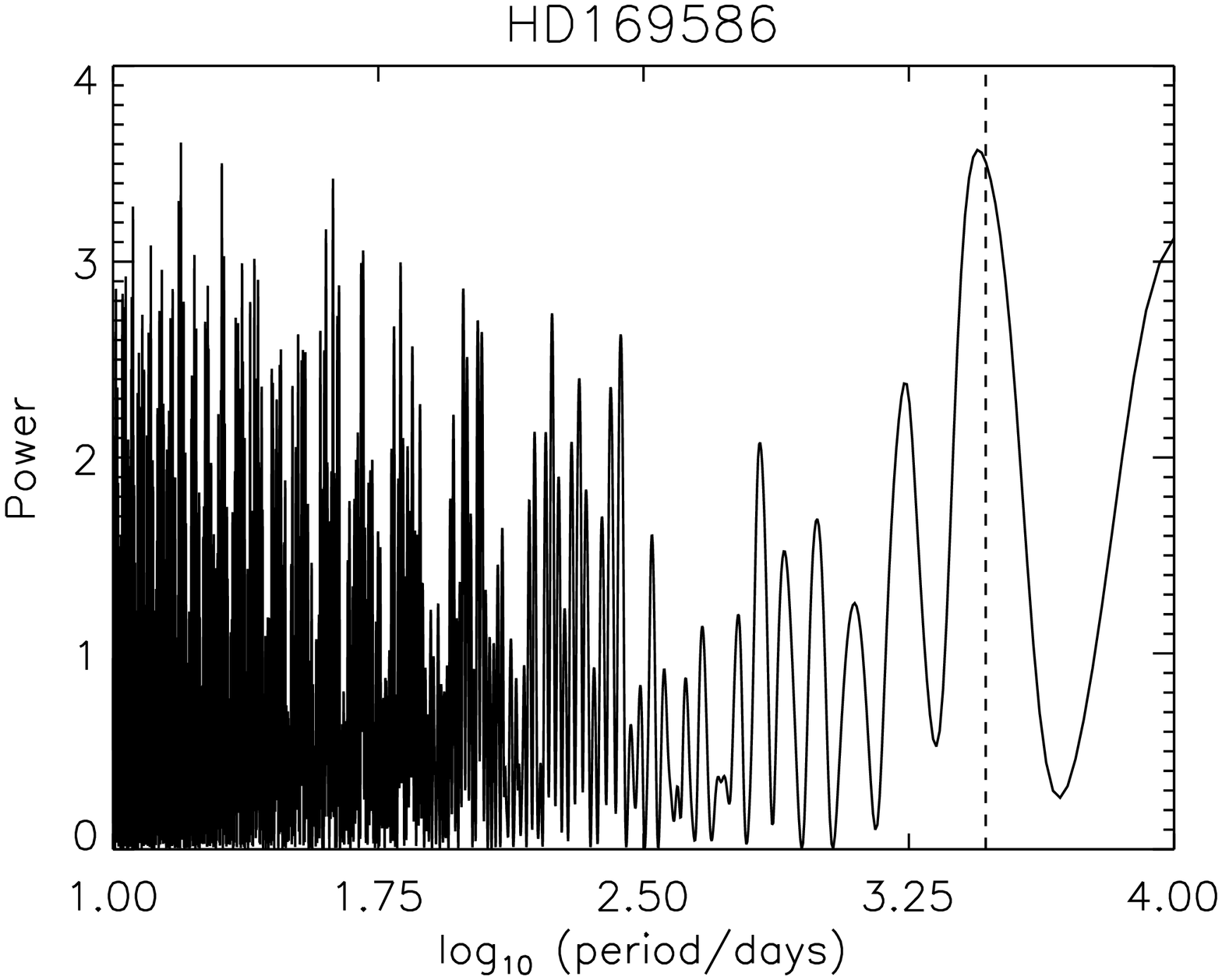,width=5.5cm,angle=90}
\epsfig{file=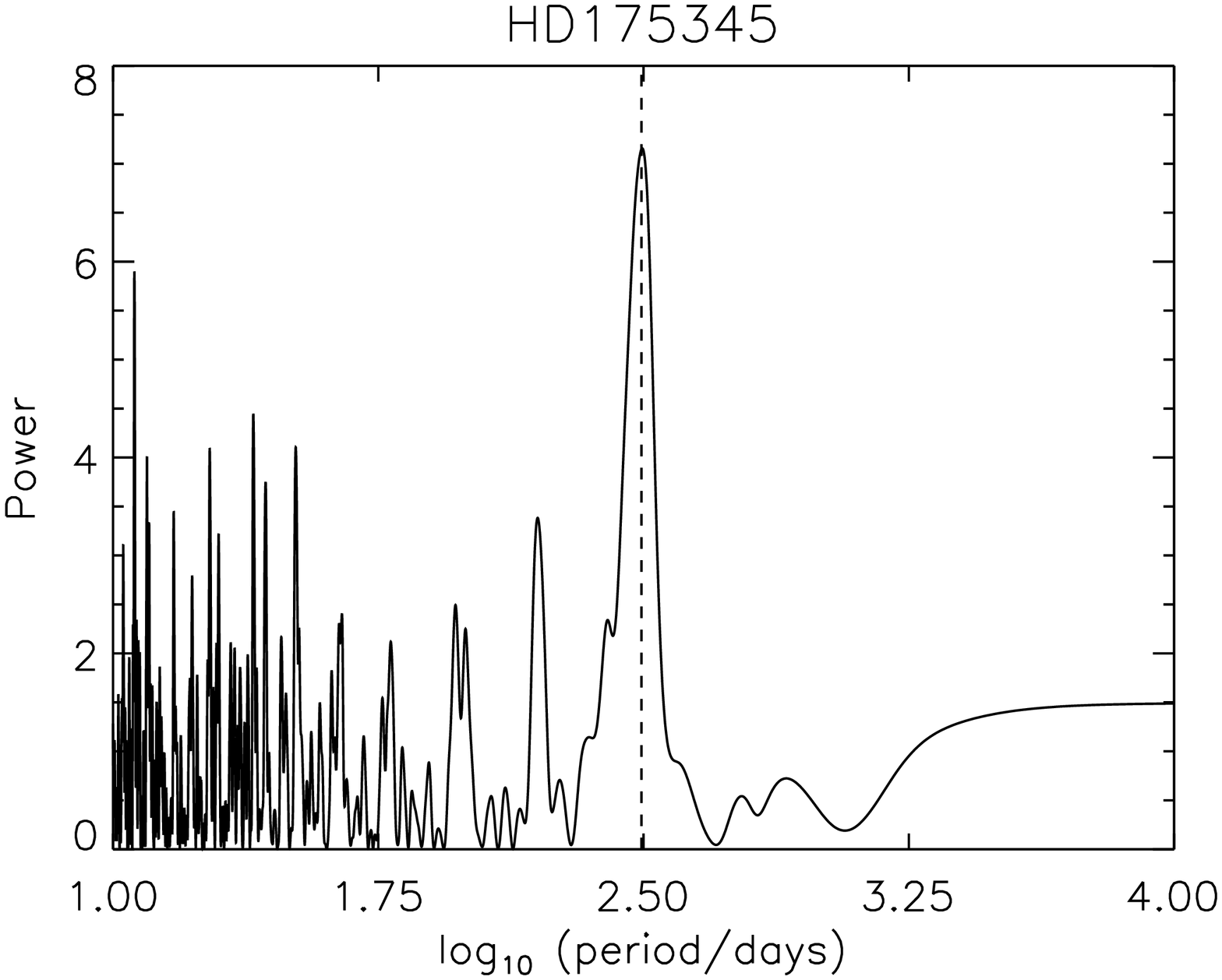,width=5.5cm,angle=90}
\caption{(cont)} \label{pgram2}
\end{figure}

\section{Discussion}
\subsection{The Binary Stars}

{\it HD\,18907}: A high proper-motion star classed as K2\,V \citep{gray06} though consistently classified as a G5\,IV prior to this (eg. \citealt*{evans57}), its colour and magnitude confirm that it is evolving away from the main sequence. Chromospherically quiet ($R'_{\rm{HK}}$=\,-5.11,
\citealt{jenkins06}), HD\,18907 is most likely an old star with an age in excess of 12\,Gyr which agrees with isochrone fits based on metallicities derived through spectral and photometric analysis - for example [Fe/H]=-0.50~dex from Bond et al. and -0.46~dex from Casagrande et al. This age/metallicity scenario translates to a stellar mass estimated to be 1.05\,M$_{\sun}$.  Limited phase coverage of the RV data means that the period isn't well constrained so the orbital solution converges to a number of fits with periods upwards of 10000\,days. The shortest period orbit translates to a secondary minimum mass ({\it M$_{s}$\,$\sin{i}$}) of 0.42\,$\pm$\,0.10\,M$_{\sun}$.

{\it HD\,25874}: Classified as a G2V by \citet{gray06} and listed by \citep{CCDM94} in the Catalogue of Components of Double and Multiple Stars (CCDM), it is identified along with a {\it V}\,=\,12 companion
(position angle 225$^\circ$, separation 29\,arcsec, 1941) as a common proper-motion pair. Jenkins et al. derive an $R'_{\rm{HK}}$ index of -4.95, suggesting an age of about 4\,Gyr, while Casagrande et al. find a 
metallicity of [Fe/H]\,=-0.02~dex, suggesting an age of 9.2\,Gyr, in agreement with the subgiant designation. Given the age-metallicity range, a mass of 1.00\,$\pm$\,0.05\,M$_{\sun}$ is inferred for the primary. 
Since the phase coverage from both the AAPS and HARPS is limited, a variety of solutions exist, however we find the best solution to the data gives rise to a very long period low-mass stellar companion, with a period of 
nearly 200~years and a minimum mass of 0.33\,$\pm$\,0.07\,M$_{\sun}$.  

{\it HD\,26491}: This is classified as a G1V by \citet{gray06}, and previously consistently classified as a G3 \citep{houk75,evans64}, and is identified by \citet{decin2000} as having a Vega-like IR excess. Bond et al. derive a spectroscopic metallicity of [Fe/H]\,=\,-0.08~dex, which is larger than the Casagrande et al. value of -0.11~dex. An age of 10.5\,Gyr is indicated by the isochrone fit which is somewhat older than the 5\,Gyr inferred from an $R'_{\rm{HK}}$ index of -4.95 (Henry et al.). Accordingly a mass of 0.97\,$\pm$\,0.05\,M$_{\sun}$ is assigned to this high-proper motion star. Given the nature of the RV variation (Fig.\,\ref{rvcurve}), orbital solutions with periods significantly longer than $\sim$ 7000\,days are possible, and in all cases the eccentricity is greater than $\sim$\,0.5. An orbital period of 9747.5\,$\pm$\,1223.2\,days ({\it e}\,=\,0.57) translates to a secondary minimum mass of 0.50\,$\pm$\,0.15\,M$_{\sun}$.

{\it HD\,39213}: The Bond et al. metallicity of this K0 dwarf \citep{houk82} is found 
to be [Fe/H]=0.20$\pm$0.07~dex. The measured $R'_{\rm{HK}}$ index of -5.10 relates to an age of 8\,Gyr and by fitting a range of isochrones
with ages from 7.5 and 10\,Gyr respectively, a mass range of 0.93\,$\pm$\,0.05\,M$_{\sun}$ is
inferred. Though the orbital parameters appear tightly defined (helped by the RV data having a phase coverage close to one cycle), the residuals are relatively high (rms=17.1\,ms$^{-1}$) and the $\chi_{v}^{2}$ of 1.97 suggests the fit is significant at under the 5\,per\,cent level. A Keplerian period of $\sim$1300\,days and eccentricity of $\sim$0.2 translate to a companion with a minimum mass of 0.07\,$\pm$\,0.01\,M$_{\sun}$ --
potentially a brown dwarf if the orbit is being seen close to edge on.

{\it HD\,42024}: Given a Casagrande et al. metallicity of 0.19 for this star, isochrones ranging in age from 2.5-4\,Gyr suggest a mass of 1.30\,$\pm$\,0.05\,M$_{\sun}$ for this F7 dwarf \citep{houk78}, though the lack of an $R'_{\rm{HK}}$ index deprives us of a secondary age indicator. Phase coverage is broad and the RV values fold convincingly around a period of 76.26\,days (Fig. \ref{rvcurve}) with well defined orbital parameters to yield a companion minimum mass of 0.066\,$\pm$\,0.003\,M$_{\sun}$, another possible brown dwarf companion. Residuals are above average with an rms for the fit of 11.8\,m s$^{-1}$.

{\it HD\,64184}: This G3 dwarf \citep{gray06} is listed in SIMBAD as a variable ({\it V}\,=\,7.49-7.55), though no variability flag is marked in HIPPARCOS. The Bond et al. and Casagrande et al. metallicities are in close agreement at -0.23 and -0.18~dex respectively, and an age between 7-10\,Gyr is indicated by the isochrone fits, somewhat older than the $\sim$4\,Gyr inferred from the $R'_{\rm{HK}}$ index (-4.88; Henry et al.). The stellar mass is estimated at 0.9\,$\pm$\,0.1\,M$_{\sun}$. The eight RV measurements fold convincingly around a period of 17.86\,days (Fig.\,\ref{rvcurve}) and the orbital parameters, which are tightly defined by the broad phase coverage, suggest a companion with an {\it M$_{s}$\,$\sin{i}$} of 0.170\,$\pm$\,0.001\,M$_{\sun}$ and an orbital separation of 0.130\,$\pm$\,0.002\,AU. With a sufficiently high orbital inclination ($\geq$87$^\circ$), the secondary could provide sufficient obscuration of the primary's surface for a variation in {\it V} of $\pm$\,0.03 magnitudes, given that a mass of 0.160\,M$_{\sun}$ would be no brighter than {\it M$_{V}\sim9.6$} at 6\,Gyr - \citep{baraffe98}.

{\it HD\,120690}: This G5+V, as classified by \citet{gray06}, has an $R'_{\rm{HK}}$ index
of -4.78 (Henry et al.), equivalent to an age of $\sim$3\,Gyr. Metallicities are found to be -0.10 and -0.08~dex from Bond et al. and Casagrande et al. respectively, suggesting an age from the isochrone fits somewhere between 6.5 and 9\, Gyr, from which a stellar mass of 0.98\,$\pm$\,0.05\,M$_{\sun}$ is inferred. While limited phase coverage of RV data means that the orbital period is only poorly constrained by periodogram analysis (1900\,$\pm$\,500\,d), a full orbital solution yields tightly defined parameters. A Keplerian period of 3799.98\,$\pm$\,18.12\,days and an eccentricity of 0.34 indicate a secondary minimum mass of 0.59\,$\pm$\,0.01\,M$_{\sun}$. With the secondary contributing upwards of $\sim$0.4\,per\,cent of the flux at 5500\,\AA, the central region over which radial velocities are determined, contamination of the primary spectrum could be a contributory factor. It could also affect the star's colour-magnitude location making it appear redder (and apparently more evolved) possibly accounting for an element of the age discrepancy mentioned above. These issues are considered further in Section\,\ref{contam}.  We do note that a second signal may be present in the data, with a period of 531\,days and a semi-amplitude of 32\,m s$^{-1}$, which if it were a genuine Doppler signal, would give rise to a planet with mass around 1\,M$_{\rm{J}}$.  The addition of this signal can serve to decrease the rms by a factor two.

{\it HD\,121384}: Classified as G8V \citep{gray06}, and listed in \citet{CCDM94}
along with a common {\it V}\,=\,13 proper-motion companion (position angle 45$^\circ$, separation 31\,arcsec, 1941), this star is identified by \citet{oud92} and by \citet{aumann91} as having a Vega-like IR excess. Bond et al. find a [Fe/H] value of -0.40$\pm$0.07~dex, in excellent agreement with the value found by Casagrande et al. of -0.39~dex, yet in good agreement within the uncertainties with Bond et al. The colour and magnitude are well fit by 7-10\,Gyr isochrones indicating that the primary component is indeed evolving away from the main sequence - a view further evidenced by the low level of $R'_{\rm{HK}}$ activity (-5.22, \citealt{henrytHK}). Accordingly its stellar mass is estimated at 0.98$\pm0.10$\,M$_{\sun}$. The 179-day period is sharply defined by the broad phase coverage and the Keplerian solution, albeit with a relatively large rms of 15.1\,m s$^{-1}$.  The data indicate there is a 0.17\,$\pm$\,0.01\,M$_{\sun}$ companion orbiting with an eccentricity of 0.84. An eccentricity of this magnitude (Fig.\,\ref{per_ecc}) places this binary in the extreme upper tail of the eccentricity distribution for systems with periods $<$\,1000\,days \citep[Fig.~6a]{duqmayor} and when combined with the window function, makes the detection of this signal from periodogram analysis alone very difficult (see Fig.~\ref{pgram}).

{\it HD\,131923}: The colour-magnitude location is well fit by 9-12\,Gyr isochrones with metallicities of -0.05 and 0.06~dex for Bond et al. and Casagrande et al. respectively.  These close to solar values suggest that this high proper-motion G4V star \citep{gray06} is starting to evolve away from the main sequence; by contrast, the age inferred from its $R'_{\rm{HK}}$ index (-4.90, \citealt{henrytHK}) is only $\sim$4\,Gyr. Given the monotonic variation in RV measurements, the period is poorly defined, though a 7496\,day Keplerian solution emerges with an eccentricity of 0.72. The rms for the fit is above average ($\sim$9\,m s$^{-1}$) and this is reflected in a $\chi_{\upsilon}^{2}$ of 21.8. With an inferred mass for the primary of 1.05\,$\pm$\,0.05\,M$_{\sun}$, the orbital parameters translate to a secondary minimum mass of $\sim$0.52 $\pm$ 0.06\,M$_{\sun}$. The star is identified in the HIPPARCOS catalogue as a `suspected non-single' object.

{\it HD\,145825}: This G3 dwarf \citep{torres06} again appears to have a metallicity consistent with solar ([Fe/H]=-0.04 and 0.12~dex from Bond et al. and Casagrande et al.) and isochrone fitting suggests an age under 3\,Gyr, which is consistent with the relatively high level of $R'_{\rm{HK}}$ activity (-4.74, \citealt{henrytHK}). Consequently the stellar mass is estimated at 1.03\,$\pm$\,0.05\,M$_{\sun}$. Recent RV measurements have improved the phase coverage, leading to a more sharply defined period of 6024\,$\pm$\,163\,days. With a minimum mass of 0.06\,$\pm$\,0.01\,M$_{\sun}$, this adds to the list of brown dwarf candidate companions in our sample.

{\it HD\,150248}: Consistently classified as a G3V \citep{gray06,houk78,evans57}, the sub-solar metallicity of [Fe/H]=-0.11~dex from Bond et al. and -0.13~dex from Casagrande et al. suggest an age from isochrone fits of $\sim$ 7\,Gyr, which contrasts with an activity-inferred age of $\sim$\,4\,Gyr ($R'_{\rm{HK}}$ activity of -4.88 from Henry et al.).  A mass of 0.93\,$\pm$\,0.05\,M$_{\sun}$ is estimated for the star. The orbital parameters are fairly well defined, with a period of 3272.4\,$\pm$\,28.7\,days and eccentricity of 0.67, relating to a companion with a minimum mass of 0.10\,$\pm$\,0.02\,M$_{\rm J}$. 

{\it HD\,156274B}: Listed as a multiple-star system in \citet{CCDM94}, comprising four known components: Gl\,666A, a G8V \citep{eggl13}, Gl666B, an M0 dwarf (separation 7.5\,arcsec, 1880), a {\it V}=12.5 companion and a 14.0 companion (respectively 279$^\circ$, 41.8\,arcsec, 1900; 30$^\circ$, 47\,arcsec, no year). Our RV measurements indicate that Gl666A is itself a spectroscopic binary. No spectral metallicity was determined but Casagrande et al. find a metal-poor value of [Fe/H]=-0.40~dex. Such a value demands isochrones of 9-10\,Gyr, though the star would appear to be no older than $\sim$6\,Gyr judging from its $R'_{\rm{HK}}$ index (-4.95, Jenkins et al.) in which case isochrones with metallicities of 0.1-0.2\,dex provide complementary fits. These two scenarios translate to a stellar mass estimated at 0.83\,$\pm$\,0.06M$_{\sun}$. The best fit period is found to be 524~years, implying a secondary minimum mass of 0.20\,$\pm$\,0.03\,M$_{\rm J}$, the fit has an rms of 5.95 m s$^{-1}$, however we note that the uncertainties on the measured quantities from the bootstrap are formal to the solution presented.  It is clear that this is the minimum best fit to the data and so the true solution could be very different and therefore properties like the time of periastron passage are essentially unconstrained. 

{\it HD\,158783}: This G3/G5 dwarf \citep{houk75} appears to have a metallicity slightly under solar (-0.05~dex from Bond et al. and 0.05~dex from Casagrande et al.) and isochrone fits demand an age of 8--10\,Gyr, significantly above the 4\,Gyr age inferred from the star's $R'_{\rm{HK}}$ index (-4.91, \citealt{henrytHK}). A stellar mass of 1.04\,$\pm$\,0.05\,M$_{\sun}$ is inferred. Orbital parameters are tightly defined given the limited phase coverage of the RV measurements and the Keplerian fit is significant at the 20\,per\,cent level. A period of 4534.78$\pm$224.92\,days and zero eccentricity translate to a secondary minimum mass of 0.20\,$\pm$\,0.02\,M$_{\sun}$.

{\it HD\,162255}: Though no $R'_{\rm{HK}}$ index is given for this G3 dwarf \citep{houk88}, isochrone fits based on $\sim$-0.01\,dex (Bond et al.) and 0.17~dex (Casagrande et al.) metallicities suggest an age ranging from 5-9\,Gyr, translating to a stellar mass of 1.12\,$\pm$\,0.08\,M$_{\sun}$. The eleven RV measurements fold convincingly around a 47.95-day period, indicating a companion minimum mass of 0.333\,$\pm$\,0.001\,M$_{\sun}$.

{\it HD\,169586}: This G0V star \citet{houk82} appears metal rich ([Fe/H]\,=\,0.32~dex, Casagrande et al.) suggesting an age around 2-4\,Gyr, which is in agreement with that derived from the $R'_{\rm{HK}}$ index (-4.92, \citealt{henrytHK}); the stellar mass is estimated at 1.25\,$\pm$\,0.05M$_{\sun}$. The acquisition of two of the most recent RV measurements has defined a relatively sharp extremum in what was originally a monotonic RV variation so that a 2935-day orbit with an eccentricity of 0.35 appears well constrained. Given that the secondary has a minimum mass of 0.68 \,$\pm$\,0.22 M$_{\sun}$ (equivalent to an {\it M$_{V}$}\,=\,9.4 companion) contamination of the primary's spectrum has almost certainly taken place, and is possibly a reason why the rms for the fit (60.2\,m s$^{-1}$) is so large, along with the large $\chi_{\upsilon}^{2}$.  Nevertheless the existing RV measurements clearly indicate that the primary has one or more companions.

{\it HD\,175345}: Listed in \citealt{CCDM94} as having a V=14.2 proper motion companion (B\,413: 252$^\circ$, 5.4\,arcsec, 1927), our measurements indicate that this G0 dwarf \citep{houk88} is itself a
spectroscopic binary. There is no $R'_{\rm{HK}}$ index for this star and the isochrone fits based on its Casagrande et al. metallicity of [Fe/H]=-0.16~dex gives an age range of 4-9\,Gyr along with an inferred
mass of 1.05\,$\pm$\,0.05\,M$_{\sun}$. There is a clear fold of the RV data around a 312-day period so orbital parameters are tightly constrained, however the residuals are relatively large (rms\,=\,14 m s$^{-1}$) making the fit statistically poor, even though  the $\chi_{\upsilon}^{2}$ is fairly low (3.44). With a minimum mass of 0.48$\pm$\,0.08\,M$_{\sun}$, (equivalent to M$_{V}$\,=\,9.5 or brighter) the secondary is again a source of spectral contamination.
\smallskip

\begin{figure}
\centering 
\epsfig{file=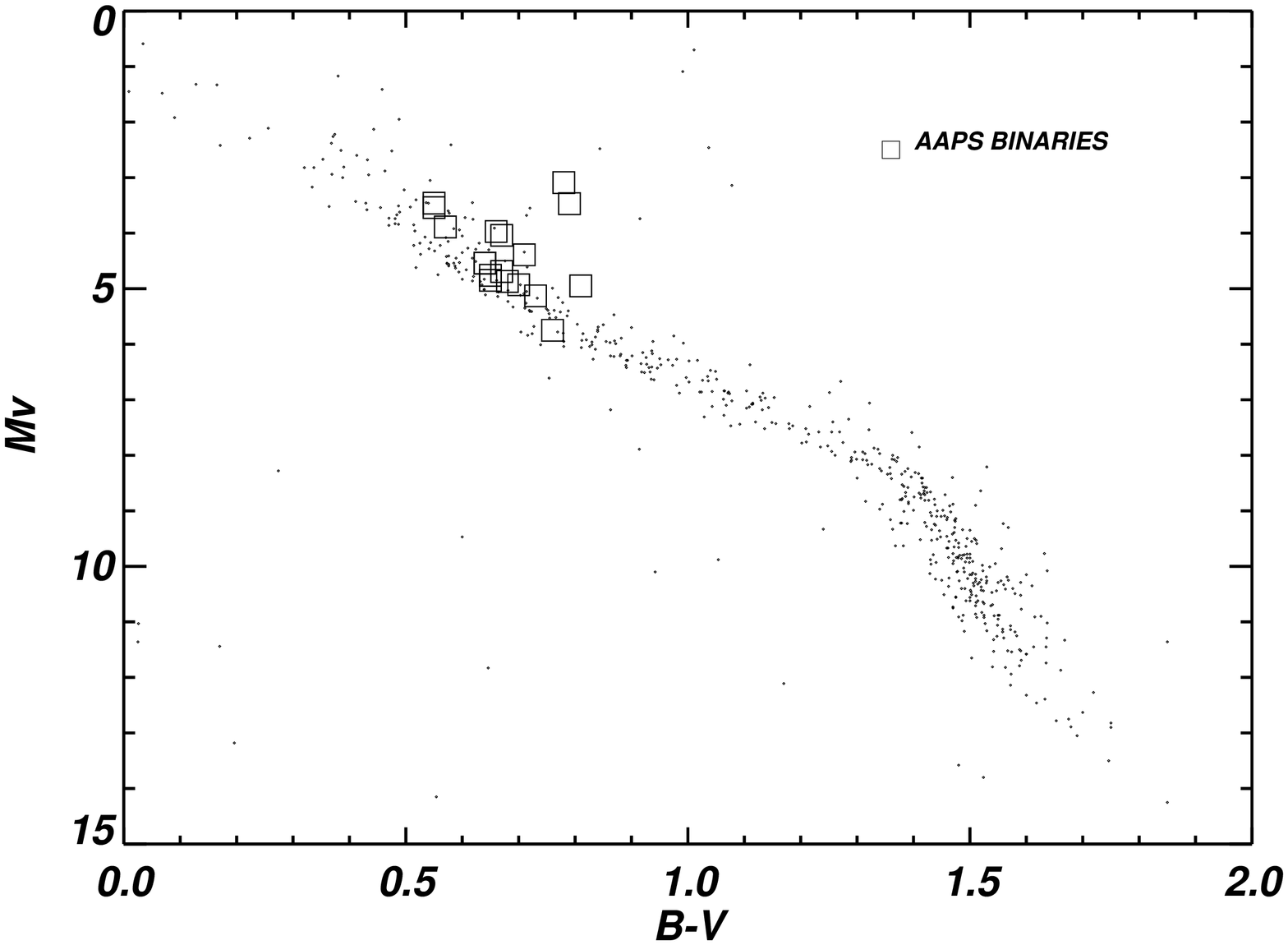,width=6cm, angle=90}
\epsfig{file=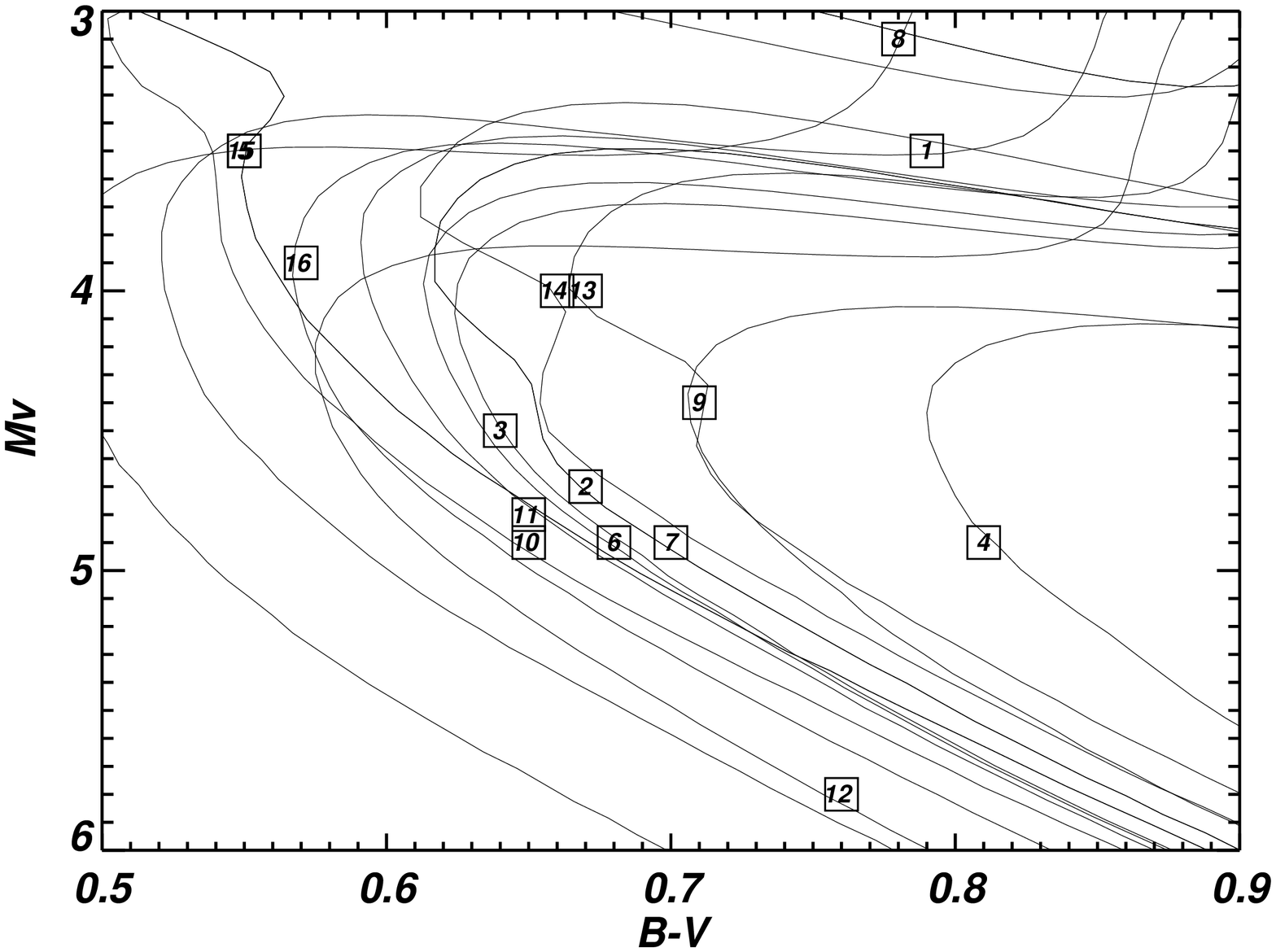, width=6cm, angle=90} \caption{(a) The AAPS
binaries are indicated on a colour-magnitude diagram constructed using
HIPPARCOS data for solar-neighbourhood stars. (b) Best-fitting
isochrones for each star; the numbers assigned to the stars are given
in Table\,\ref{primaries}} \label{hr}
\end{figure}

\begin{figure}
\centering 
\vspace{0.5cm}
\epsfig{file=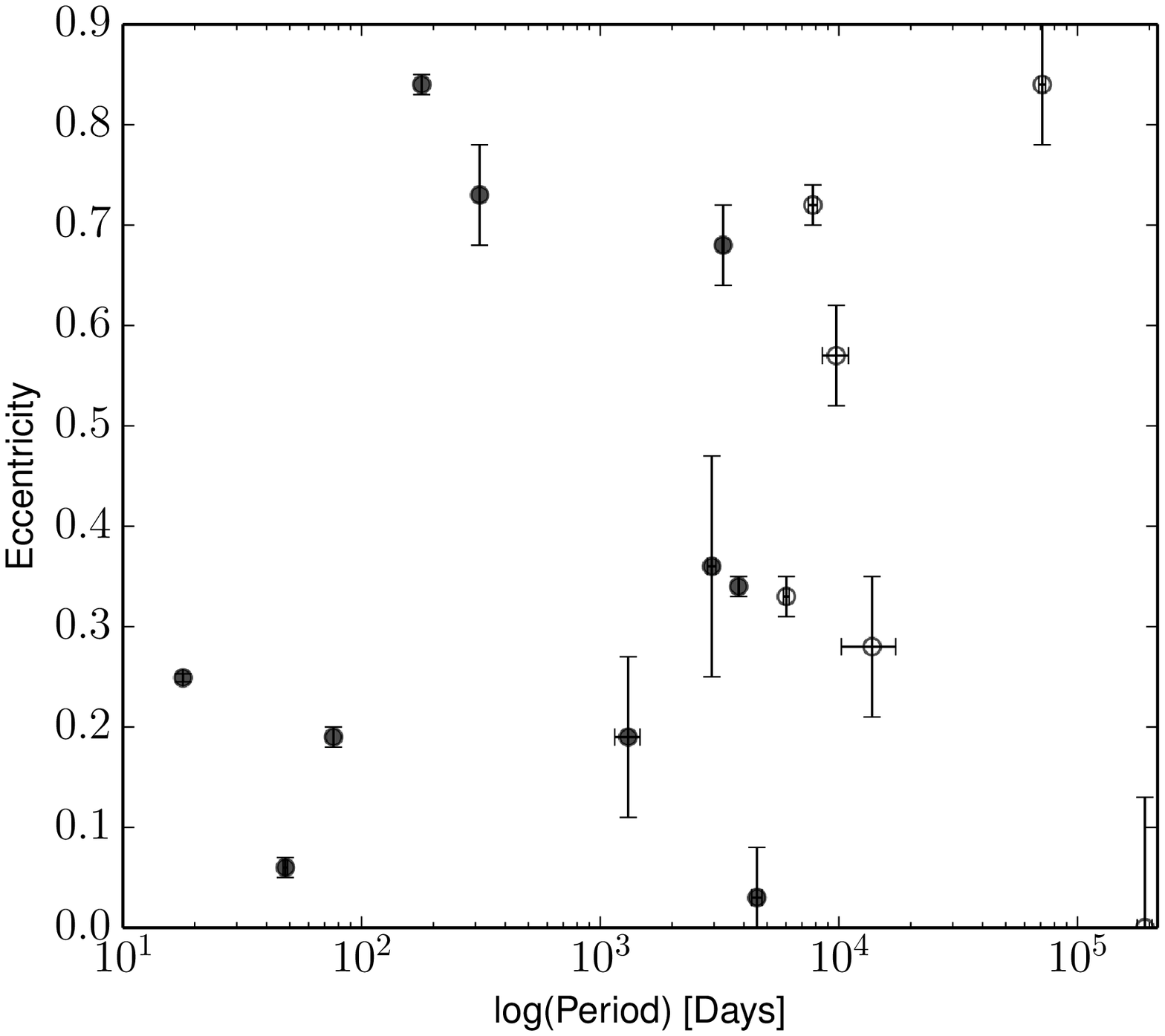,scale=0.7} 
\caption{A plot of eccentricity versus period for the AAPS binaries. The binaries
represented by filled circles are from Table\,\ref{orbit}\,(i) and have
well constrained Keplerian periods; open circles are used for binaries
with poorly constrained periods - given in Table\,\ref{orbit}\,(ii).} \label{per_ecc}
\end{figure}

The locations of these stars on a HIPPARCOS-based HR diagram, along with a summary of the isochrone fits, are shown in Fig.\,\ref{hr}\,a\,\&\,b, respectively. We also show the distribution of eccentricity 
versus period in Fig.\,\ref{per_ecc}.  We find that the companions to HD121384 and HD175345 have very high eccentricities for companions with orbital periods below 1000 days, placing them in the extreme upper tail 
of the distribution in this parameter space.  We also find a few of the longer period companions that have not been fully constrained yet due to the limited baseline of the data have moderate to high 
eccentricity.  Although the periods of the orbits could be significantly longer, the eccentricities are rather well constrained with the current data in hand, assuming a single Keplerian fit.  The one 
obvious exception is HD156274B which currently only exhibits a linear trend with velocity over time, and although we fit this will a circular model, this could very well be highly eccentric. 

\subsection{High Contrast Observations}

A number of the binaries we have discovered in this work have been followed up using adaptive optics systems 
to search for direct confirmation of the companions.  Obtaining orbital motion from direct images of low-mass 
companions to bright Sun-like stars, especially when coupled with RV information, can yield dynamical 
masses for the companions (\citealp{liu07}).  Combining dynamical masses with photometric colours and 
spectra can allow evolutionary and atmospheric properties to be well constrained and models to be tested 
(e.g. \citealp{dupuy09}).

In \citet{jenkins10} we observed two of the host stars we report new binaries for in this work, HD25874 and HD145825.  
In that work we found contrast ratios of greater than 11 magnitudes at separations of only 0.5$''$ using the VLT NAOS CONICA 
instrument (\citealp{rousset03}) in Simultaneous Differential Imaging mode.  Although a tentative 
detection of the companion around HD25874 was discussed, further analysis revealed this to be a probable 
artifact of the reduction and analysis procedure, and therefore no companion detection was conclusively 
made for either of these stars with mid-T dwarf masses of around 50 M$_{\rm J }$ or so.  Some of the other stars we report 
companions for are included in our ongoing NACO/NICI imaging campaign.

\subsection{HIPPARCOS Astrometry}

\begin{figure}
\centering 
\epsfig{file=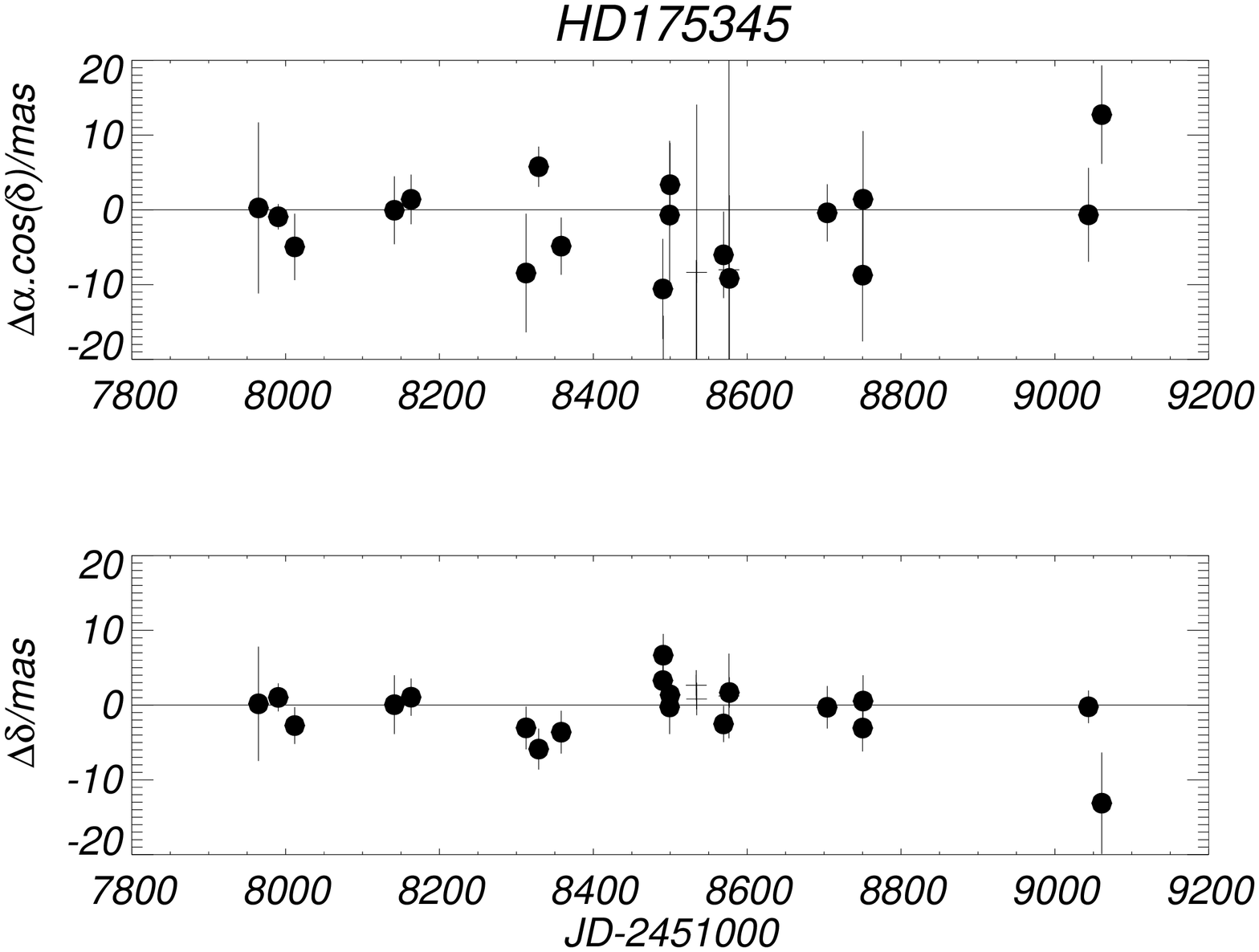,width=6cm, angle=90} \caption{An example astrometric
plot of the positional residuals and associated errors for HD\,175345.} \label{astrometry}
\end{figure}

Out of these sixteen binaries, four (HD\,39213, HD\,120690, HD\,121384, and 
HD\,175345) have well constrained periods in the range
0.5--6\,years, making them suitable candidates for analysis of their
HIPPARCOS astrometry. In order to determine if there are any
astrometric signatures that would allow us to place additional constraints
on our orbital parameters, we extracted the astrometric data
from the HIPPARCOS database \citep{hipparcos07} and derived positional
residuals ($\alpha\cos{\delta}$, $\delta$). In Fig.\,\ref{astrometry}
we show these with associated errors for HD\,175345. 

There are no significant variations obvious in the plots for the four
candidates, though in the case of HD\,175345, periodogram analysis
reveals significant power at around 320\,days. (c.f. a 312-day period from
RV analysis). Failure, however, of these data to fold around either
period suggests that the spectral power owes more to sampling of the
data (which have a strong 300-day element) than to any possible
astrometric signature. The significance of the astrometric variations
is clearly low and we do not attempt further to constrain our orbital
parameters using these data.

\subsection{Secondary Flux Contamination}\label{contam}

\begin{figure}
\includegraphics[scale=0.45]{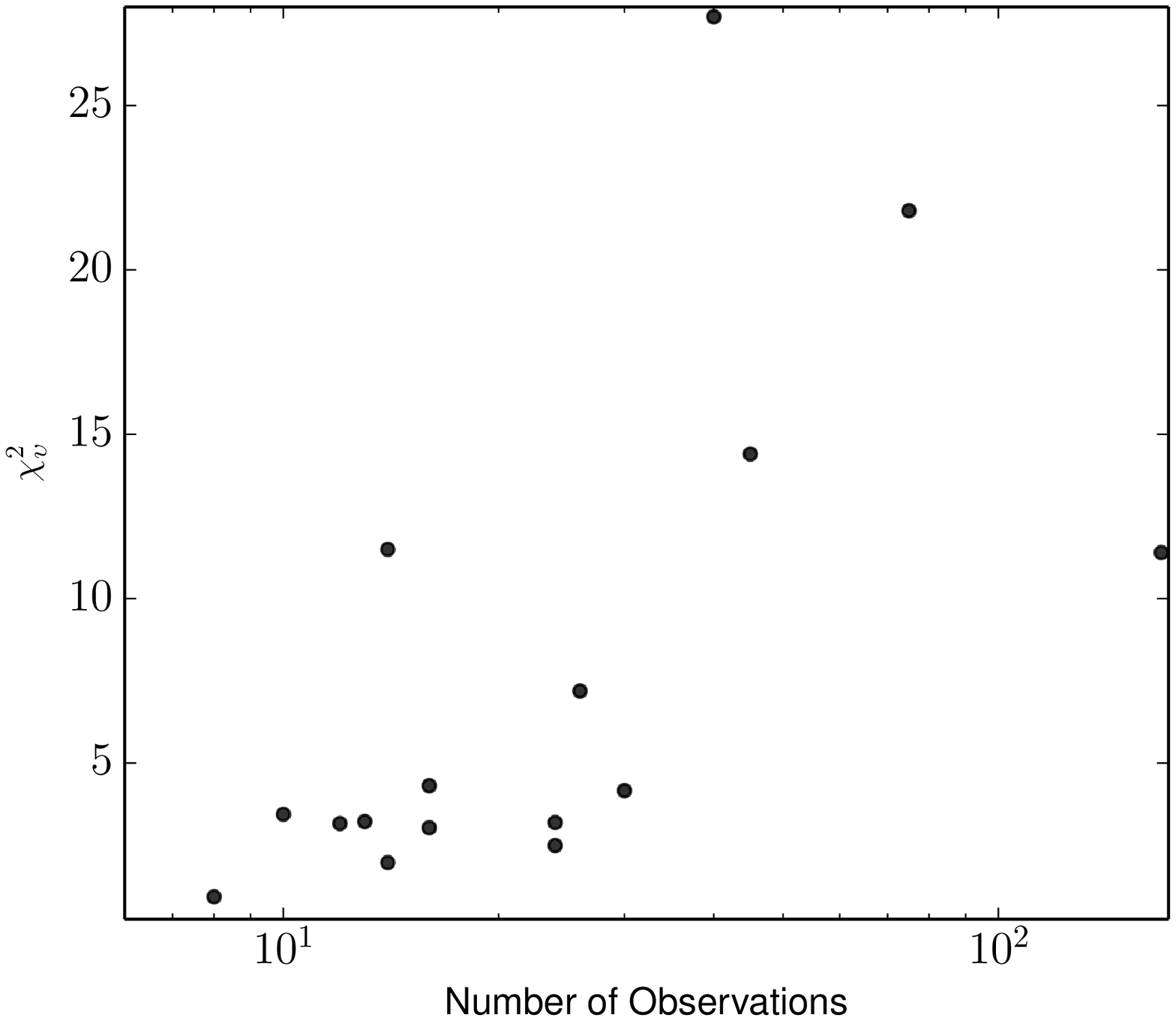}
\hspace{-1.2cm}\includegraphics[scale=0.45]{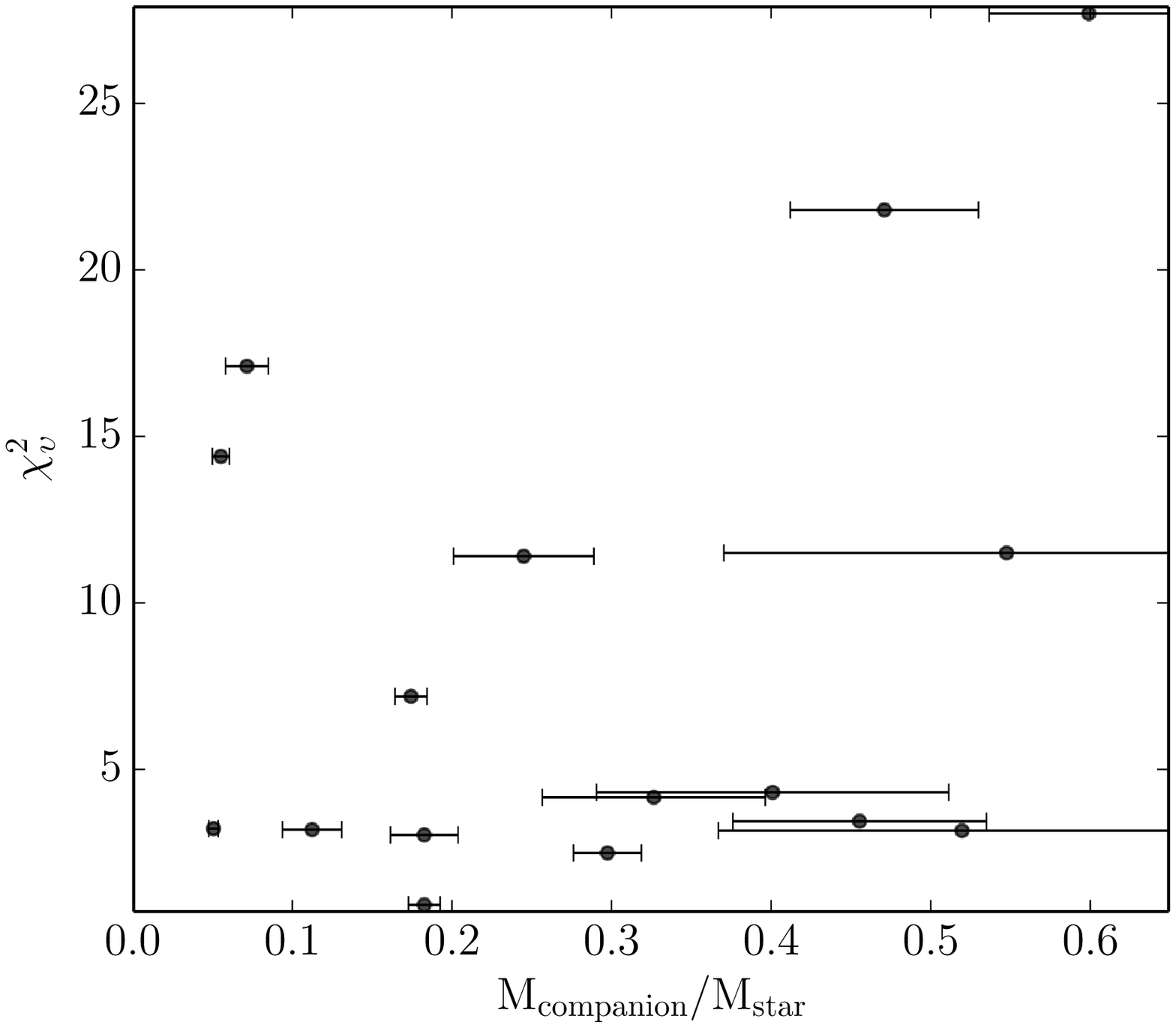}
\vspace{-3.2cm}
\caption{Quality of orbital fit (in terms of the statistical significance attributed to the
$\chi_{v}^{2}$ value) versus (a) the number of observations per star (b) the minimum mass ratio of the system.} \label{chi}
\end{figure}

In those cases where the orbits are fully constrained and the secondary minimum mass is $\sim$0.5M$_{\sun}$ (HD\,18907, HD\,26491, 
HD\,120690, HD\,131923, HD\,169586 and HD\,175345) the contribution of the
secondary to the overall flux is at least 1\,per\,cent and the signature of this contribution is found in the quality of the orbital
solutions: the rms scatter of these stars is generally much higher than the internal measurement uncertainty. The reason for this is that flux
from the secondary, which is contaminating the primary's spectrum, will be associated with a different radial velocity at each of the
subsequent observations from that when the template spectrum was acquired. The radial velocity fitting process relies on the assumption
that the primary's spectrum is modified only by the Doppler shift of the primary and the spectrograph PSF variation. Consequently, less than optimal
solutions can be expected when a faint secondary contribution to the primary's spectrum is present at variable Doppler shifts. In essence,
the fitting process matching the observed and synthetic spectra would be expected to generate larger measurement errors, and this is the case
particularly for HD\,169586. Fig.\,\ref{chi} shows how the quality of the orbital solutions - measured by the significance of
the fit - varies with the number of observations (plot\,{\it a}) and the binary minimum-mass ratio (plot\,{\it b}).  While no correlation
appears to be evident between the reduced Chi-squared and companion mass ratio, in general, the statistical significance of the fits 
appear to indicate a marked deterioration as the number of data points increase.  This could reflect the fact that these are the systems 
exhibiting the most stellar flux contamination and therefore they required more observations in order to better constrain their orbits, 
or, another possibility is that these systems also contain additional companions, either brown dwarfs or planets, that are 
giving rise to mixed signals that are being manifest once enough RV measurements have been acquired.

Given the precision with which radial velocities can be obtained by the AAPS, we could speculate that where low M$\sin{i}$ values yield
statistically poor fits (HD\,39213, HD\,121384), the orbital inclinations are low and that the companions
have masses high enough for spectral contamination to be taking place. The large errors in RV measurements for HD\,39213 and HD\,156274 lend
credence to this scenario, though this isn't the only explanation. It is also possible that these are multiple systems for which a
double-star solution simply isn't appropriate. Moreover, enhanced activity in any of these stars (bearing in mind that the $R'_{\rm{HK}}$
index is merely a `snapshot' measurement) would mean that the internal errors would not properly reflect the uncertainty in the RV
measurements. This is why it is pragmatic to use relatively inactive stars in Doppler searches. For the targets considered here, enhanced
activity is unlikely to be the cause: \citet{henrytHK} estimate that 90\,percent of the time a single $R'_{\rm{HK}}$ measurement for a solar-type
star is sufficient to identify correctly if it has an activity greater, or less, than $R'_{\rm{HK}}$\,=\,-4.75, and all but one of the stars
have $R'_{\rm{HK}}$ indices lower than this. Where the measurement errors are low but the orbital solutions are statistically weak (HD\,121384
and HD\, 162255), the most convincing explanation is that these systems comprise more than a single companion.

\begin{figure}
\centering 
\epsfig{file=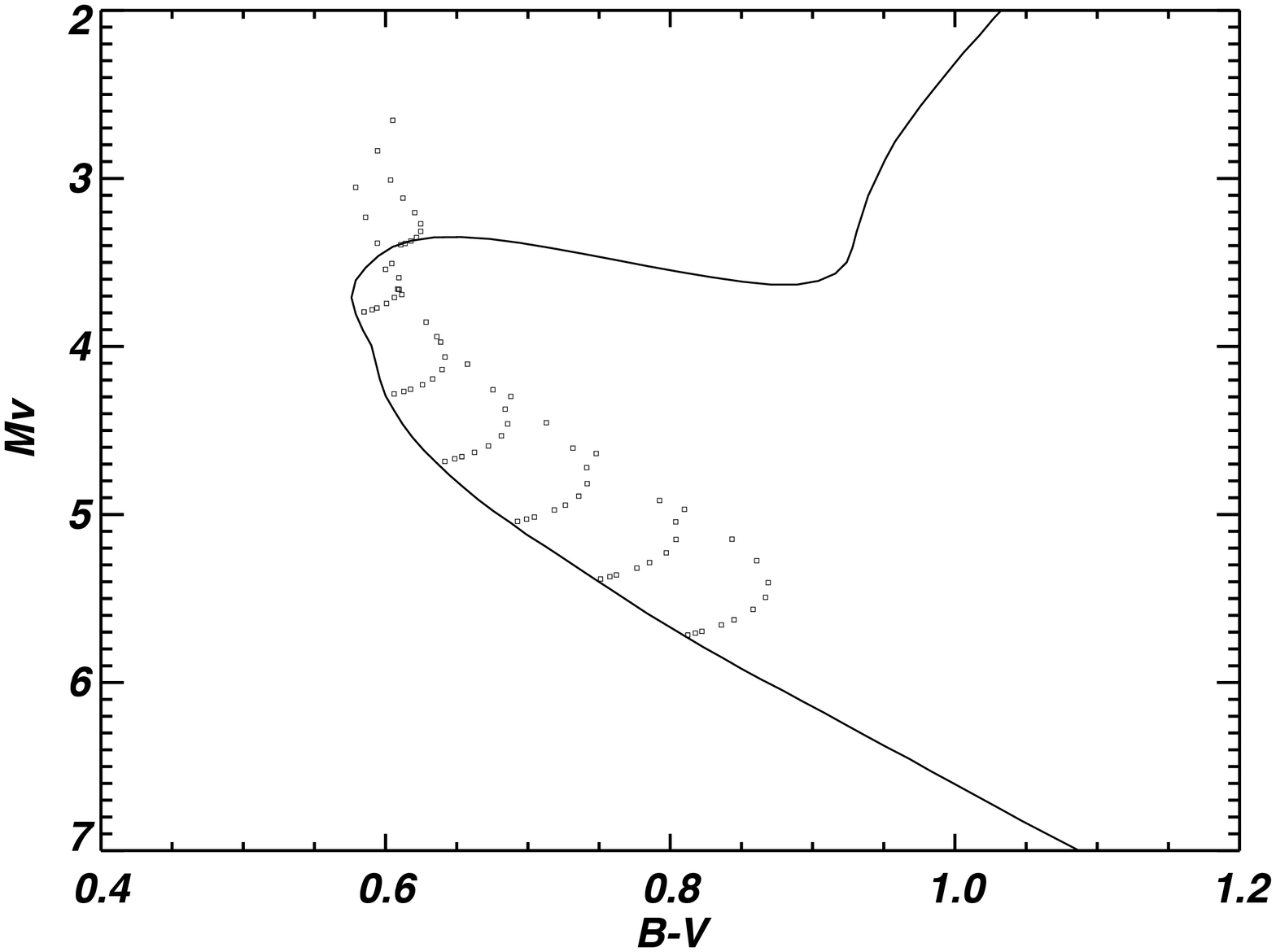,width=6cm, angle=90} \caption{Effect
of binarity on the colour and magnitude at 5.5\,Gyr for solar
metallicity primaries ranging in mass from 0.88-1.20M$_{\sun}$.
Secondary mass values increase in 0.05M$_{p}$ increments from
0.5-1.0\,M$_{p}$.} \label{binseq}
\end{figure}

The effect of binarity on the colour and magnitude of a primary of any age and metallicity can be modelled easily using the Yonsei-Yale
isochrone data. Fig.\,\ref{binseq} shows the variation in colour and magnitude for a 5.5\,Gyr, solar metallicity isochrone ranging in mass
from 0.88-1.20\,M$_{\sun}$ (solid curve). The effect of adding a secondary is marked by the dotted curves at various mass intervals, with 
each point representing a secondary companion that increases in mass from 0.5-1.0\,M$_{p}$ in steps of 0.05\,M$_{p}$. 

For companions below 0.55\,M$_{p}$ the effect of binarity on the colour-magnitude location
of the primary is negligible. The effect of a 0.70\,M$_{p}$ companion is to make the primary appear redder and brighter by around 0.05 and 
0.3 magnitudes respectively, for a solar-mass star, with slighter larger and smaller values for less massive and more massive primary stars, respectively. 
Such an effect generally makes the unresolved pair appear older and/or more metal rich; however if the primary itself is evolving
away from the main sequence, the companion will make the pair appear `bluer'. These effects will complicate the process of mass estimation.
For example, the presence of a 0.70M$_{p}$ companion translates effectively to a systematic error of $\sim$\,0.04M$_{\sun}$ in the mass
of the primary. This is of the order of the error in mass due to the age-metallicity uncertainties. The orbital solutions indicate binary
mass ratios generally significantly less than unity (HD\,169586 being the exception) so that the effect of binarity on the determination of
the primary mass (and by extension the secondary mass) is negligible. Clearly, for many of the binaries in our sample, the uncertainty in
secondary mass is due principally to poorly constrained orbital parameters.

\subsection{Mass Distribution}\label{mass}

The binaries and planetary companions to solar-type stars reported by the AAPS provide an opportunity to examine, from a single radial
velocity survey, the distribution of M$\sin{i}$ values for a mass regime extending from Jovian through brown dwarf to sub solar in value.

In order to derive a more meaningful distribution of minimum-mass ratios, we need to impose a period cut-off on both the planetary and
binary companions so that we count only those companions within a certain distance of the primary stars. As a rule, planetary candidates
are announced when the phase coverage of the RV data are close to one orbital period. For the AAPS, which has been operating since 1998, we
can say that the inventory of exoplanet candidates orbiting with periods up to 12\,years (i.e. out to around the orbit of Jupiter in the 
solar system) is reasonably complete down to the level permitted by a Doppler precision of $\sim$3ms$^{-1}$, i.e. complete for Jupiter-mass 
objects and above. Brown dwarf and low mass stellar companions induce larger reflex velocities making them easier to detect over a greater 
range of distances and periods. In order to compare directly with \citet{duqmayor}, six out of the sixteen binaries have periods greater than 
12\,years and these need to be excluded from our count.

\begin{figure}
\centering \epsfig{file=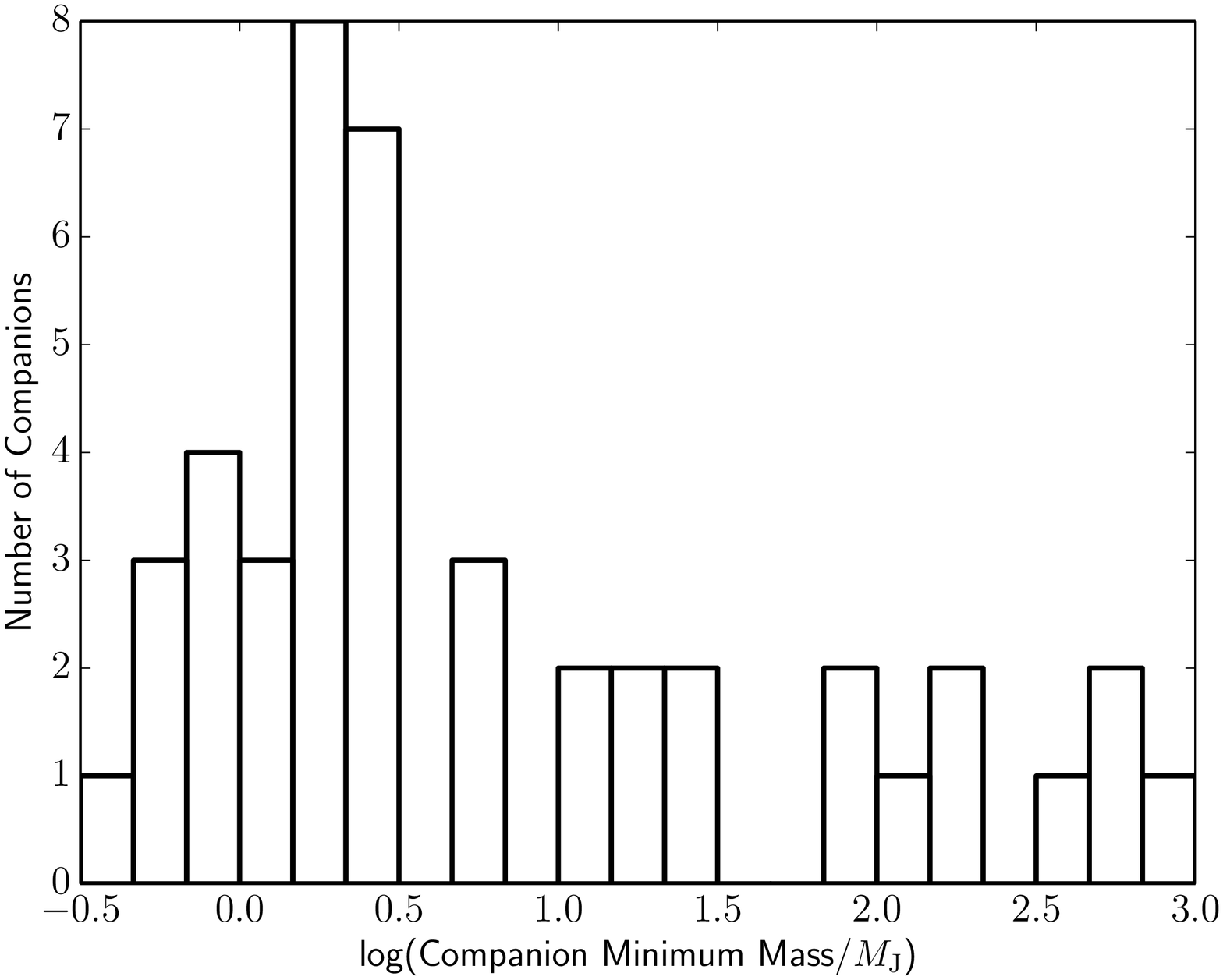,width=10cm, angle=90}
\caption{The raw distribution of companion minimum masses uncovered by the AAPS with periods less than 12 years. }\label{mass_func}
\end{figure}

Ten remaining binaries out of a sample of 178 stars is around half of that expected from the period distribution given in Fig.~13 of
\citet{raghavan10}, when normalised by the multiplicity fraction, though there are several reasons for this. 
First, our sample excludes all {\it known} short-period spectroscopic binaries
with separations less than 2\,arcseconds;  second, all binaries beyond 2\,arcseconds detected using Hipparcos data are excluded from our sample; third, `double-lined' spectrum 
binaries are immediately removed as soon as they are recognized; fourth, our requirement for the AAPS target stars to have an $R'_{\rm{HK}}$ index 
below -4.5 has the effect of filtering out some short-period, chromospherically active binaries -- RS\,CVn/W\,UMa types for example -- though 
admittedly these are few in number. The observed distribution in minimum masses, corrected for completeness for periods up to 12\,years, is shown 
in Fig.\,\ref{mass_func}.

The main features seen in this {\it P}$<$ 12\,yr distribution are (i) a sub-stellar companion mass function rising strongly below 10\,M$_{\rm
J}$, (ii) a comparatively flat distribution of stellar companions, and (iii) a region from $\sim$20-70\,M$_{\rm J}$ (corresponding to the brown dwarf
regime) where relatively few objects are found despite a selection bias in the observations making them easier to detect than planets.  Note that 
although objects do exist in this part of the parameter space (e.g. \citealp{jenkins09}), the term brown dwarf desert was given to highlight the relative 
paucity in comparison to planets and stellar objects (\citealp{marcy00}).   Such
features accord with RV observations elsewhere: the CORALIE, Keck, and Lick surveys all report the same form of sub-stellar function while the
`flat' stellar distribution mirrors that seen in \citet[Fig.~11]{duqmayor} This similarity comes about despite the fact
that the various RV surveys work with different samples and operating strategies. The form of the distribution of planetary and stellar
companions is considered to reflect the different formation mechanisms for these two populations (respectively accretion in dissipative
circumstellar disks and gravitational instabilities in collapsing cloud fragments) and their consequent evolution. The relatively small number
of brown dwarf companions has been noted elsewhere (for example \citealt{halb2000,butler2001,chrisphd}) and may be a reflection of a
formation mechanism different again from that of stars or planets, though \citet{armbonnell2002} argue that its existence is a consequence
of orbital migration of brown dwarfs within an evolving proto-stellar disk.

The question arises as to what effect a correction for inclination would have on the observed distribution. The simplest (crudest)
correction is to scale masses up by a factor of 1/$<\sin{i}>$. This doesn't alter the form of the distribution, nor in this case the total number
of objects that can be regarded as brown dwarf candidates - just those objects out of the survey that can be regarded as such. Clearly, a
knowledge of the precise form of the mass distribution for this (or any) RV survey is precluded until inclinations can be determined
accurately via sub-milli-arcsecond astrometric surveys such as GAIA \citep{sozzetti_gaia} and SIM \citep{sozzetti_sim}, though it would
take a remarkable confluence of inclinations for the objects uncovered by the AAPS to alter the underlying distribution of masses.  Indeed, 
recent attempts have been made to recover the 'true' mass distribution (\citealp{lopez12}), with very few of the sub-10\,M$_{\rm J}$ objects moving 
above this mass limit.

\subsection{Metallicity-Mass Distribution}

One of the most interesting features to emerge from the early study of exoplanets, 
is the dependence of gas giants to be found orbiting stars with super-solar 
metallicities (\citealp{gonzalez97,fischer05,sousa11}).  This 
result is a key prediction of the core accretion scenario for planet formation (\citealp{ida04,mordasini12}).  
However, it seems that this bias towards the most metal-rich stars is only found 
for gas giants and not lower-mass rocky planets (\citealp{udry07,buchhave12,jenkins13b}).  
Therefore, given there is a clear mass dependence as a function of metallicity, it is 
interesting to test what the metallicity distribution looks like for 
binaries drawn from a representative sample.

\begin{figure}
\centering
\vspace{1cm}\epsfig{file=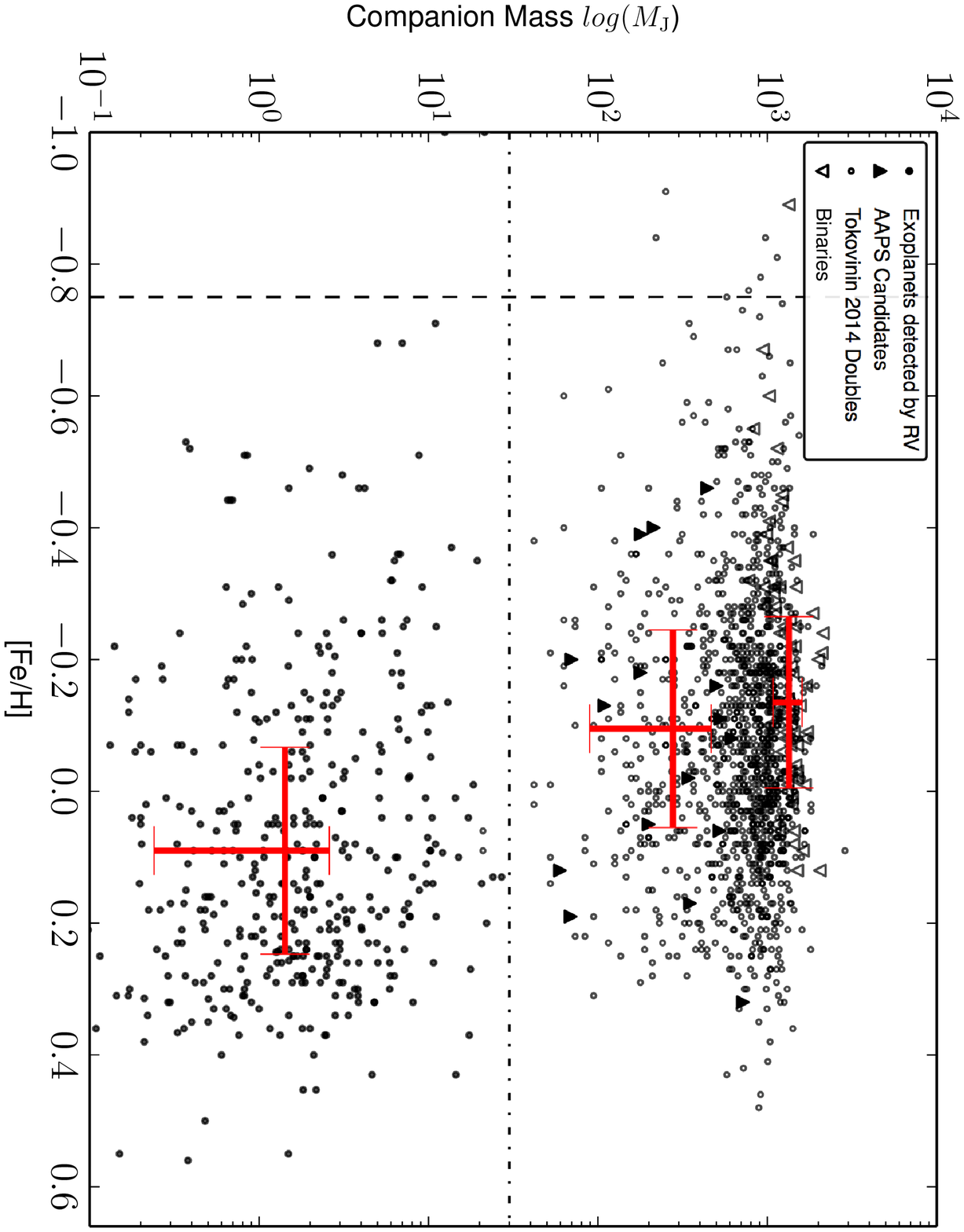,scale=0.75,angle=0}
\vspace{-3cm}
\caption{Metallicity-Mass distribution. Filled circles correspond to exoplanets with $1.5\, M_{\rm J}<M< 5.0\, M_{\rm J}$, discovered using the RV technique, open triangles correspond to binary stars from \citet{halbwachs12} and \citet{duqmayor}, filled triangles show our AAPS candidates, and open circles show the distribution for doubles reported by \citet{tokovinin14}. The metallicities were obtained from \citet{casagrande11}.  The horizontal and vertical lines mark the approximate planet-brown dwarf boundary and the extreme lower tail of the Tokovinin binary distribution, respectively.  The cross-hairs mark the sample medians for the planets, AAPS candidates, and the Tokovinin et al. binaries, increasing in mass respectively.}\label{mass_met}
\end{figure}

In Fig.~\ref{mass_met} we show the distribution in the metallicity-mass plane of planets, brown dwarfs, and stellar binaries that have 
been detected mostly by the radial velocity method, with a large clutch of the stellar binaries being 
drawn from the sample of F,G, and K stars from Tokovinin et al. (2014).  The iron abundances used in this plot were taken from 
high-resolution spectral analysis where possible, generally from the published papers for the detected exoplanets, 
with the Tokovinin et al. primary star metallicities being drawn from \citet{casagrande11}.  The giant planet metallicity 
bias discussed above is evident here, where the sample mean cross-hairs are clearly offset from the sample means at higher 
masses (i.e. above the horizontal dot-dashed line).  The brown dwarf and stellar binaries have mean values in good agreement with 
each other, both with sub-solar values, in comparison to the exoplanet primary mean distribution that is significantly above 
the solar value.  

\begin{figure}
\centering 
\vspace{1cm}\epsfig{file=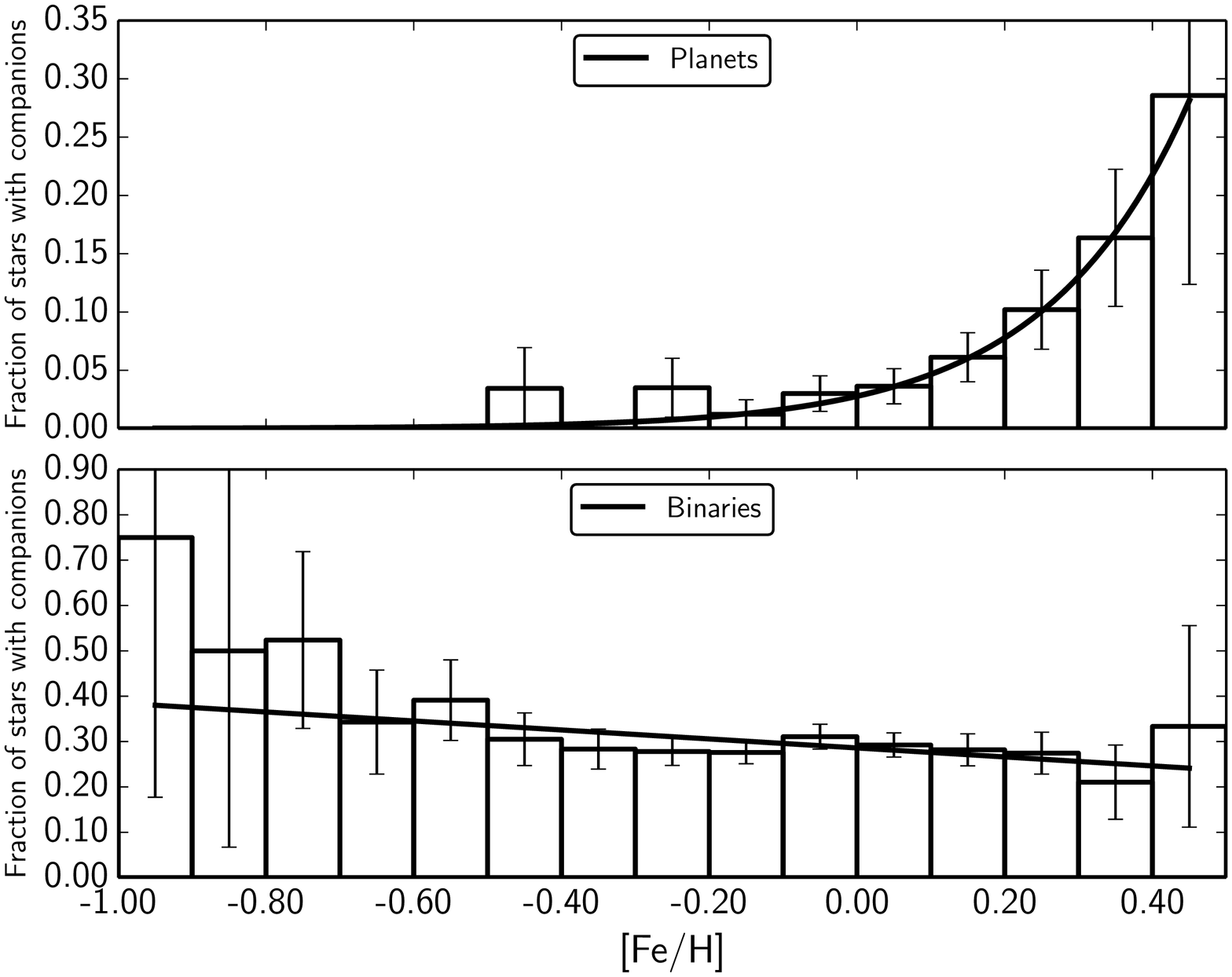,width=13cm, angle=90}
\caption{The metallicity distribution of exoplanet (upper panel) and binary (lower panel) companions.  The best fit model distributions are also shown 
in the plots, where a power-law is used for the exoplanet sample, and a linear function is used to model the binary distribution.
}\label{met_fraction}
\end{figure}

In Fig.~\ref{met_fraction} we show the distribution of exoplanets and binaries as a function of primary star metallicity.  We used 
the large sample of \citet{fischer05} for the exoplanet distribution and again the \citet{tokovinin14} sample for the 
binary population.  The biases in both of these samples are discussed in each of the works, yet they are large enough and have been 
examined well enough that they can be thought of as good representations of their respective field populations.  The full Tokovinin 
et al. sample was not included since we wanted to maintain metallicity homogeneity and also we aimed for a direct overlap in orbital 
separation with the exoplanet sample, meaning we only included binary companions with orbital periods out to 4~years.  For metallicity 
homogeneity, we cross-matched the sub-4~year sample with the \citet{casagrande11} catalogue of metallicities, leaving a total of 
874 binaries or multiple stars, out of the complete 3936 sample.

As was shown in \citet{fischer05}, the distribution of exoplanets follows a power law where the percentage of stars with giant planets 
increases as a function of metallicity.  After constructing a similar histogram 
of values to that in Fischer \& Valenti, we find a power law described by an amplitude of 0.028$\pm$0.002 and an index of 2.23$\pm$0.09, 
which we represent by the dashed black curve in the plot.  Beyond around a solar metallicity, the increase 
in planet hosting fraction rises steeply, possibly accelerating beyond a value of +0.2~dex in metallicity (\citealp{sousa11}).

The distribution of binaries on the other hand is extremely flat across all the metallicity range, within the uncertainties, with the fraction found 
to be 43$\pm$4\%, in excellent agreement with \citet{raghavan10}.  The best fit weighted linear function is shown in the figure and has values for the gradient ($b$) and offset 
($a$) of 0.286$\pm$0.011 and -0.099$\pm$0.044, respectively.  These parameters are fairly insensitive to different bin widths, therefore, it is clear 
that the binarity fraction as a function of metallicity is significantly different to the planetary system fraction.   

We investigated if there is any dependence on the binary fraction with orbital separation by constructing the same distribution on the sample of binaries 
with orbital periods longer than 4~years, and on the full sample regardless of orbital period.  We found no significant differences between the 
distributions, however we do note a small drop in the fraction of binaries in the metal-poor regime (-0.8$\le$~[Fe/H]~$\le$-0.5) for the longer 
period sample.  Although it still agrees with a flat distribution within the uncertainties, in the future it may be worthwhile to revisit this regime 
with more binaries in these bins to see if this drop in fraction becomes significant, which would indicate there is a dependence of the binary 
fraction with separation as a function of metallicity.

\begin{figure}
\centering
\epsfig{file=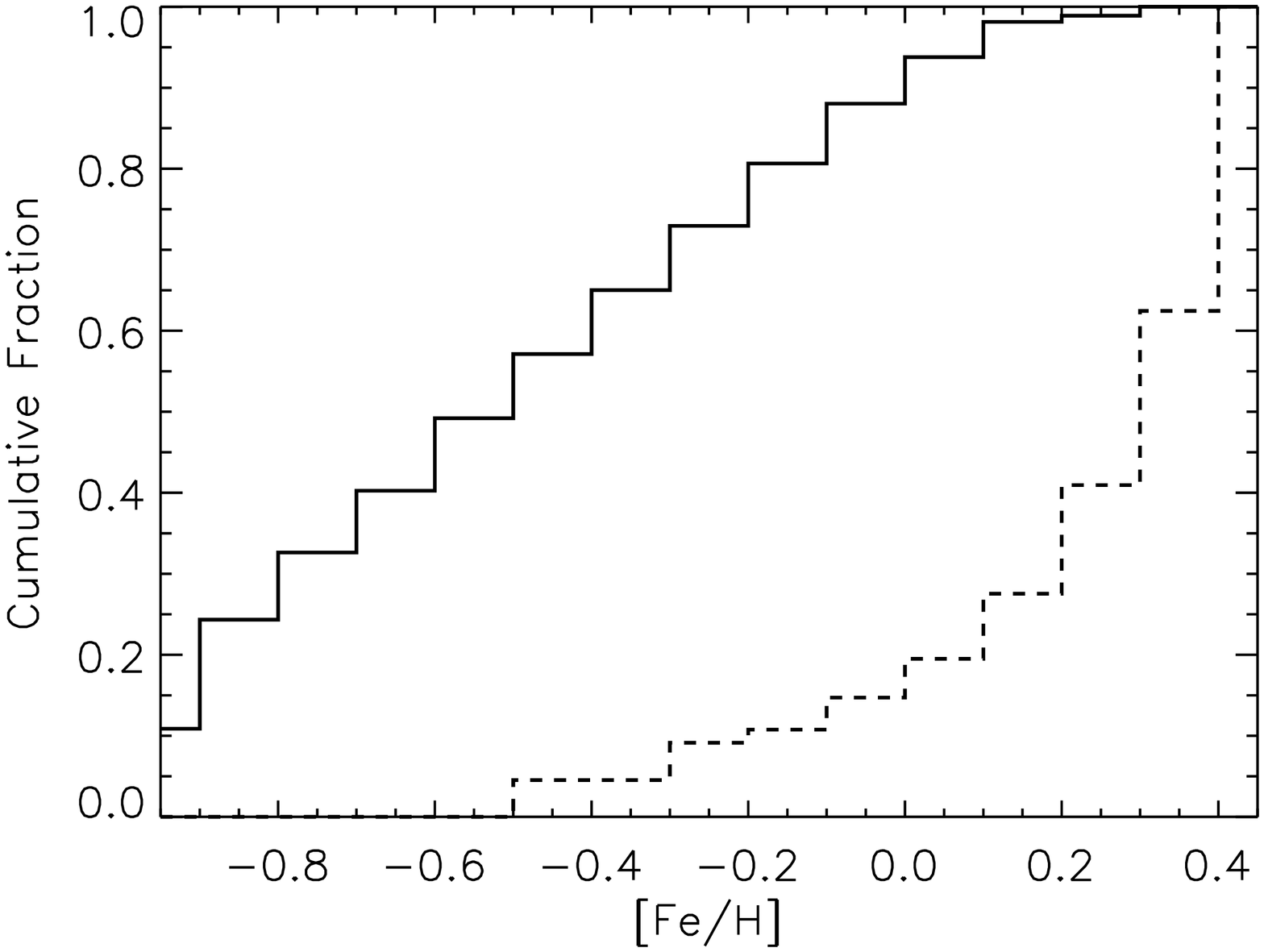,scale=0.5,angle=90}
\caption{The cumulative fraction of the binary fraction distribution (solid curve) and the exoplanet fraction distribution (dashed curve).}\label{cumulative}
\end{figure}

In Fig.~\ref{cumulative} we show the cumulative fraction between the binary fraction and the exoplanet fraction as a function of metallicity.  
The early rise at low metallicities in the binary fraction is apparent, along with the rapid rise at the high metallicity end for the exoplanet 
fraction.  We find that the largest disparity between the two populations occurs around 0.15~dex in metallicity, a little over solar metallicity, where 
the planet fraction begins to significantly increase.  A Kolmogorov-Smirnov test yields a D-statistic of 0.733 here, relating to a probability 
of the null hypothesis of 2.377$\times10^{-2}$\%.

\citet{gao14} show that the fraction of binaries with orbital periods of less than 1000~days dramatically falls in the metal-rich regime 
when they split their SDSS and LAMOST FGK star samples up into three different metallicity bins.  They find a total binary fraction of 43$\pm$2\%, 
again in excellent agreement with what we find here, but they calculate binary fractions of 56$\pm$5\%, 56$\pm$3\%, and 30$\pm$5.7\% for 
metallicities ([Fe/H]) of $<$-1.1, -1.1 to $<$-0.6, and $\ge$-0.6, respectively, indicating a drop in the fraction of binaries in the most 
metal-rich region.  Although this is an indication of a higher fraction of binaries near our metallicity lower limit, within the uncertainties 
the distribution is still flat (similar to the result reported in \citealp{raghavan10}).  Therefore, we can not confirm if such a binary fraction 
change exists, a change that is also recovered by some models (e.g. \citealp{machida09}).

\cite{bate14} studied the effects of changes of the metallicity on their star formation models, assuming the dust opacity scales 
linearly with metallicity across a range of metallicities from 1/3rd solar to three times solar metallicity.  He finds no significant 
changes in the multiplicity fraction with metallicity, suggesting gas opacity does not overtly change the large-scale properties 
of star formation from the fragmentation of giant molecular clouds.  Furthermore, he also finds no dependence with metallicity on 
the orbital separation of binaries and higher order multiples, as we find here.  Given that recent works have shown that the 
cooling times in proto-planetary disks increase as a function of metallicity, meaning suppression of disk fragmentation in 
the super metal-rich regime (e.g. \citealp{cai06}), we might expect that if FGK star secondaries formed primarily through 
fragmentation of the proto-planetary disk, there would be a strong dependence of the binary fraction with metallicity.  As 
this does not appear to be the case, then a flat distribution of binary fraction with metallicity suggests that these close binaries 
predominantly form through fragmentation of the giant molecular cloud that also formed the primary star.

\subsection{Summary}\label{summary}

Our target sample of 178 solar-type stars has revealed that $\sim$10\,
percent are spectroscopic binaries. Orbital solutions indicate that
two systems potentially have brown dwarf companions and another two
could have eccentricities that place them in the extreme upper tail of
the eccentricity distribution for binaries with periods less than
1000\,days. The systems with the largest quantity of data points appear to generate
the least robust orbital solutions, which could owe to secondary flux contamination of the template spectra, and 
hence the necessity to garner more data to constrain their solutions. 
When the radial velocity measurement errors are low, yet the Keplerian solutions have low significance, the most 
likely scenario is a multiple-star system. HIPPARCOS astrometry was examined in an attempt to constrain our orbital parameters, 
however no significant astrometric variation could be discerned in the positional residuals. The distribution of companion 
masses was examined for both the binaries and candidate exoplanets detected by the AAPS. For periods up to 12\,years the 
`steep' planetary and `flat' binary mass distributions mirror those seen by other surveys. Over a time scale equivalent to one 
orbital period of Jupiter, upwards of 30 exoplanet detections can be expected from our original sample of 178 stars. The 
discovery of these sixteen AAPS binaries from a sample of solar-type stars selected to have no
resolvable or known SB companions is a reminder that the data for even the relatively bright southern stars remain far from complete.

Finally, analysis of the metallicity-mass plane from planetary companions all the way up to stellar companions 
reveals a stark difference in the mean metallicities of these populations, with planets orbiting stars more metal-rich than 
stellar companions, in general.  The fraction that host these companions as a function of metallicity is also different, 
with the binary fraction being found to be flat (43$\pm$4\%) in the metallicity range between -1.0 to 0.6~dex.  
This is in contrast to the fraction of stars hosting giant planetary systems, which has a very low fraction until around a 
solar metallicity, at which point the fraction rises steeply following a power law.  The flat binary fraction across this wide 
range of metallicities is in agreement with recent hydrodynamical simulations of star formation through fragmentation 
of giant molecular clouds.  Such a result suggests this is the dominate formation mechanism for FGK-type binaries and 
not fragmentation of the proto-planetary disk that was left over from the formation of the primary star.

\section{Acknowledgements}

We acknowledge funding by CATA-Basal grant (PB06, Conicyt), from Fondecyt through grant 3110004, and partial support from Centro de Astrof\'\i sica FONDAP 15010003, the  GEMINI-CONICYT FUND and from the Comit\'e Mixto ESO-GOBIERNO DE CHILE. (JSJ). We thank the support of CONICYT-PFCHA/Doctorado Nacional-Chile (MD). We further acknowledge Australian governmental support of the Anglo-Australian Telescope as part of the Department of Industry and Science (CGT, BC) and NSF grant AST-9988087~(RPB). We also thank David Soderblom for granting the use of Ca~{\small II} H and K spectra taken at CTIO. This research has made use of the SIMBAD database, operated at CDS, Strasbourg, France.

\begin{table}
\centering
\begin{minipage}{300mm}
\epsfig{file=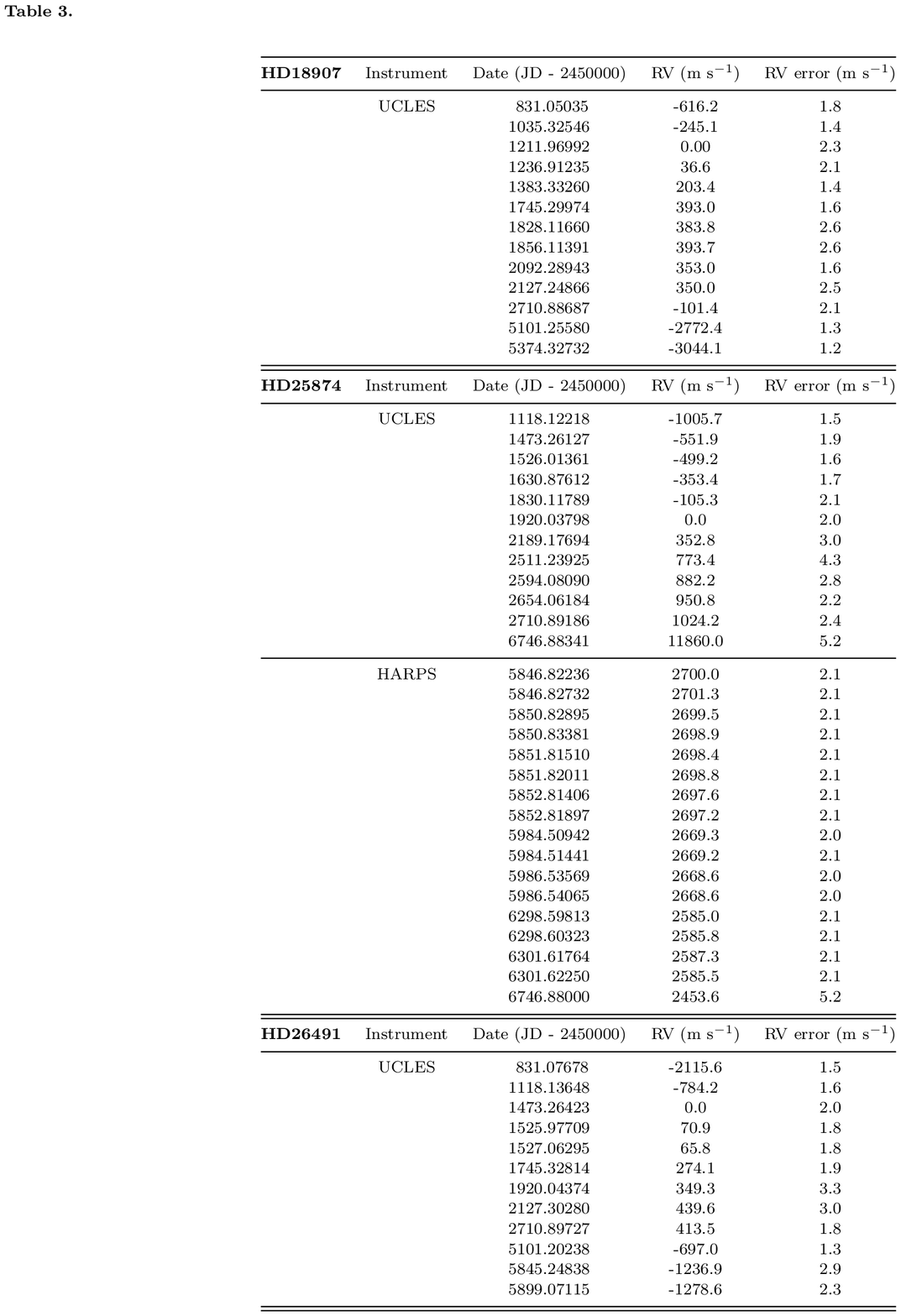,width=18cm, angle=0} \caption{}
\label{rvdata}
\end{minipage}
\end{table}
\begin{table}
\centering
\setcounter{table}{1}
\begin{minipage}{300mm}
\epsfig{file=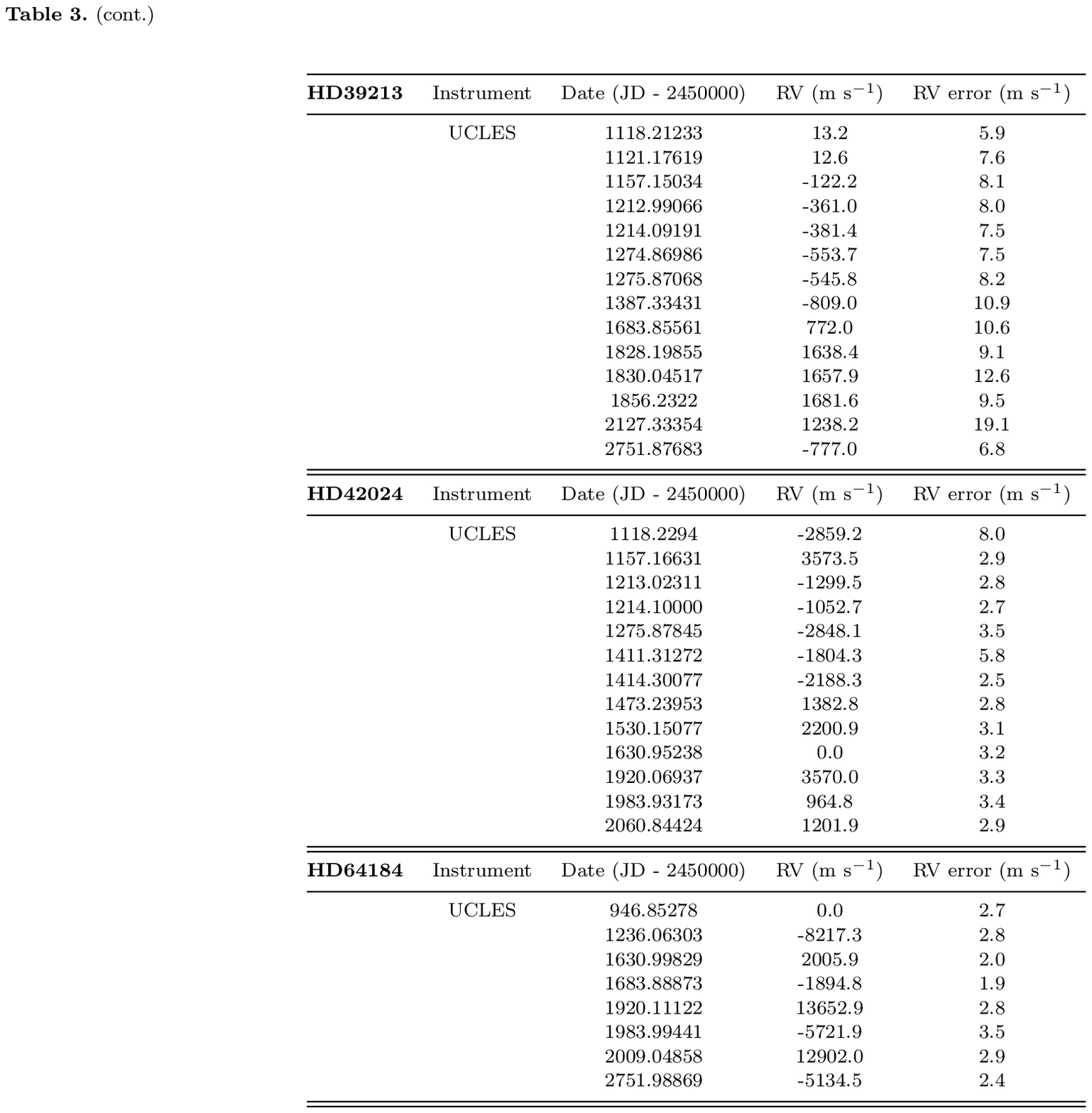,width=18cm, angle=0} 
\label{}
\end{minipage}
\end{table}
\begin{table}
\centering
\setcounter{table}{1}
\begin{minipage}{300mm}
\epsfig{file=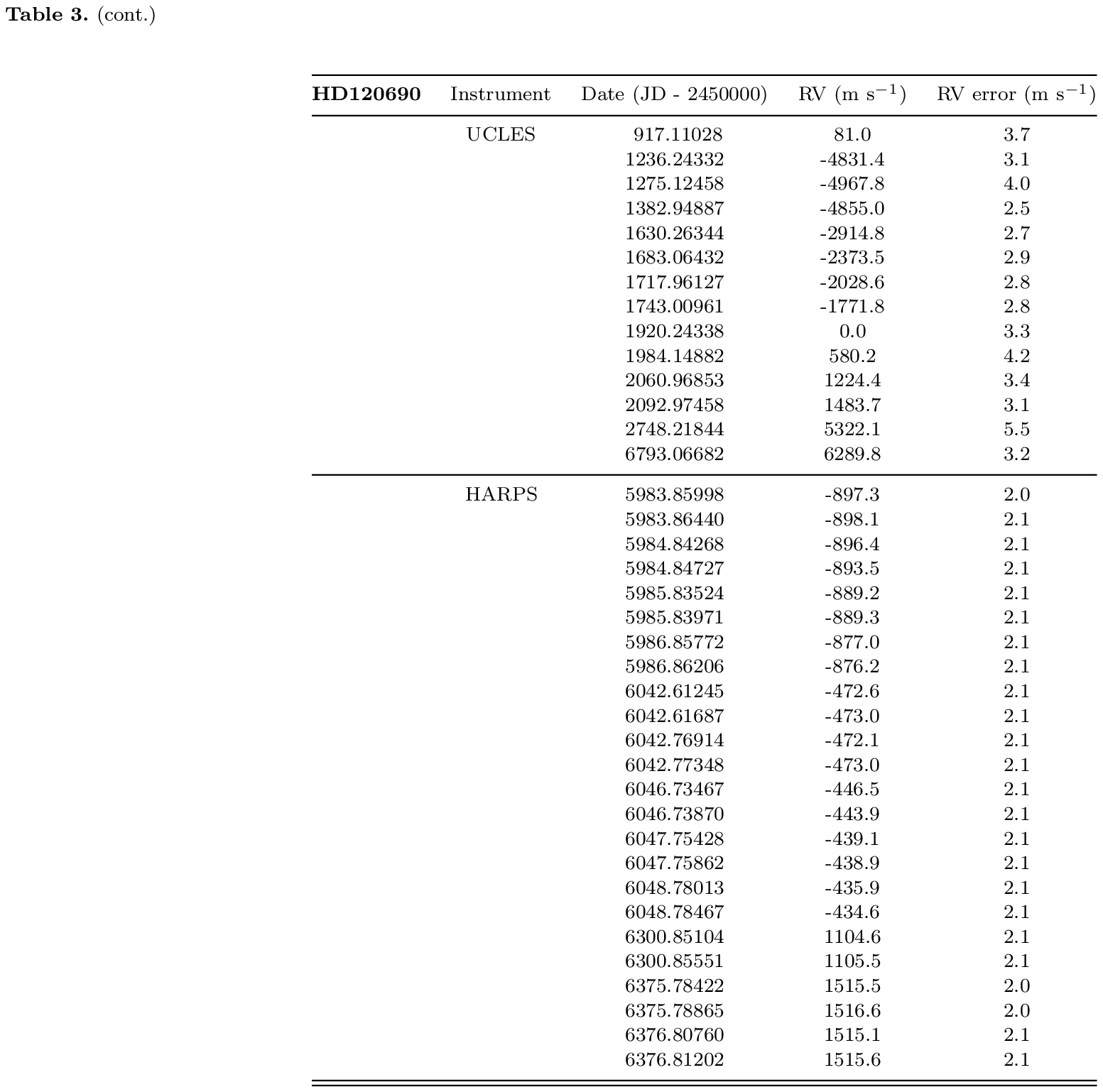,width=18cm, angle=0} \caption{}
\label{}
\end{minipage}
\end{table}
\begin{table}
\centering
\setcounter{table}{1}
\begin{minipage}{300mm}
\epsfig{file=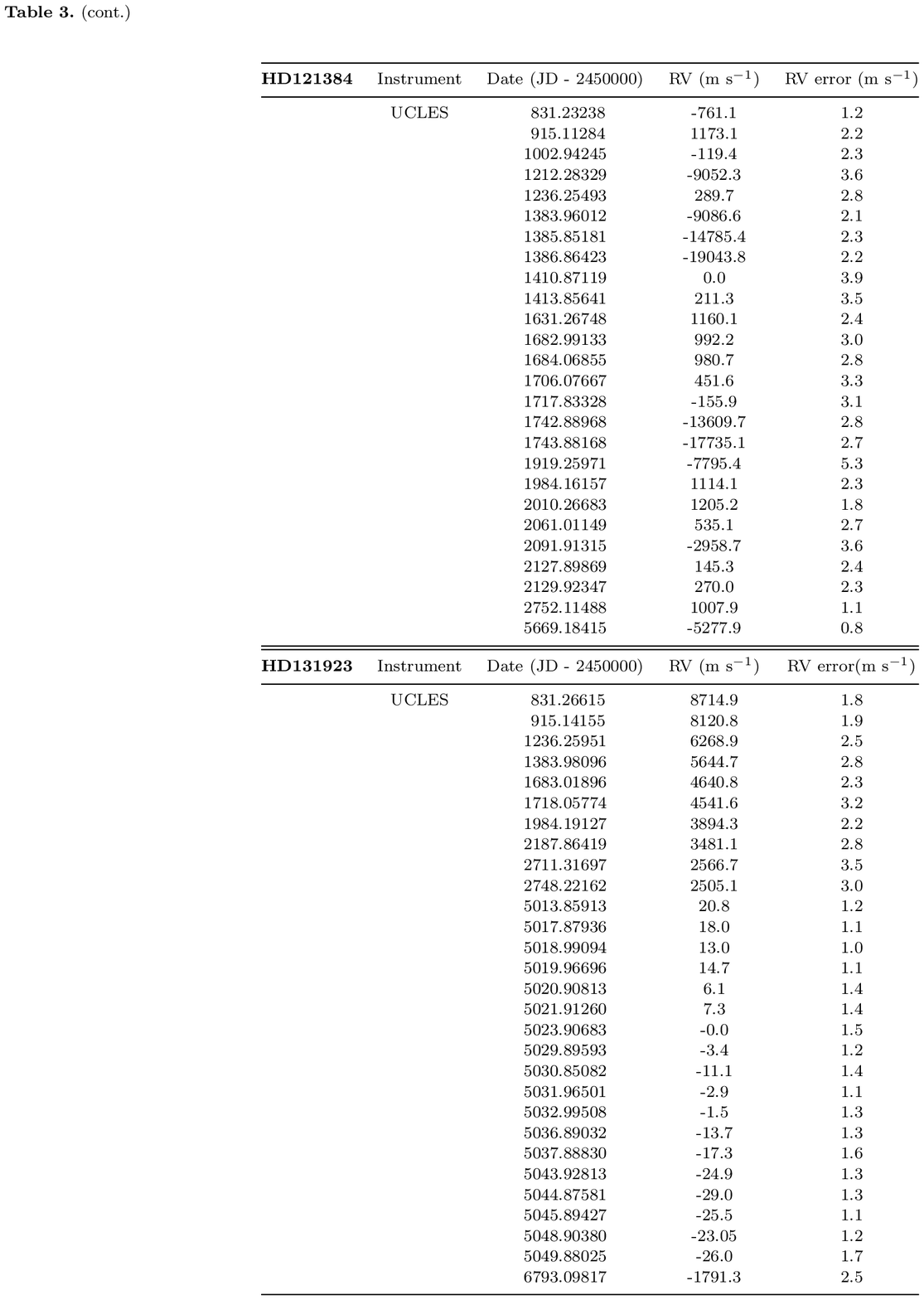,width=18cm, angle=0} \caption{}
\label{}
\end{minipage}
\end{table}
\begin{table}
\centering
\setcounter{table}{1}
\begin{minipage}{300mm}
\epsfig{file=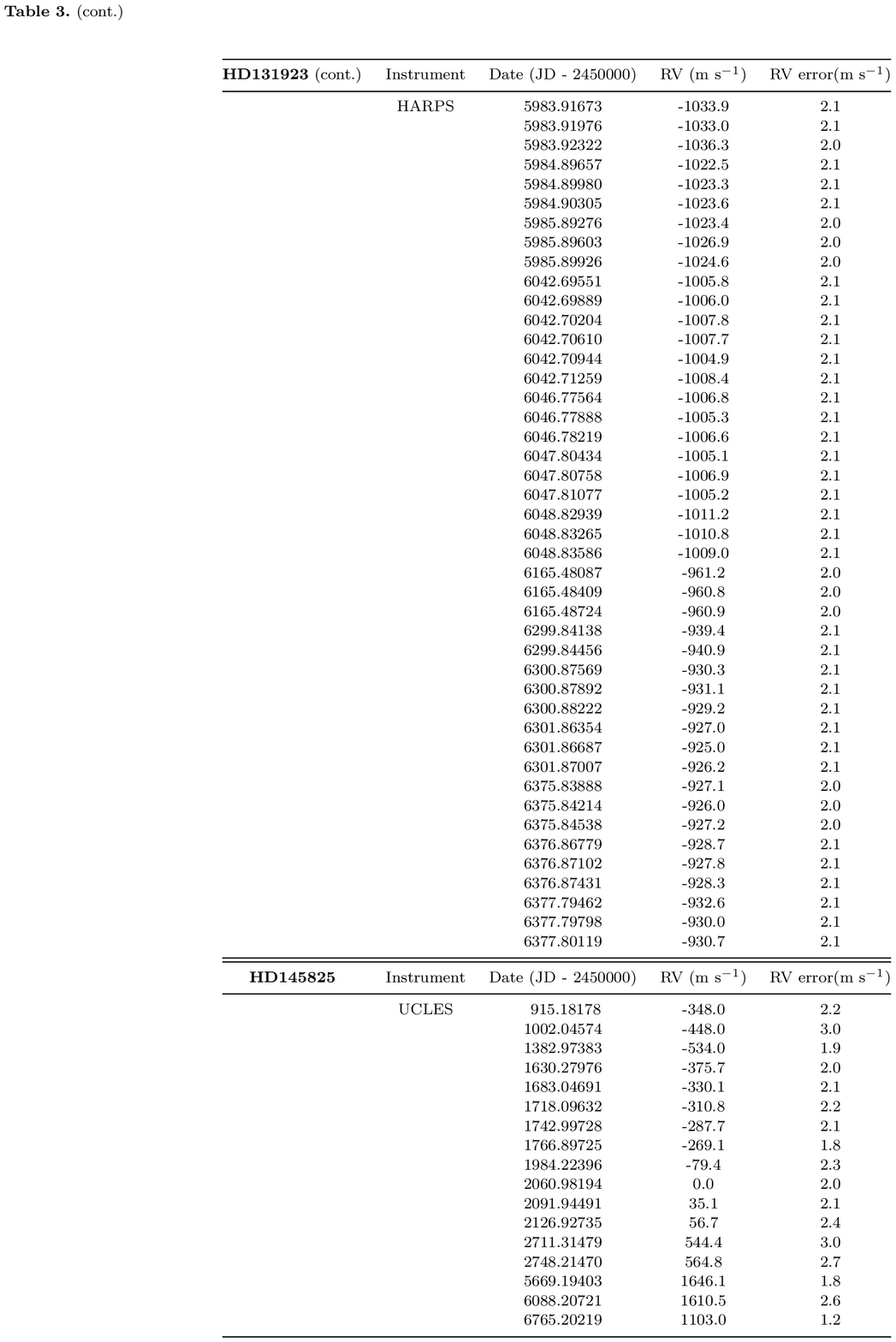,width=18cm, angle=0} \caption{}
\label{}
\end{minipage}
\end{table}
\begin{table}
\centering
\setcounter{table}{1}
\begin{minipage}{300mm}
\epsfig{file=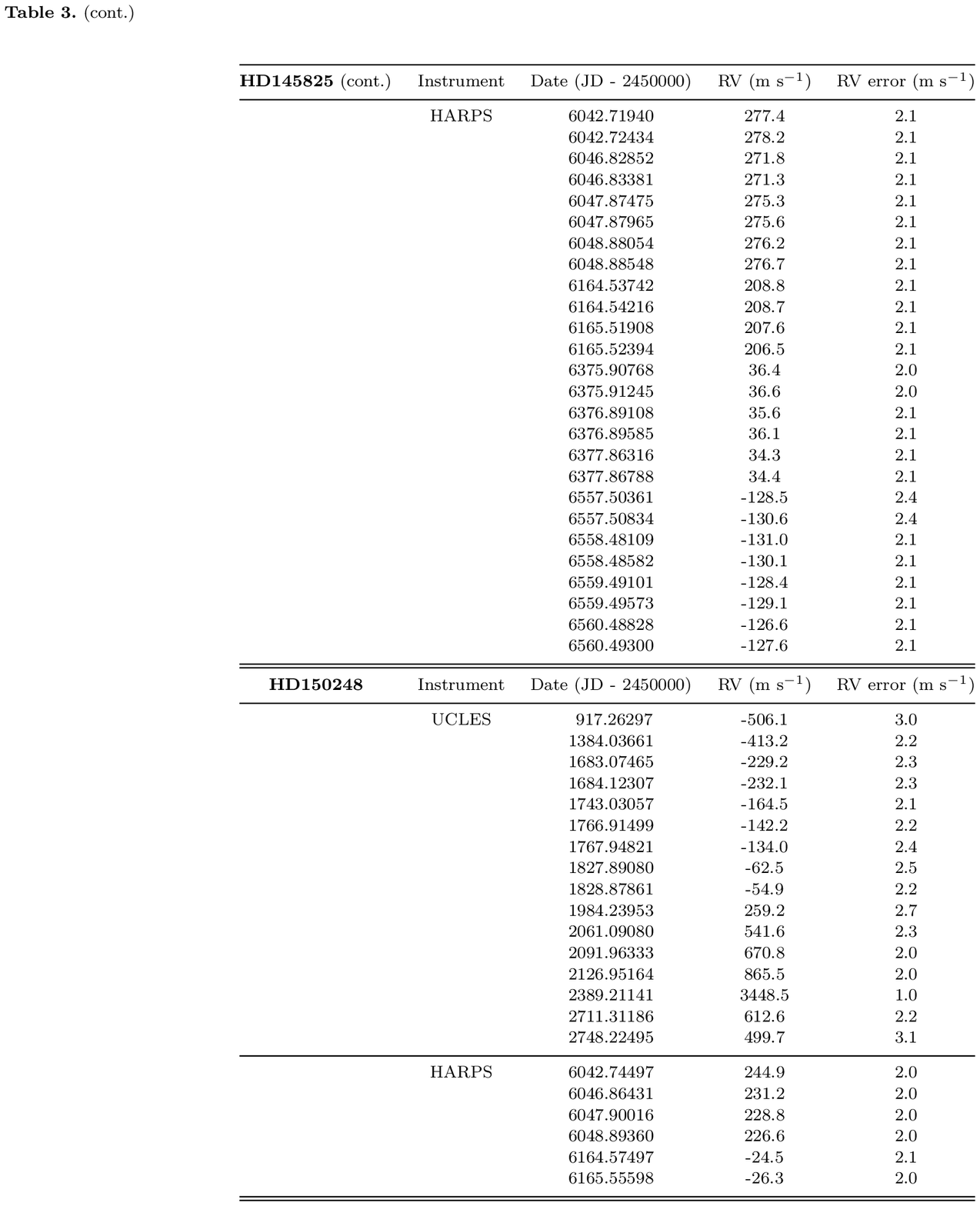,width=18cm, angle=0} \caption{}
\label{}
\end{minipage}
\end{table}
\begin{table}
\centering
\setcounter{table}{1}
\begin{minipage}{300mm}
\epsfig{file=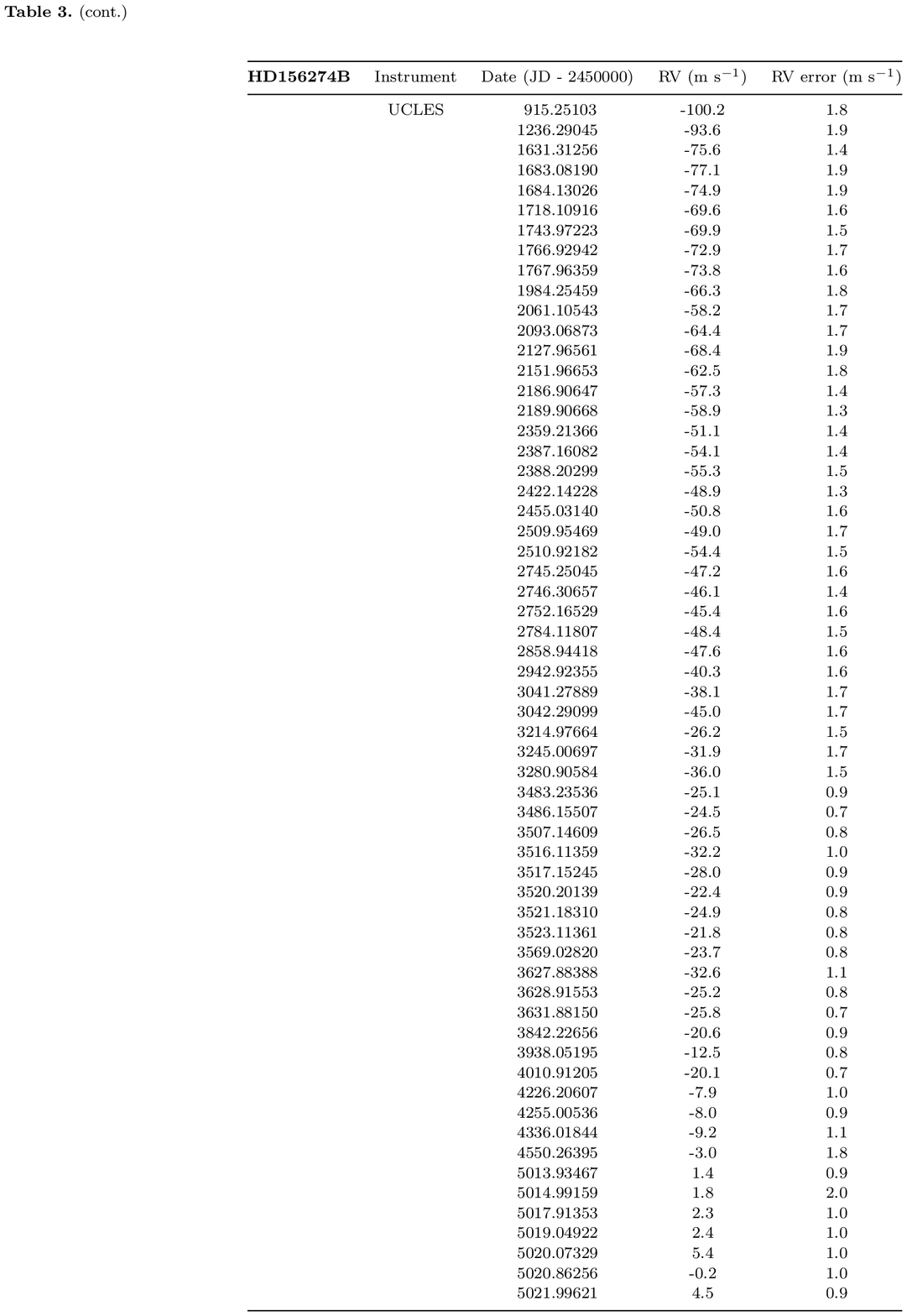,width=18cm, angle=0} \caption{}
\label{}
\end{minipage}
\end{table}
\begin{table}
\centering
\setcounter{table}{1}
\begin{minipage}{300mm}
\epsfig{file=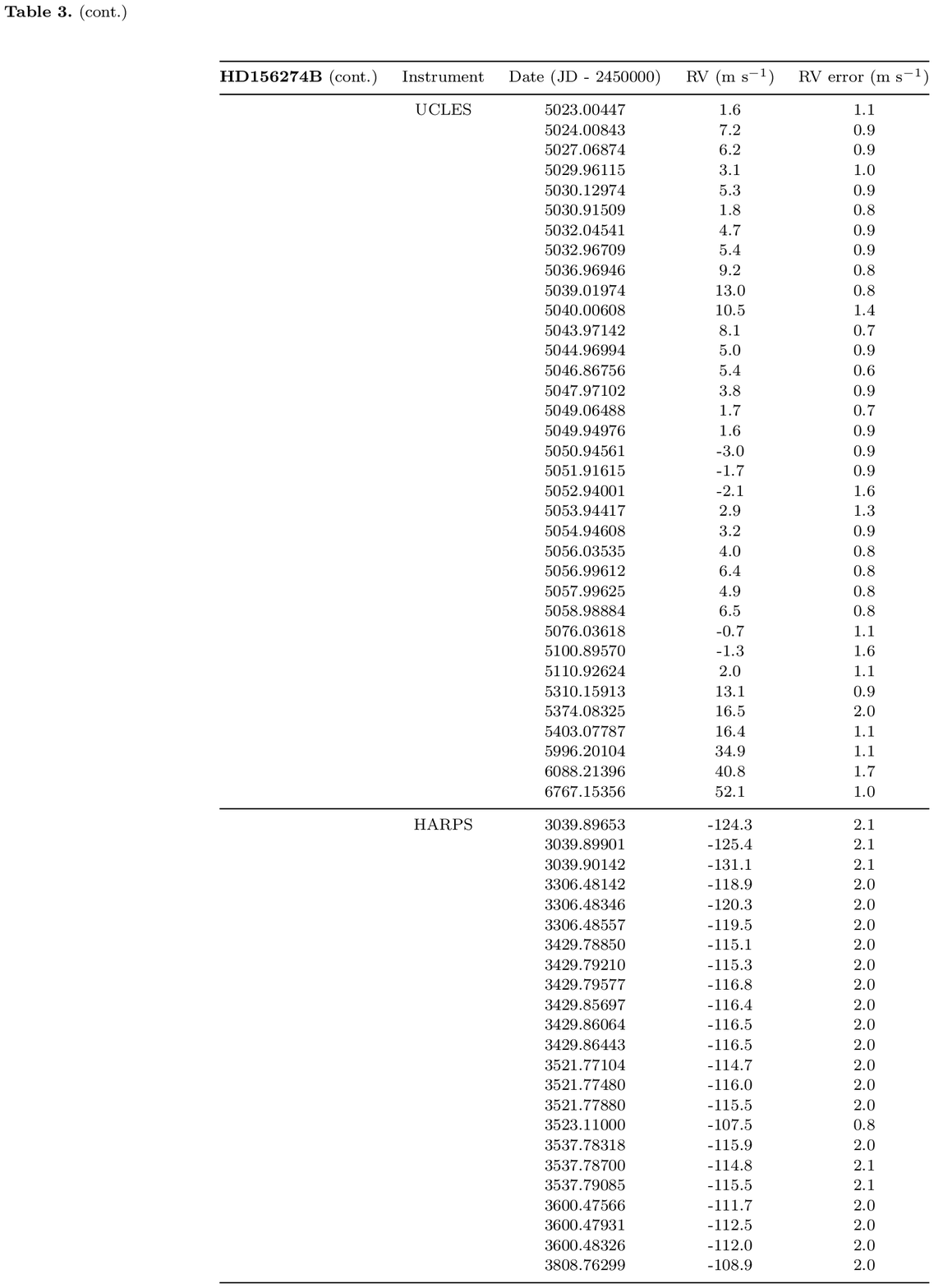,width=18cm, angle=0} \caption{}
\label{}
\end{minipage}
\end{table}
\begin{table}
\centering
\setcounter{table}{1}
\begin{minipage}{300mm}
\epsfig{file=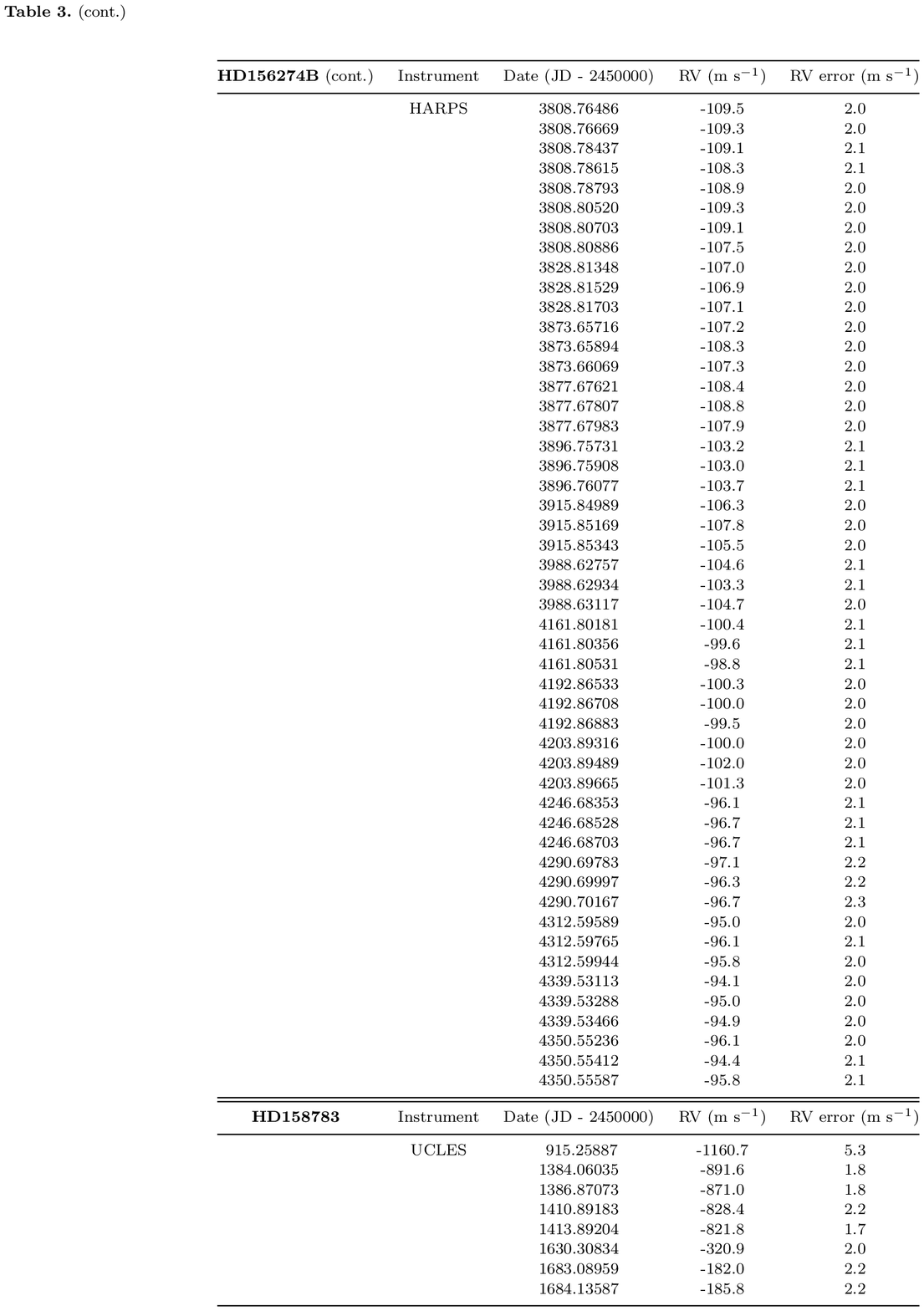,width=18cm, angle=0} \caption{}
\label{}
\end{minipage}
\end{table}
\begin{table}
\centering
\setcounter{table}{1}
\begin{minipage}{300mm}
\epsfig{file=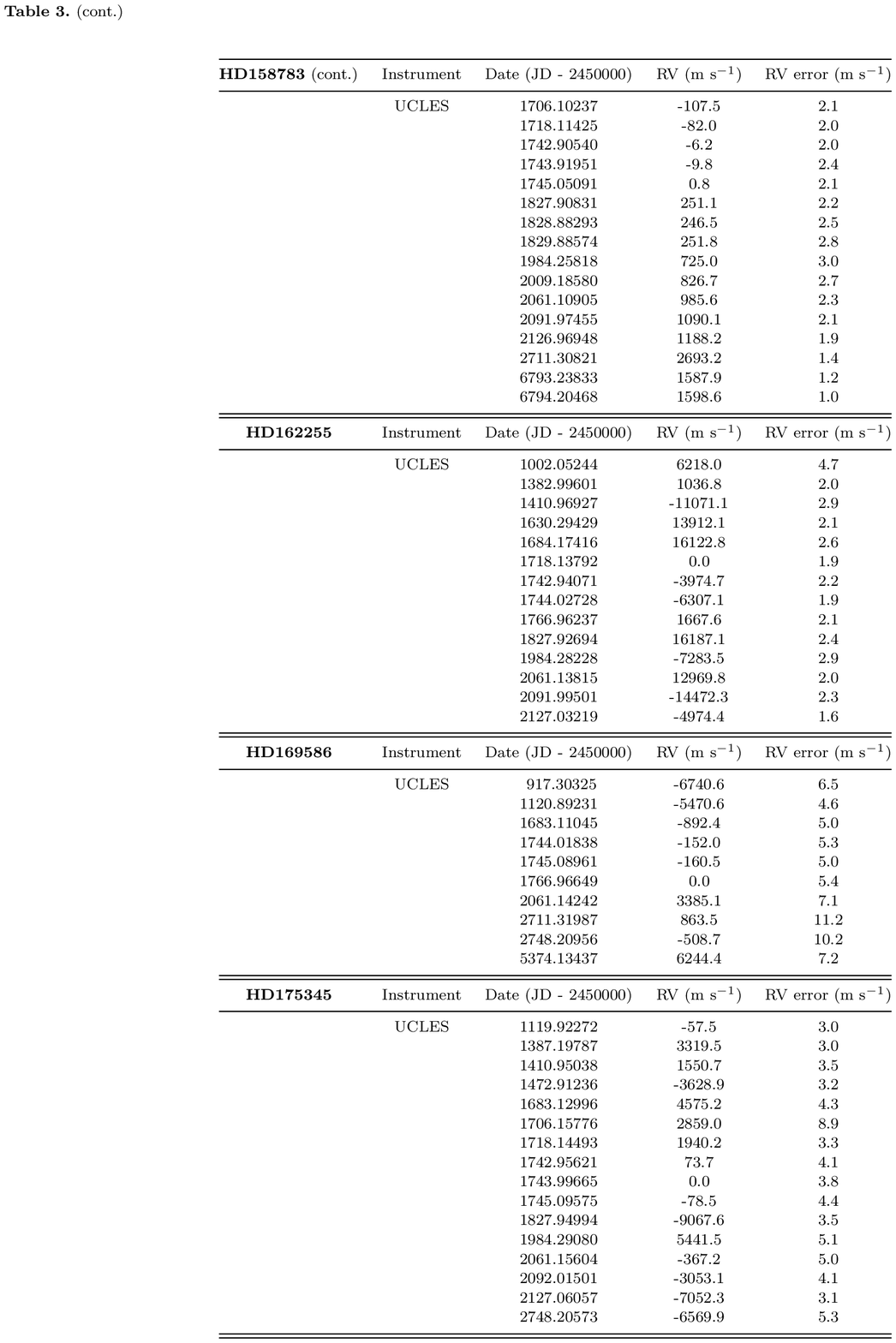,width=18cm, angle=0} \caption{}
\label{}
\end{minipage}
\end{table}

\end{document}